\documentclass[twocolumn]{aastex7}

\usepackage{threeparttable}
\usepackage{color}
\usepackage{amsmath}
\usepackage{ulem}
\usepackage{hyperref}

\definecolor{ForestGreen}{RGB}{34,139,34}
\def\tw#1 {{\textcolor{ForestGreen}{#1}}\ }
\def\blue#1 {{\textcolor{blue}{#1}}\ }

\definecolor{ForestGreen}{RGB}{34,139,34}
\def\tw#1 {{\textcolor{ForestGreen}{#1}}\ }
\def\blue#1 {{\textcolor{blue}{#1}}\ }
\def\red#1 {{\textcolor{red}{#1}}\ }

\received{\today}
\submitjournal{ApJS}

\shorttitle{ULTIMATE deblending I}
\shortauthors{Sun et al.}
\usepackage{amsmath} % for \dfrac{}{}

\begin{document}

\title{ULTIMATE-deblending I. A 50-band Ultraviolet to Mid-infrared Photometric Catalog Combining Space- and Ground-based Data in the JWST/PRIMER survey }   
%The ULTIMATE deblending catalog of the JWST/PRIMER survey. I. A 50-band UV-to-MIR photometric catalog combining space- and ground-based data
\author[0009-0007-0241-0213,sname=Sun,gname=Hanwen]{Hanwen Sun}
\affiliation{School of Astronomy and Space Science, Nanjing University, Nanjing 210093, China}
\affiliation{Key Laboratory of Modern Astronomy and Astrophysics, Nanjing University, Ministry of Education, Nanjing 210093, China}
\email{hanwensun@smail.nju.edu.cn}

\author[0000-0002-2504-2421,sname=Wang,gname=Tao]{Tao Wang}
\affiliation{School of Astronomy and Space Science, Nanjing University, Nanjing 210093, China}
\affiliation{Key Laboratory of Modern Astronomy and Astrophysics, Nanjing University, Ministry of Education, Nanjing 210093, China}
\email[show]{taowang@nju.edu.cn}

\author[0000-0002-8046-984X,sname=Xu,gname=Ke]{Ke Xu}
\affiliation{School of Astronomy and Space Science, Nanjing University, Nanjing 210093, China}
\affiliation{Key Laboratory of Modern Astronomy and Astrophysics, Nanjing University, Ministry of Education, Nanjing 210093, China}
\email{kexu@smail.nju.edu.cn}

\author[0000-0002-7631-647X,sname=Elbaz,gname=David]{David Elbaz}
\affiliation{Université Paris-Saclay, Université Paris Cité, CEA, CNRS, AIM, 91191, Gif-sur-Yvette, France}
\email{delbaz@cea.fr}

\author[0000-0001-6870-8900,sname=Merlin,gname=Emiliano]{Emiliano Merlin}
\affiliation{INAF – Osservatorio Astronomico di Roma, Via Frascati 33, 00078
Monteporzio Catone, Rome, Italy}
\email{emiliano.merlin@inaf.it}

\author[0000-0003-0202-0534,sname=Cheng,gname=Cheng]{Cheng Cheng}
\affiliation{Chinese Academy of Sciences South America Center for Astronomy, National Astronomical Observatories, CAS, Beijing 100101, People’s Republic of China}
\email{chengcheng@nao.cas.cn}

\author[0000-0002-3331-9590,sname=Daddi,gname=Emanuele]{Emanuele Daddi}
\affiliation{Université Paris-Saclay, Université Paris Cité, CEA, CNRS, AIM, 91191, Gif-sur-Yvette, France}
\email{emanuele.daddi@cea.fr}

\author[0000-0002-8412-7951,sname=Jin,gname=Shuowen]{Shuowen Jin}
\affiliation{Cosmic Dawn Center (DAWN), Denmark}
\affiliation{DTU-Space, Technical University of Denmark, Elektrovej 327, DK2800 Kgs. Lyngby, Denmark}
\email{shuowen.jin@gmail.com}

\author[0000-0003-2588-1265,sname=Wang,gname=Wei-hao]{Wei-hao Wang}
\affiliation{Academia Sinica Institute of Astronomy and Astrophysics (ASIAA), No. 1, Section 4, Roosevelt Rd., Taipei 10617, Taiwan}
\email{whwang@asiaa.sinica.edu.tw}

\author[0009-0001-7925-4876,sname=Chen,gname=Longyue]{Longyue Chen}
\affiliation{School of Astronomy and Space Science, Nanjing University, Nanjing 210093, China}
\affiliation{Key Laboratory of Modern Astronomy and Astrophysics, Nanjing University, Ministry of Education, Nanjing 210093, China}
\email{652023260002@smail.nju.edu.cn}

%\author{Mark Dickinson}
%\affiliation{NSF’s National Optical-Infrared Astronomy Research Laboratory, 950 North Cherry Avenue, Tucson, AZ 85719, USA}
%\email{mark.dickinson@NOIRLAB.EDU}

\author[0000-0003-3820-2823,sname=Fontana,gname=Adriano]{Adriano Fontana}
\affiliation{INAF – Osservatorio Astronomico di Roma, Via Frascati 33, 00078
Monteporzio Catone, Rome, Italy}
\email{adriano.fontana@oa-roma.inaf.it}

\author[0000-0003-1262-7719,sname=Gao,gname=Zhen-Kai]{Zhen-Kai Gao}
\affiliation{Academia Sinica Institute of Astronomy and Astrophysics (ASIAA), No. 1, Section 4, Roosevelt Rd., Taipei 10617, Taiwan}
\affiliation{Graduate Institute of Astronomy, National Central University, 300 Zhongda Road, Zhongli, Taoyuan 32001, Taiwan}
\email{zkgao@asiaa.sinica.edu.tw}

\author{Jiasheng Huang}
\affiliation{Chinese Academy of Sciences South America Center for Astronomy, National Astronomical Observatories, CAS, Beijing 100101, People’s Republic of China}
\email{jhuang@nao.cas.cn}

\author[0000-0002-6777-6490,sname=Magnelli,gname=Benjamin]{Benjamin Magnelli}
\affiliation{Université Paris-Saclay, Université Paris Cité, CEA, CNRS, AIM, 91191, Gif-sur-Yvette, France}
\email{Benjamin.magnelli@cea.fr}

\author[0009-0006-9595-4393,sname=Sangalli,gname=Valentina]{Valentina Sangalli}
\affiliation{Université Paris-Saclay, Université Paris Cité, CEA, CNRS, AIM, 91191, Gif-sur-Yvette, France}
\email{Valentina.SANGALLI@cea.fr}

\author[0000-0002-1010-7763,sname=Yijun,gname=Wang]{Yijun Wang}
\affiliation{School of Astronomy and Space Science, Nanjing University, Nanjing 210093, China}
\affiliation{Key Laboratory of Modern Astronomy and Astrophysics, Nanjing University, Ministry of Education, Nanjing 210093, China}
\email{wangyijun@nju.edu.cn}

\author[0009-0008-4971-035X,sname=Yang,gname=Tiancheng]{Tiancheng Yang}
\affiliation{School of Astronomy and Space Science, Nanjing University, Nanjing 210093, China}
\affiliation{Key Laboratory of Modern Astronomy and Astrophysics, Nanjing University, Ministry of Education, Nanjing 210093, China}
\email{652023260012@smail.nju.edu.cn}

\author[0000-0001-5757-5719,sname=Zhang,gname=Yuheng]{Yuheng Zhang}
\affiliation{School of Astronomy and Space Science, Nanjing University, Nanjing 210093, China}
\affiliation{Key Laboratory of Modern Astronomy and Astrophysics, Nanjing University, Ministry of Education, Nanjing 210093, China}
\email{yuheng@nju.edu.cn}

\author[0000-0003-1687-9665,sname=Zhou,gname=Luwenjia]{Luwenjia Zhou}
\affiliation{School of Astronomy and Space Science, Nanjing University, Nanjing 210093, China}
\affiliation{Key Laboratory of Modern Astronomy and Astrophysics, Nanjing University, Ministry of Education, Nanjing 210093, China}
\email{wenjia@nju.edu.cn}

\begin{abstract}

Our understanding of the early Universe has long been limited by biased galaxy samples selected through various color criteria. With deep JWST infrared imaging, mass-complete galaxy samples can now be studied up to $z \sim 8$ for the first time. However, recent work has revealed systematic uncertainties in measuring the physical properties of galaxies based solely on JWST/NIRCam and Hubble Space Telescope (HST) photometry, due to their limited wavelength coverage. This highlights the need for supplementary data, particularly in the rest-frame UV and near-infrared. 
Here we present the ULTIMATE-deblending project, which will eventually deliver self-consistent UV to Radio photometry for galaxies detected in deep JWST surveys, including both NIRCam and MIRI data. 
In this first paper, we release a 50-band photometric catalog spanning CFHT/U to JWST/MIRI F1800W, covering a total of 627.1 arcmin$^2$ across two JWST/PRIMER fields. We detail the reduction of the JWST imaging data, the photometric procedures, and the spectral-energy-distribution-fitting methodology used to derive the galaxy properties. Compared with photometry including only HST and JWST bands, the inclusion of deblended low-resolution photometry from ground-based telescopes 
improves the accuracy of photometric redshifts by $\sim$40\%, while reducing the outlier fraction by $\sim$60\%. This galaxy sample can serve as a key reference for statistical studies of galaxy formation and evolution in the early Universe. 
The UV-to-MIR catalogs and JWST mosaics from the ULTIMATE-deblending project have been made publicly available.
%(CFHT, Subaru, VISTA, UKIRT, and Magellan)
% by providing additional wavelength coverage in the rest frame UV to optical and additional narrow- and medium-bands to identify the emission lines and breaks. 
%These improved photometric redshifts are essential for the accurate determination of the number density of high-redshift galaxies. 

%\tw{may emphasize the bestup to XX bands}
\end{abstract}

%% Keywords should appear after the \end{abstract} command. 
%% See the online documentation for the full list of available subject
%% keywords and the rules for their use.
%\keywords{editorials, notices --- miscellaneous --- catalogs --- surveys}

%% From the front matter, we move on to the body of the paper.
%% Observe the use of the LaTeX \label
%% command after the \subsection to give a symbolic KEY to the
%% subsection for cross-referencing in a \ref command.
%% You can use LaTeX's \ref and \label commands to keep track of
%% cross-references to sections, equations, tables, and figures.
%% That way, if you change the order of any elements, LaTeX will
%% automatically renumber them.
%%
%% We recommend that authors also use the natbib \citep
%% and \citet commands to identify citations. The citations are
%% tied to the reference list via symbolic KEYs. The KEY corresponds
%% to the KEY in the \bibitem in the reference list below. 

%\tableofcontents

%%%%%%%%%%%%%%%%%%%%%%%%%%%%%%%%%%%%%%%%%%%%%%%%%%%%%%%%%%%%%%%%%%%%%%%%%%%%%%%%
\section{Introduction}
\label{sec:intro}
%%%%%%%%%%%%%%%%%%%%%%%%%%%%%%%%%%%%%%%%%%%%%%%%%%%%%%%%%%%%%%%%%%%%%%%%%%%%%%%%

%\blue{To put this catalog in a larger context, better start with the fact that previous census of (mass-complete samples) galaxies can only be done at $z < 3$ (with H or Ks-band selected samples, e.g., CANDELS or COSMOS (UltraVista), now with JWST we can push forward to $z \sim 8$. But the problem is the limited wavelength coverage of JWST/NIRCam... With PRIMER, we can really build a reference sample of high-z galaxies, by providing probably the most complete wavelength coverage to date.}

Deep blank-field surveys have transformed our understanding of galaxy formation and evolution by providing unbiased selections of galaxies in the distant Universe. Before the launch of the James Webb Space Telescope (JWST), deep surveys based on Hubble Space Telescope \citep[\textrm{HST; e.g., the CANDELS survey},][]{Galametz2013,Nayyeri2017} or ground-based telescopes \citep[\textrm{e.g., the UltraVISTA survey},][]{McCracken2012} can only provide unbiased mass-complete censuses at $z < 3$ and $\log ({M_{\rm{*}}}/{M_ \odot }) > 9$. At $z>3$, presence of a significant population of massive old and/or dusty galaxies that are HST-dark or near-infrared (NIR)-dark~\citep{Huang2011,Franco2018,WangT2019,Yamaguchi2019} requires deep imaging at longer wavelengths to obtain a mass-complete sample of galaxies. Meanwhile, imaging data with higher resolution is also required considering the high rate of foreground contaminants toward high-redshift galaxies~\citep{Liu2026}.
%  galaxies massive dusty galaxies can be so red that they can not be detected even in the deepest HST surveys \citep{Huang2011,Franco2018,WangT2019,Yamaguchi2019}, 
%while the low-mass galaxies can also be missed due to their faintness. 
Recently, JWST has led us into a new era of extragalactic astronomy, opening rest-frame NIR windows on galaxies at $z>3$ with unprecedented depth and resolution. With large JWST programs such as PRIMER \citep{Dunlop2021}, CEERS \citep{Finkelstein2023}, JADES \citep{Eisenstein2023}, UNCOVER \citep{Rachel2024}, and COSMOS-Web \citep{Casey2023}, mass-complete samples of high-redshift galaxies have been systematically obtained and analyzed \citep[e.g.][]{Papovich2023, Kartaltepe2023, YangG2023, Weibel2024, Shuntov2025} that are (mostly) based on catalogs incorporating JWST/NIRCam and HST data \citep[e.g.][]{Merlin2024}. In addition to extending the selection of a mass-complete sample of normal galaxies to higher redshifts, JWST observations also help to reveal new galaxy populations, most notably the Little Red Dots (LRDs) \citep[e.g.][]{PerezGonzalez2024, Kocevski2025, Labbe2025}. 

However, JWST surveys themselves have limited wavelength coverage, as they are mostly focused on NIR bands. Without comprehensive multiband constraints on galaxy spectral energy distributions (SEDs) from rest-frame UV to far-infrared (FIR) and radio, the estimates of photometric redshifts and physical properties of galaxies from JWST surveys remain highly uncertain~\citep{Song2023,WangBJ2024,WangT2025}. In rare cases, the most obscured galaxies at $z \gtrsim 6$ may still be missed in deep JWST NIR surveys~\citep{SunFW2025}. A key challenge in combining the multiband data from different telescopes is their different angular resolutions, ranging from $\sim 0.04$ arcseconds of JWST/NIRCam to $\sim 40$ arcseconds of Herschel/SPIRE. The ASTRODEEP team has developed a deblending framework with template-fitting photometry using high-resolution HST data as priors \citep{Merlin2015, Merlin2016}, and has applied it to self-consistently combine HST and ground-based data in the GOODS-South field \citep{Merlin2021} and the Frontier Fields survey around local galaxy clusters \citep{Lotz2014,Koekemoer2014AAS,Merlin2016B, Criscienzo2017}. Moreover, the super-deblending FIR photometry of GOODS-North \citep{Liu2018} and COSMOS \citep{Jin2018} has also proven the great potential of template-fitting multiband photometry in revealing the blended fluxes at the low-resolution FIR to radio bands. To fully exploit the new JWST datasets, it is therefore essential to perform deblending photometry from UV to radio for a mass-complete sample of JWST-detected galaxies.

Among the recent JWST surveys, the PRIMER survey \citep{Dunlop2021} has provided one of the deepest JWST datasets with the largest survey area in the PRIMER-COSMOS and PRIMER-UDS deep fields. In particular, the deep and homogeneous JWST/MIRI coverage of the two PRIMER fields has been proven to be essential for robust stellar mass constraints of massive galaxies in the early Universe \citep{WangT2025}. In addition to JWST, both PRIMER fields are covered by abundant multiwavelength data collected from a number of deep surveys in the last decade, including CFHT \citep{WangWH2022}, Subaru \citep{Aihara2022}, Magellan \citep{Straatman2016}, VISTA \citep{McCracken2012}, UKIRT (O.Almaini et al., W.Hartley et al., in preparation), Spitzer \citep{Steinhardt2014,Mehta2018}, Herschel, SCUBA-2 \citep{WangWH2017, GaoZK2024}, VLA \citep{Smolvcic2017} and MeerKAT \citep{Hale2025}. 
These rich multi-wavelength data, combined with the new JWST observations, provide a comprehensive sampling of galaxy SEDs from UV to Radio. This enables accurate constraints on their physical properties, including stellar masses, star formation rates traced by both UV and infrared emission, and emission line strengths inferred from narrow- and medium-band data.  
%a chance to obtain the full and accurate views of high-redshift galaxies, including their young stars from UV to optical data, stellar masses from NIR to MIR data, dust and AGN emissions from FIR to Radio data, and the emission lines identified from the narrow- and medium-bands data. 
%However, these multi-wavelength data have different resolutions ranging from $\sim 0.04$ arcseconds of JWST/NIRCam to $\sim 40$ arcseconds of Herschel/SPIRE, inhabiting the direct combination of these data from different telescopes.
%Especially, in the Radio and Far-Infrared (FIR) bands, the PRIMER fields can benefit from the deep MeerKAT data at $\sim 1.2-1.3$GHz from the MeerKAT International GHz Tiered Extragalactic Exploration Survey (MIGHTEE) \citep{Hale2025}, the deep SCUBA-2 observations at $450{\rm{\mu m}}$ and $850{\rm{\mu m}}$ from the SCUBA-2 Ultra Deep Imaging EAO Survey (STUDIES) \citep{WangWH2017, GaoZK2024}, and the deep Herschel/PACS and SPIRE data between $100-500{\rm{\mu m}}$.

Here we present the ULTIMATE-deblending project, which will eventually provide photometric catalogs across UV to radio for JWST deep fields (starting with the two PRIMER fields). In this first paper of the ULTIMATE-deblending project, we present the reduction of JWST data in the two PRIMER fields and the construction of the UV to mid-infrared (MIR) catalog, which combine the high-resolution HST and JWST/NIRCam data with the low-resolution data from JWST/MIRI, Spitzer/IRAC, and ground-based surveys, deblended using the template-fitting technique developed by the ASTRODEEP team \citep{Merlin2015, Merlin2016}. These additional data from low-resolution bands can help to improve the accuracy of photometric redshifts by providing additional wavelength coverage in UV to MIR (low-resolution data in the UV-optical is especially helpful since the Lyman break cannot be constrained by JWST/F090W at $z \lesssim 4.5$) and additional narrow and medium bands that emphasize the breaks and emission lines. All JWST mosaics and the multi-wavelength catalog presented in this paper are publicly available on \dataset[http://www.taoofcosmos.space/ultimate/]{http://www.taoofcosmos.space/ultimate/.}

In future works of the ULTIMATE-deblending project, we will present a deblended FIR to radio catalog in the two PRIMER fields (H.Sun et al., in prep) and then extend our project to other JWST deep fields. We aim to present state-of-the-art catalogs with the most complete wavelength coverage to deliver accurate photometric redshifts and physical properties constrained by the full UV to Radio SED fitting. These catalogs will directly provide robust abundance and properties of massive galaxies up to $z\sim8.5$, enabling accurate studies of galaxy formation and evolution in the early Universe. Moreover, this accurate and mass-complete reference sample of high-redshift galaxies bears potential application with machine learning algorithms \citep{Masters2015, Wright2020, Roster2025, WangZH2026} to improve the modeling of galaxies in surveys with fewer bands or lower depth, such as the Euclid survey \citep{Euclid2025} with shallower data but a much larger field of view than JWST.

This paper is organized as follows. Section~\ref{sec:data} lists the UV to MIR dataset collected in our work and the reduction of JWST data. Section~\ref{sec:detection} presents the methods for source detection and multiwavelength photometry. Section~\ref{sec:SED} shows the details of the SED fitting, which is used to derive photometric redshifts and physical properties. Section~\ref{sec:discussion} discusses that the low-resolution ground-based data can significantly improve the accuracy of photometric redshifts. All magnitudes presented in this paper are given in the AB system \citep{Oke1983}. We assume a cosmological model with ${H_0}{\rm{ = }}70~{\rm{km}} \cdot {{\rm{s}}^{ - 1}} \cdot {\rm{Mp}}{{\rm{c}}^{ - 1}}$, ${\Omega _{\rm{m}}} = 0.3$ and ${\Omega _\Lambda } = 0.7$, and adopt a \citet{Kroupa2001} initial mass function during the SED fitting.

%%%%%%%%%%%%%%%%%%%%%%%%%%%%%%%%%%%%%%%%%%%%%%%%%%%%%%%%%%%%%%%%%%%%%%%%%%%%%%%%
\section{Data}
\label{sec:data}

\begin{figure*}[htbp]
\centering
\includegraphics[width=1\textwidth]{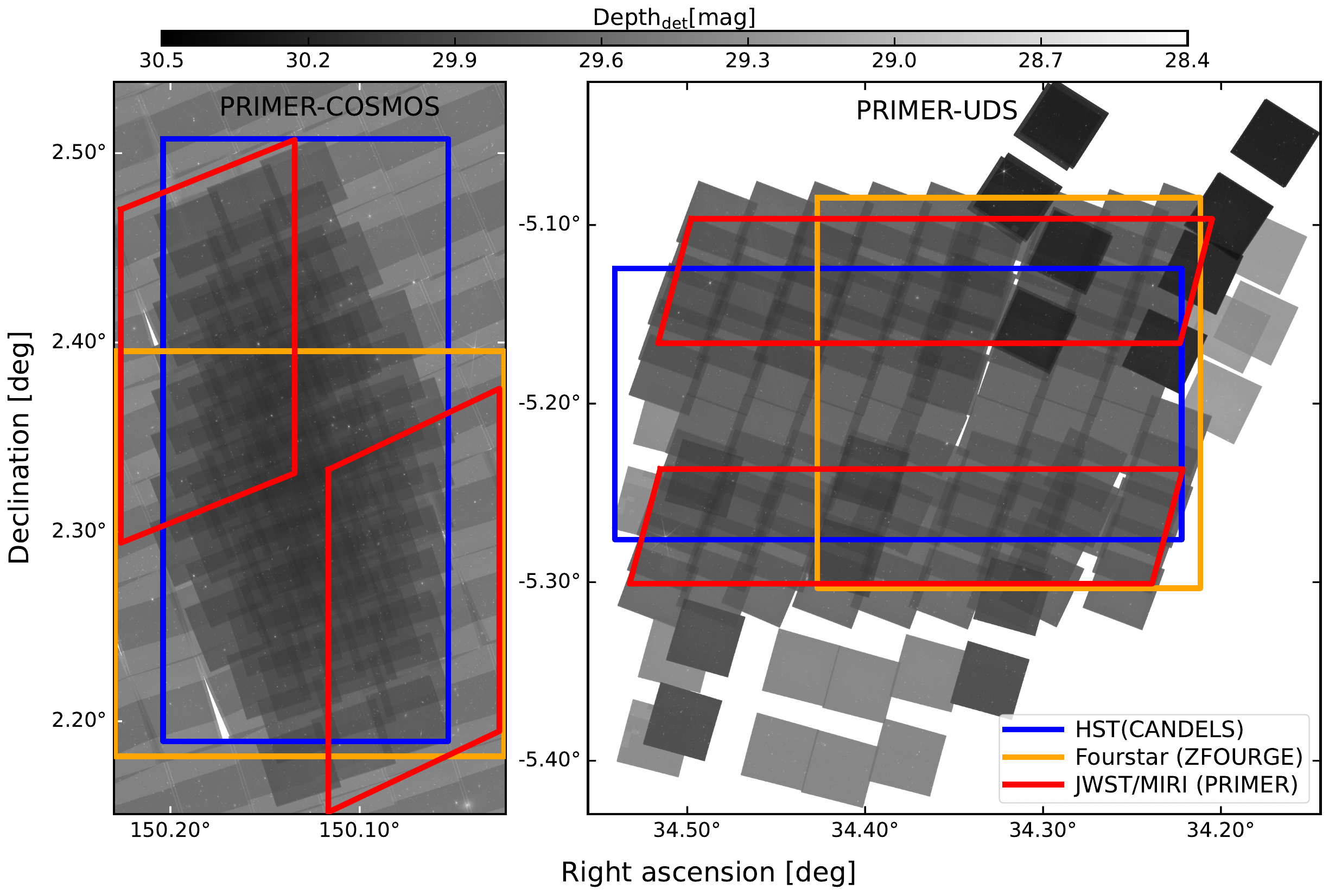}
\caption{\label{Fig:coverage}\textbf{The coverage of the key surveys within PRIMER-COSMOS (left panel) and PRIMER-UDS (right panel).} The background grayscale image presents the 3$\sigma$ limiting depth (in 0.2" diameter apertures) for point sources of the stacked JWST/NIRCam image, which is used for source detection. The blue, orange, and red lines show the coverage of the HST data from the CANDELS survey \citep{Galametz2013,Nayyeri2017}, the Magellan/Fourstar medium-bands data from the ZFOURGE survey \citep{Straatman2016}, and the JWST/MIRI data taken with the PRIMER survey. The boundary of our mosaics in PRIMER-COSMOS is determined by the coverage of the JWST/MIRI data from PRIMER \citep{Dunlop2021} and the deep SCUBA-2 data from STUDIES \citep{WangWH2017,GaoZK2024}.} 
\end{figure*}
%%%%%%%%%%%%%%%%%%%%%%%%%%%%%%%%%%%%%%%%%%%%%%%%%%%%%%%%%%%%%%%%%%%%%%%%%%%%%%%%
\subsection{Overview of the included data}

\begin{figure*}[htbp]
\centering
\includegraphics[width=1\textwidth]{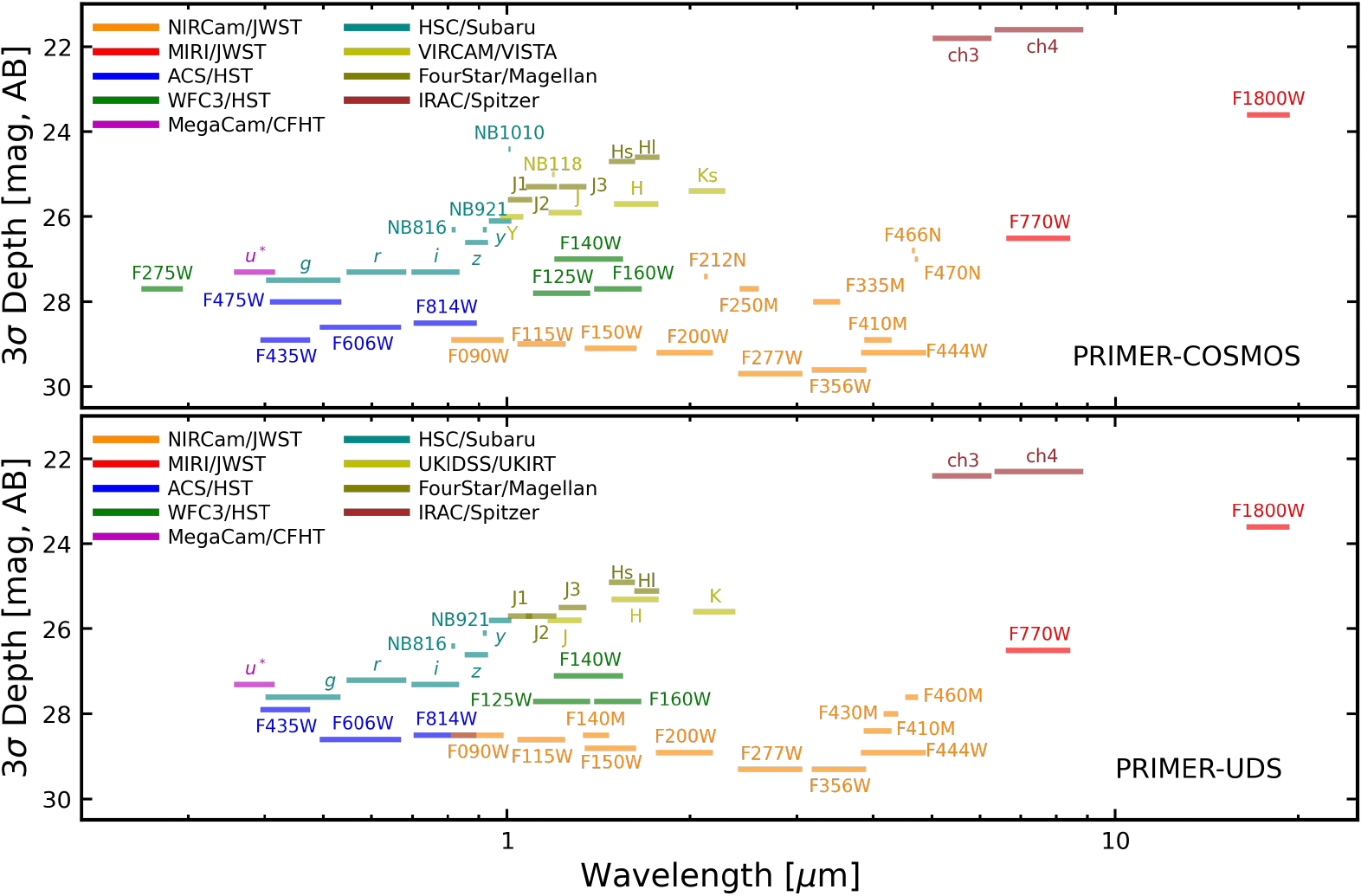}
\caption{\label{Fig:depth}\textbf{The 3 $\sigma$ depth and wavelength range of the multiband data used by PRIMER-COSMOS (upper panel) and PRIMER-UDS (lower panel).} All depths of point sources are measured using empty circular apertures and have been corrected to consider the aperture losses. We use the apertures with the highest sensitivity (among 0.2", 0.5", and Kron apertures) for JWST and HST data, and apertures with $d = 3.5 \times {\rm{FWH}}{{\rm{M}}_{{\rm{PSF}}}}$ for the other low-resolution data.}
\end{figure*}

This work focuses on the PRIMER-COSMOS and PRIMER-UDS fields. These two fields are covered by the deep JWST/NIRCam and MIRI observations from the Public Release IMaging for Extragalactic Research \citep[\textrm{PRIMER},][]{Dunlop2021} survey.
%as well as the deep SCUBA-2/450 ${\rm{\mu m}}$ and 850 ${\rm{\mu m}}$ data from the SCUBA-2 Ultra Deep Imaging EAO Survey (STUDIES, \citep{WangWH2017,GaoZK2024}). Taken together, these datasets 
Moreover, we further collect other available data in these two fields that have sufficiently useful depth or wavelength coverage, and have area overlap with our JWST mosaics (For example, we do not use the data from VIDEO \citep{Jarvis2013} because the new NIR images released by the UKIDSS data release 11 (O.Almaini et al., W.Hartley et al., in preparation) have provided deeper NIR data in the same bands with UKIRT). The combination of these data yields a 50-band dataset to construct a multiwavelength photometric catalog. Figure~\ref{Fig:coverage} presents the coverage of our survey, which covers 287.3 arcmin$^2$ area in PRIMER-COSMOS and 339.8 arcmin$^2$ area in PRIMER-UDS with the combined JWST dataset shown in Section~\ref{sec:JWST_dataset}. Figure~\ref{Fig:depth} presents the 3 $\sigma$ depth of point sources for all 50 bands used by our work, and the detailed depths, areas, and effective wavelengths are listed in Table~\ref{tab:band_infos}.

\subsubsection{JWST data}
\label{sec:JWST_dataset}
The PRIMER survey \citep{Dunlop2021} provides large and deep JWST/NIRCam (F090W, F115W, F150W, F200W, F277W, F356W, F410M, and F444W) and MIRI (F770W and F1800W) data for PRIMER-COSMOS and PRIMER-UDS. Additionally, we also collect JWST data from many other surveys that overlap with PRIMER-COSMOS or PRIMER-UDS: the COSMOS-Web survey \citep{Casey2023} provides more exposures for the entire PRIMER-COSMOS field with F115W, F150W, F277W, F444W, and F770W (partly covered); the JWST Emission Line Survey \citep[\textrm{JELS},][]{Duncan2025} provides narrow-band observations (F212N, F466N, and F470N) in the central 63 arcmin$^2$ area of PRIMER-COSMOS; the PANORAMIC survey \citep{Williams2025} and the BEACON survey \citep{Morishita2025} provide additional NIRCam imaging data for both PRIMER-COSMOS and PRIMER-UDS; the JWST project GO1840 \citep{Alvarez-Marquez2021} provides medium bands (F250M and F335M) observations for two pointings in PRIMER-COSMOS. All of these JWST data are reduced by our modified JWST pipeline (Section~\ref{sec:data_reduction}), and more JWST data observed after 2025 March will be added in a future data release.
%By May 2026, both PRIMER-COSMOS and PRIMER-UDS will also have the large ($\sim 378~{\rm{arcmi}}{{\rm{n}}^2}$) JWST data in 8 medium-bands from the MINERVA survey \citep{Muzzin2025}, and we plan to combine these new data into our project in the future data release. 

\subsubsection{Archival Ultraviolet to Mid-infrared data}
Both PRIMER-COSMOS and PRIMER-UDS have been observed with HST/WFC3 (F125W and F160W) and ACS (F606W and F814W) as part of the Cosmic Assembly Near-infrared Deep Extragalactic Legacy Survey \citep[\textrm{CANDELS},][]{Galametz2013,Nayyeri2017}. We use the HST mosaics from The DAWN JWST Archive (DJA, v7.0), which additionally combines the HST data from the Cosmic Evolution Survey \citep{Scoville2007} and the UVCANDELS survey \citep{Mehta2024}.

Apart from the high-resolution data from JWST and HST, multiband data from ground-based telescopes and Spitzer are also used by our catalog with deblending photometry (Section~\ref{sec:tphot_photometry}), including the ${u^*}$-band data taken with CFHT from the MegaCam Ultra-deep Survey \citep[\textrm{MUSUBI},][]{WangWH2022}; eight bands ($g$, $r$, $i$, $z$, $y$, NB816, NB921, and NB1010) from the third data release of the Hyper Suprime-Cam Subaru Strategic Program \citep{Aihara2022}; five medium bands (Magellan/Fourstar J1, J2, J3, Hl, and Hs) taken with the ZFOURGE survey \citep{Straatman2016}, which covers one-half of both PRIMER-COSMOS and PRIMER-UDS; NIR images ($Y$, NB118, $J$, $H$, and $Ks$) for PRIMER-COSMOS from the fifth data release of the UltraVISTA near-infrared imaging survey \citep{McCracken2012}; NIR images ($J$, $H$, and $K$) for PRIMER-UDS from the UKIDSS Ultra-Deep Survey data release 11 (O.Almaini et al., W.Hartley et al., in preparation); and the Spitzer/IRAC 5.8 ${\rm{\mu m}}$ and 8.0 ${\rm{\mu m}}$ mosaics reduced by the SPLASH team \citep{Steinhardt2014,Mehta2018}, which was observed as part of the S-COSMOS~\citep{Sanders2007} and SpUDS~\citep{SpUDS2020} survey.

\begin{table*}[!tbh]\footnotesize
\centering
\begin{minipage}[center]{\textwidth}
\centering
\caption{UV to MIR data used in the photometric catalog}
\label{tab:band_infos}
\begin{tabular}{lcccccccc}
\hline\hline
Instrument & Band & ${\lambda _{{\rm{eff}}}}$ & Bandwidth$^a$ & ${\rm{Depth}}_{{\rm{COS}}}^b$ & ${\rm{Area}}_{{\rm{COS}}}$ & ${\rm{Depth}}_{{\rm{UDS}}}^b$ & ${\rm{Area}}_{{\rm{UDS}}}$ & FWHM\\
/Telescope &  &  [\AA{}] & [\AA{}] & (mag) & (arcmin$^2$) & (mag) & (arcmin$^2$) & (\arcsec) \\
\hline
%print(f"AREA:{s*0.04*0.04/3600} arcmin2")
NIRCam & F090W & 8985 & 1773 & 28.9/27.9/27.1$^c$ & 146.2 & 28.5/27.5/26.8 & 268.1 & 0.033 \\
/JWST & F115W & 11434 & 2055 & 29.0/28.0/27.1$^d$ & 287.5 & 28.6/27.6/26.9 & 282.1 & 0.040 \\
 & F140M & 14023 & 1367 & - & - & 28.5/27.6/26.9 & 76.1 & 0.048 \\
 & F150W & 14873 & 2890 & 29.1/28.1/27.3 & 287.2 & 28.8/27.8/27.1 & 281.8 & 0.050 \\
 & F200W & 19680 & 4190 & 29.2/28.2/27.4 & 146.3 & 28.9/27.9/27.2 & 264.0 & 0.066 \\
 & F212N & 21212 & 274 & 27.4/26.4/25.6 & 63.1 & - & - &0.072 \\
 & F250M & 25006 & 1783 & 27.7/26.9/26.1 & 10.1 & - & - &0.085 \\
 & F277W & 27279 & 6615 & 29.7/28.7/27.9 & 287.3 & 29.3/28.4/27.7 & 287.5 & 0.092 \\
 & F335M & 33537 & 3389 & 28.0/27.1/26.4 & 10.1 & -& -& 0.111 \\
 & F356W & 35287 & 7239 & 29.6/28.7/28.0 & 147.4 & 29.3/28.5/27.8 & 287.9 & 0.116 \\
 & F410M & 40723 & 4263 & 28.9/28.1/27.5 & 147.4 & 28.4/27.8/27.2 & 255.9 & 0.137 \\
 & F430M & 42784 & 2295 & - & - & 28.0/27.2/26.7 & 81.0 & 0.144 \\
 & F444W & 43504 & 10676 & 29.2/28.5/27.9 & 286.9 & 28.9/28.2/27.5 & 287.6 & 0.145 \\
 & F460M & 46270 & 2309 & - & - & 27.6/26.9/26.4 & 78.7 & 0.157 \\
 & F466N & 46540 & 535 & 26.8/26.2/25.7 & 63.0 & - & - &0.158 \\
 & F470N & 47078 & 510 & 27.0/26.4/25.8 & 63.0 & - & - &0.160 \\
\hline
MIRI & F770W & 75225 & 18278 & 26.5/26.4/25.8 & 194.3 & 26.5/26.4/25.8 & 128.7 & 0.269 \\
/JWST & F1800W & 178734 & 29146 & 23.1/23.6/23.6 & 111.2 & 23.0/23.6/23.6 & 126.9 & 0.591 \\
\hline
ACS & F435W & 4342 & 822 & 28.9/27.9/27.2 & 126.8 & 27.9/26.9/26.3 & 185.2 & 0.08 \\
/HST & F475W & 4709 & 1272 & 28.0/27.1/26.5 & 85.9 & - & - & 0.08 \\
 & F606W & 5809 & 1772 & 28.6/27.8/27.1 & 225.5 & 28.6/27.8/27.1 & 192.6 & 0.08 \\
 & F814W & 7973 & 1889 & 28.5/27.6/27.0 & Full & 28.5/27.6/26.9 & 204.7 & 0.09 \\
\hline
WFC3 & F275W & 2720 & 424 & 27.7/26.8/26.1 & 87.8 & - & - & 0.07 \\
/HST & F125W & 12364 & 2674 & 27.8/27.4/26.8 & 182.2 & 27.7/27.4/26.8 & 177.3 & 0.16 \\
 & F140W & 13735 & 3570 & 27.0/26.6/26.1 & 135.5 & 27.1/26.7/26.2 & 123.1 & 0.17 \\
 & F160W & 15278 & 2750 & 27.7/27.3/26.8 & 199.2 & 27.7/27.4/26.9 & 187.3 & 0.17 \\
\hline
MegaCam & $u^*$ & 3858 & 598 & 27.3 & Full & 27.3 & Full & 0.86 \\
/CFHT &  &  &  &  &  \\
\hline
HSC & $g$ & 4670 & 1303 & 27.5 & Full & 27.6 & Full & 0.83 \\
/Subaru & $r$ & 6136 & 1382 & 27.3 & Full & 27.2 & Full & 0.77 \\
 & $i$ & 7657 & 1383 & 27.3 & Full & 27.3 & Full & 0.66 \\
 & NB816 & 8168 & 112 & 26.3 & Full & 26.4 & Full & 0.70 \\
 & $z$ & 8920 & 781 & 26.6 & Full & 26.6 & Full & 0.78 \\
 & NB921 & 9202 & 134 & 26.3 & Full & 26.1 & Full & 0.67 \\
 & $y$ & 9760 & 824 & 26.1 & Full & 25.8 & Full & 0.70 \\
 & NB1010 & 10097 & 97 & 24.4 & Full & - & - & 0.77 \\
\hline
VIRCAM & Y & 10193 & 880 & 26.0 & Full & - & - & 0.82 \\
/VISTA & NB118 & 11907 & 117 & 25.0 & Full & - & - & 0.75 \\
 & J & 12485 & 1554 & 25.9 & Full & - & - & 0.79 \\
 & H & 16354 & 2720 & 25.7 & Full & - & - & 0.76 \\
 & Ks & 21376 & 2914 & 25.4 & Full & - & - & 0.75 \\
\hline
UKIDSS & J & 12483 & 1590 & - & - & 25.8 & Full & 0.83 \\
/UKIRT & H & 16313 & 2920 & - & - & 25.3 & Full & 0.85 \\
 & K & 22010 & 3510 & - & - & 25.6 & Full & 0.85 \\
\hline
Fourstar & J1 & 10521 & 956 & 25.6 & 158.3 & 25.7 & 165.9 & 0.59 \\
/Magellan & J2 & 11407 & 1330 & 25.3 & 155.5 & 25.7 & 160.2 & 0.57 \\
 & J3 & 12835 & 1325 & 25.3 & 152.8 & 25.5 & 164.6 & 0.54 \\
 & Hs & 15479 & 1543 & 24.7 & 153.6 & 24.9 & 161.5 & 0.55 \\
 & Hl & 17001 & 1565 & 24.6 & 153.7 & 25.1 & 162.8 & 0.62 \\
\hline
IRAC & ch3 & 56281 & 12561 & 21.8 & Full & 22.4 & Full & 1.9 \\
/Spitzer & ch4 & 75890 & 25289 & 21.6 & Full & 22.3 & Full & 2.0 \\
\hline
\end{tabular}
\begin{flushleft}
{\sc Note.} --- 
($a$) The effective width of the filter.
($b$) The average 3$\sigma$ depth for point sources.
($c$) The depths of high-resolution images are given with the 0.2", 0.5", and Kron apertures with aperture corrections. 
($d$) For the bands with supplementary coverage from COSMOS-Web (F115W, F150W, F277W, F444W, and F770W) at the edge of the image, we only show the average depth of the area covered by the PRIMER survey.
\end{flushleft}
\end{minipage}
\end{table*}

%%%%%%%%%%%%%%%%%%%%%%%%%%%%%%%%%%%%%%%%%%%%%%%%%%%%%%%%%%%%%%%%%
\subsection{Reduction of JWST data}
\label{sec:data_reduction}
The reduction of all JWST imaging data is performed using our modified version of the JWST Calibration Pipeline. Data collected before 2024 August are reduced using the JWST Calibration Pipeline v1.13.4 \citep{Bushouse2024} with the Calibration Reference Data System (CRDS) pipeline mapping 1241, while data collected after 2024 August are reduced using the JWST Calibration Pipeline v1.19.1 \citep{Bushouse2025} with the CRDS pipeline mapping 1413. We have tested that different versions of the JWST Calibration Pipeline would not yield any systematic bias in the photometric results, and the mosaics reduced by our modified pipeline based on these two versions have comparable quality with a difference in depth smaller than 0.05 magnitude (because we have solved most of the problems of the old pipeline with our modifications). Therefore, we decided to keep the old mosaics reduced by the JWST Calibration Pipeline v1.13.4.

For JWST/NIRCam images, we first run the default stage 1 pipeline to perform detector-level corrections, during which we set ``jump.expand\_large\_events = True" and increased the expand factor to mask the ``snowballs"\footnote{Description of snowballs can be found at \dataset[jwst-docs.stsci.edu/known-issues/shower-and-snowball-artifacts]{https://jwst-docs.stsci.edu/known-issues/shower-and-snowball-artifacts}.} around large cosmic ray events. Based on the output of stage 1 (i.e. *\_rate.fits), we further subtract the scattered light (``claws" and ``wisps") using the templates provided by the JWST User Documentation\footnote{The templates of claws and wisps provided by the JWST user documentation are available at \dataset[stsci.app.box.com/s/1bymvf1lkrqbdn9rnkluzqk30e8o2bne]{https://stsci.app.box.com/s/1bymvf1lkrqbdn9rnkluzqk30e8o2bne}.}. In addition to this correction, we also adopt the methods given by \citet{Duncan2025} to subtract or mask the unstable wisps that vary between different pointings.

Stage 2 of the JWST Calibration Pipeline performs the instrument-level calibrations for individual exposures. After running the default stage 2 pipeline, we further run a set of additional calibrations for the output "cal" files: Firstly, we subtract the background and detect sources from each individual exposure using photutils v1.5.0 \citep{Bradley2022}. In order to accurately mask the extended structures of bright sources before the subtraction of stripes while avoiding masking the noise, we set a relatively high detection threshold and then extended the range of each detected bright source by 2 arcseconds on the segmentation map, which is then used for masking of sources. Secondly, following \citet{Bagley2023}, we remove the stripe-like 1/f noise by subtracting a median value for each row and column in each amplifier. For some exposures of PRIMER-UDS, we also subtract the slanted stripes by fitting them with cubic polynomials. Thirdly, we identify the remaining cosmic rays and hot pixels \citep[e.g.][]{Rieke2023} with their sharper edges compared to the point sources and set the DQ (the Data Quality extension in the JWST FITS file for a single exposure) values of the affected pixels to one (so they can be ignored by the pipeline). Lastly, we visually identify and mask the remaining noise structures on the image, such as the ``Dragon's Breath"\footnote{Description of the "Dragon's Breath" can be found at \dataset[jwst-docs.stsci.edu/known-issues/nircam-known-issues/nircam-scattered-light-artifacts]{https://jwst-docs.stsci.edu/known-issues/nircam-known-issues/nircam-scattered-light-artifacts}.} and the trajectories of nearby celestial bodies.

\begin{figure*}[]
\centering
\includegraphics[width=0.45\textwidth]{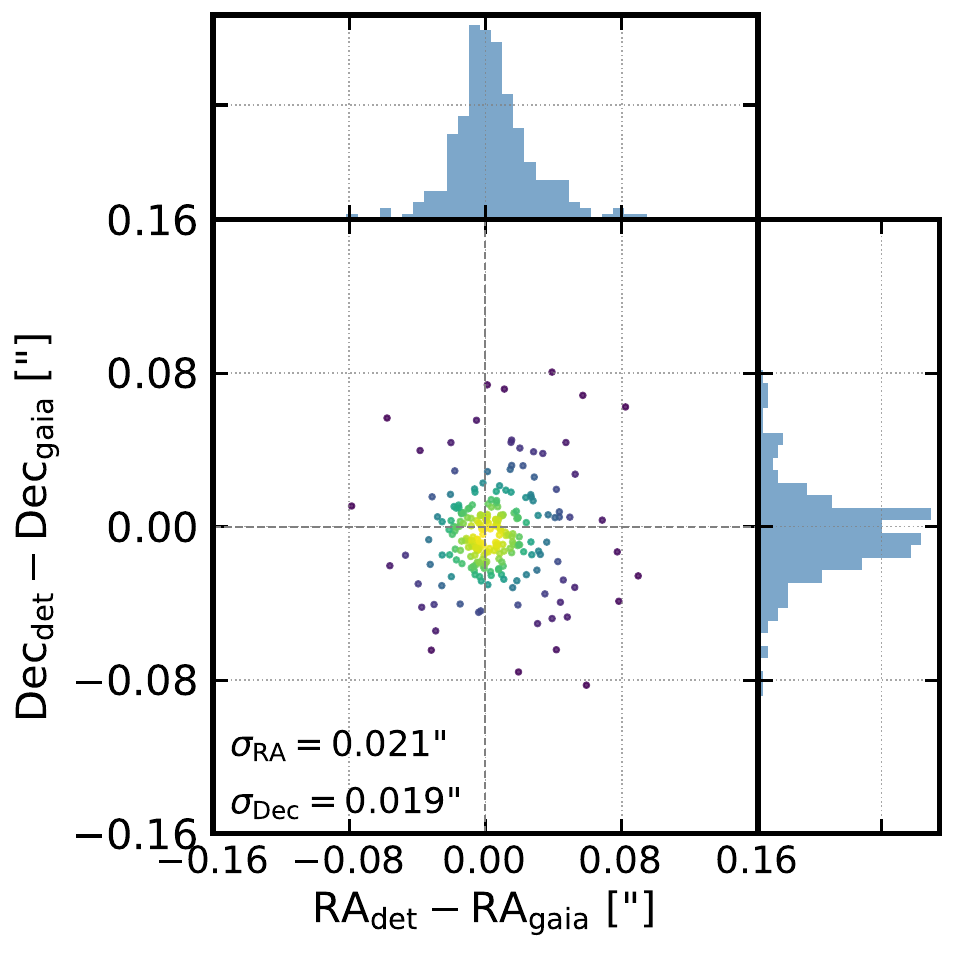}
\includegraphics[width=0.45\textwidth]{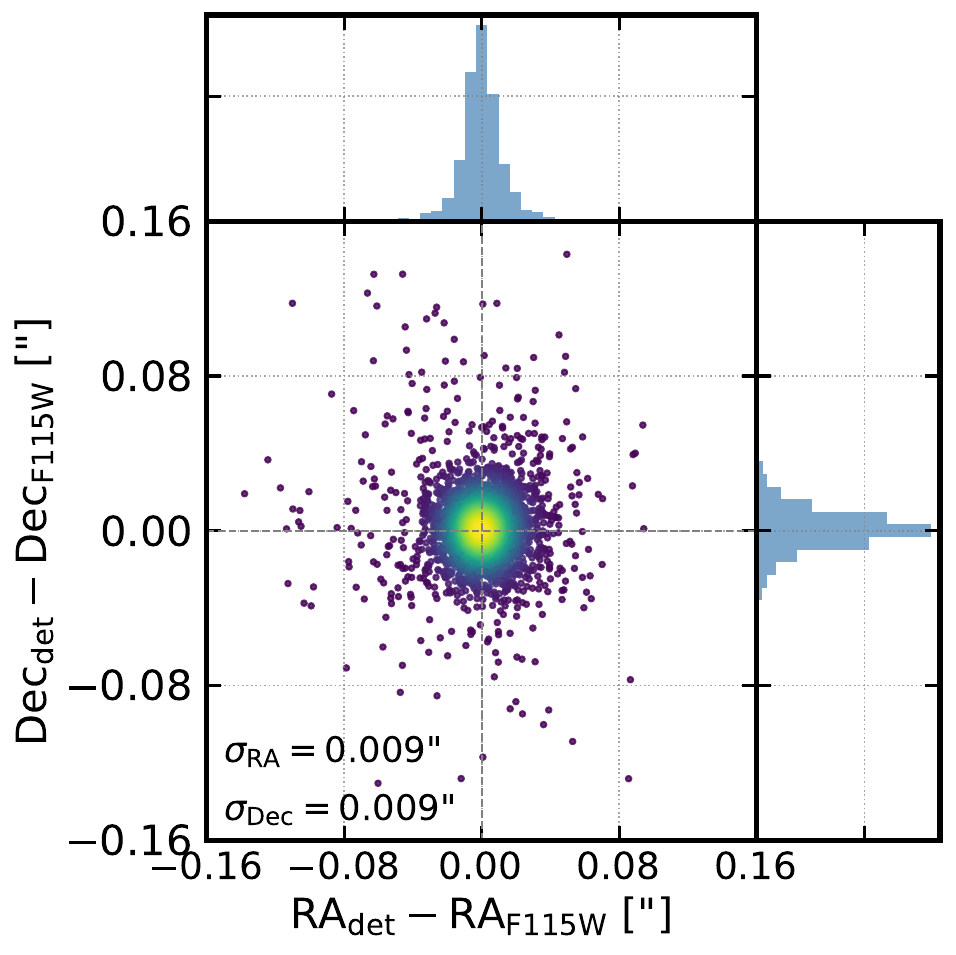}
\caption{\label{Fig:astrometry} \textbf{Validation of our astrometric alignment.} The left panel shows the astrometric difference between our detection image and gaia DR3 \citep{Gaia2018}, while the right panel shows the astrometric difference between the detection image (stack of all NIRCam/LW images) and the F115W image, which is used as a representative example to show the relative astrometric accuracy among different NIRCam bands. Sources from both PRIMER-COSMOS and PRIMER-UDS are included in this figure. Outliers in the right panel include galaxies with different morphologies at different bands.}
\end{figure*}

Stage 3 of the JWST Calibration Pipeline combines all available single exposures in one field into a large and deep image. During the ``TweakReg" step of stage 3, we use an astrometric reference catalog based on the HST/F160W or F814W images from the Grizli Image Release v7.0. For a few areas without any HST data, we use the positions reported by the COSMOS2020 \citep{Weaver2022} and the FENIKS catalog \citep{Zaidi2024} instead. All of these reference catalogs have been aligned with the third data release of Gaia \citep[\textrm{Gaia DR3},][]{Gaia2018}. Figure~\ref{Fig:astrometry} shows the results of our astrometric alignment, where the astrometric scatter is ${\sigma _{{\rm{RA}}}} \approx {\sigma _{{\rm{Dec}}}} \approx 0.02''$ between our mosaics and Gaia DR3, and ${\sigma _{{\rm{RA}}}} \approx {\sigma _{{\rm{Dec}}}} \approx 0.009''$ across the different NIRCam bands. Then, we run the remaining part of stage 3 with default parameters. The pixel scales of our final mosaics are set to be $0.04''/{\rm{pixel}}$ for JWST/NIRCam images and $0.08''/{\rm{pixel}}$ for JWST/MIRI images.

For the JWST/MIRI images, in addition to the modifications mentioned above, we make the following improvements to the JWST Calibration Pipeline as shown in \citet{WangT2025}. Firstly, the default flat fields provided by the CRDS are replaced by supersky flat fields, which are constructed for each MIRI pointing by stacking the output of stage 1 from all pointings observed within 5 days. Secondly, we perform the source masking for MIRI images in an iterative way: we first construct the flat fields and run stage 2 without source masking, and then perform source masking using the output of our first iteration and repeat those operations for a second time.

%%%%%%%%%%%%%%%%%%%%%%%%%%%%%%%%%%%%%%%%%%%%%%%%%%%%%%%%%%%%%%%%%%%%%%%%%%%%%%%%
\section{Source Detection AND PHOTOMETRY}
\label{sec:detection}
%%%%%%%%%%%%%%%%%%%%%%%%%%%%%%%%%%%%%%%%%%%%%%%%%%%%%%%%%%%%%%%%%%%%%%%%%%%%%%%%
\subsection{Point-spread Function Matching}
\label{sec:psf_match}
\begin{figure*}[]
\centering
\includegraphics[width=0.7\textwidth]{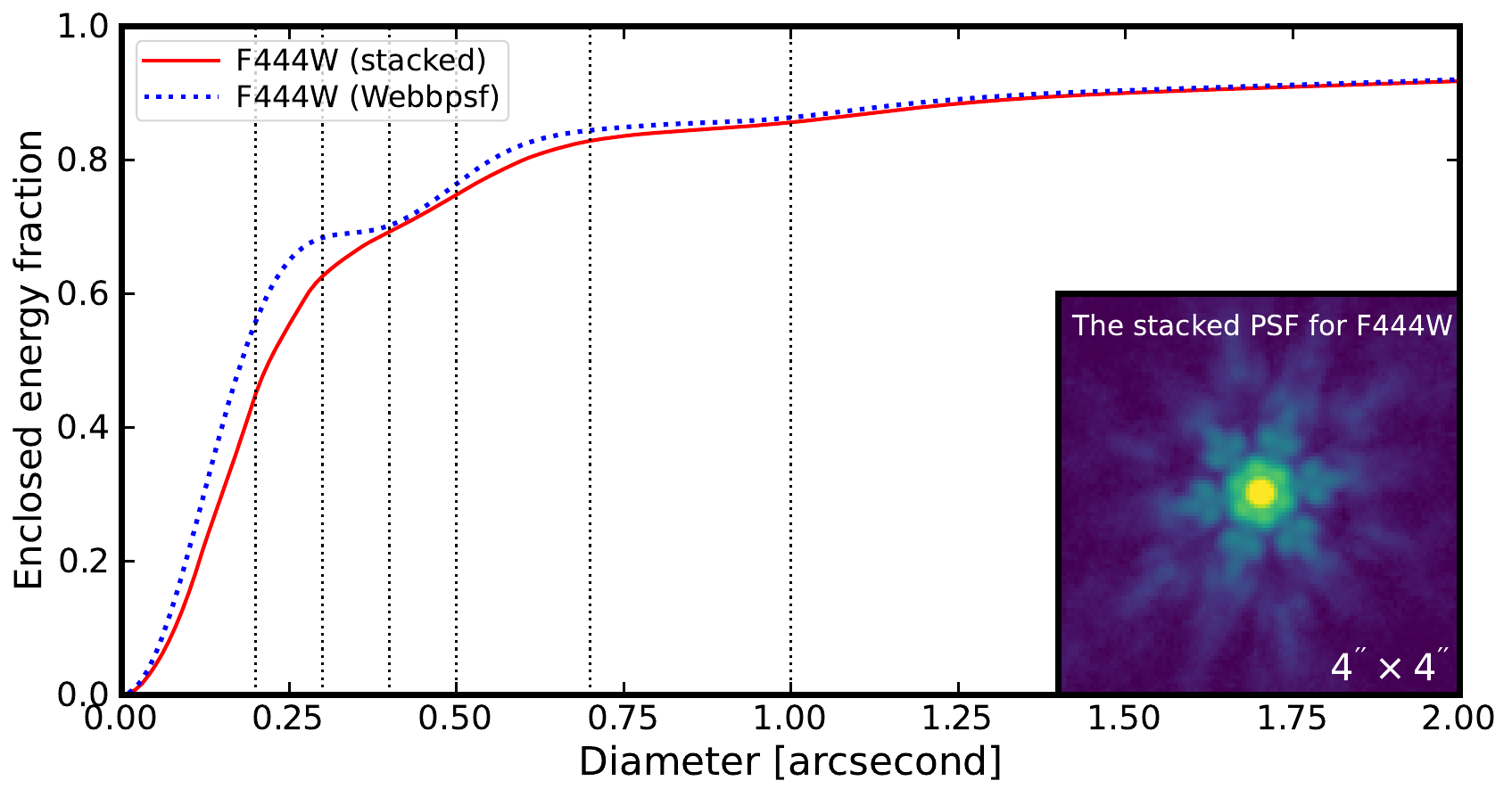}
\caption{\label{Fig:psf} \textbf{Example of the stacked PSF for NIRCam/F444W.} The vertical dotted lines show the diameters of the six circular apertures used for photometry (0.2", 0.3", 0.4", 0.5", 0.7", and 1.0"). Compared to the results from Webbpsf \citep{Perrin2015,Osborne2018}, the enclosed energy of the stacked F444W PSF can be significantly lower at $d<0.4''$. Similar bias has also been reported by \citet{Ji2024}.}
\end{figure*}

Before source detection, we perform point spread function (PSF) matching to ensure that all images used for detection have the same resolution. To get the effective PSF for each JWST and HST filter, we stack $4'' \times 4''$ cutouts of point sources from our real mosaics. These point sources are selected as the sources with the highest fraction of flux density within a small ($d = 0.2''$) aperture (${f_{0.2''}}/{f_{4''}}$). The fluxes outside the $4'' \times 4''$ cutouts are corrected using the model PSFs from the Python package Webbpsf v1.1.1 \citep{Perrin2015,Osborne2018}, whose enclosed energy should only be inaccurate in the center \citep{Ji2024}. An example of a stacked PSF is shown in Figure~\ref{Fig:psf}.

For all HST and JWST images whose resolutions are better than the JWST/NIRCam F444W image, we use photutils v1.5.0 \citep{Bradley2022} with the CosineBellWindow to obtain the convolution kernels that match their effective PSF to that of F444W. The convolution is performed using the "convolve\_fft" function from Astropy v5.0 \citep{Astropy2022}. Based on this convolution, source detection will be performed at the F444W resolution (Section~\ref{sec:dual_mode_detection}). We do not perform PSF matching for the JWST/NIRCam F410M, F430M, F460M, F466N, and F470N images because their resolutions are already comparable (difference of PSF FWHM within $10\%$) to F444W. And the larger PSFs of HST/WFC3 mosaics are considered using an additional correction during aperture photometry (Section~\ref{sec:aperture_photometry}).

\subsection{"Cold" and "Hot" Dual mode Detection}
\label{sec:dual_mode_detection}
After the PSF match, we make an inverse-variance-weighted stack of six JWST/NIRCam longwave (LW) images (F277W, F356W, F410M, F430M, F444W, and F460M) as the detection image. The stacking of all LW images enables the deepest detection for the average sample of galaxies. A few faint and red galaxies may have a lower signal-to-noise ratio (SNR) in these stacks than they would in individual filters, but they would likely be detected only in one or two bands and their photometric redshifts and physical properties cannot be constrained.
%Only the data collected before August 2024 (when we perform source detection) are used to generate this detection image (Because of the same reason mentioned above, although new data can enable deeper detection, we can not properly constrain the properties of these newly detected sources).
All saturated stars on this detection image are visually identified and filled so that they can be detected as one single star in our catalog. Then, we perform source detection by running Source-Extractor v2.25.0 \citep{Bertin1996} in dual "cold" and "hot" modes to reach the limiting depth of the detection image without oversplitting the bright sources \citep[e.g.][]{Galametz2013,Guo2013}. This is necessary for our deblending photometry, which requires accurate measurements of the bright sources to reduce the flux contamination caused by them. The key parameters for the dual-mode detections are listed in Table ~\ref{Tab:Detection}. Firstly, we run cold mode detection to detect 20059 and 28017 bright sources in PRIMER-COSMOS and PRIMER-UDS, respectively. Then, 138609 (176214) faint sources are detected by the hot mode in PRIMER-COSMOS (UDS). Lastly, we exclude all hot-mode sources that are located within the Kron aperture of a cold-mode source and reallocate the corresponding pixels on the segmentation map to this cold-mode source. After this exclusion, our final catalog consists of 135855 and 172384 sources in PRIMER-COSMOS and PRIMER-UDS, respectively. 

\begin{figure}[]
\centering
\includegraphics[width=0.48\textwidth]{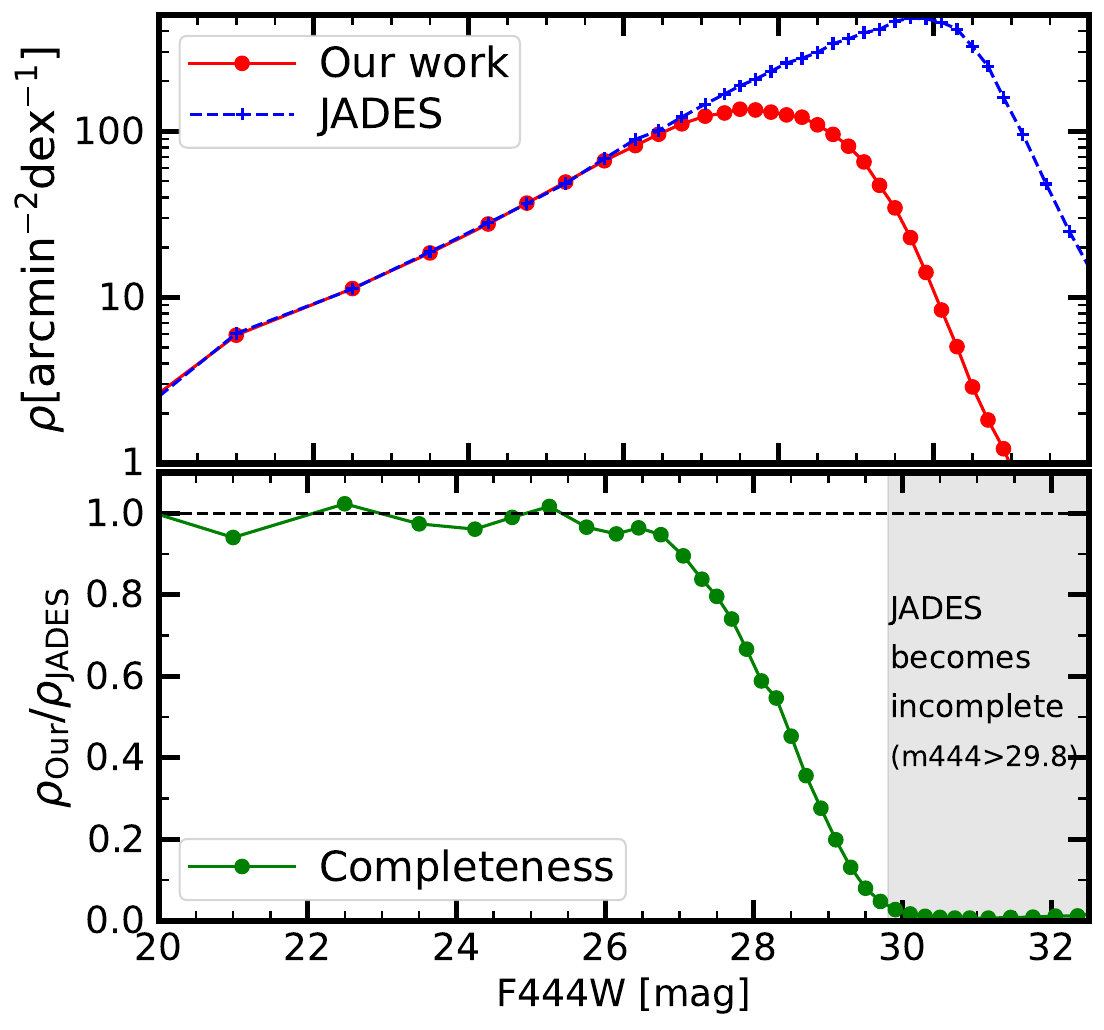}
\caption{\label{Fig:completeness} \textbf{Source number counts and detection completeness.} The red line in the upper panel presents the source number count density as a function of F444W magnitude. The blue line shows the number count density from the JADES survey \citep{Rieke2023, Eisenstein2023}, whose $5\sigma$ limiting magnitude in F444W is 29.8. The lower panel shows the ratio between the number count density of our work and JADES, which presents the detection completeness of our catalog.}
\end{figure}

The number count density of our catalog as a function of F444W magnitude is shown in Figure~\ref{Fig:completeness}, which is consistent with the JWST Advanced Deep Extragalactic Survey (JADES) \citep{Rieke2023,Eisenstein2023} at the bright end. This number count from JADES is calculated using the deepest region in GOODS-South, whose $5\sigma$ limiting magnitude in F444W reaches 29.8.
Since the JWST mosaics from JADES are much deeper than PRIMER, the detection completeness of our work can be inferred from the number count density ratio between our catalog and the JADES catalog. According to this ratio, our catalog reaches $\sim 80\%$ ($50 \%$) completeness at ${\rm{ma}}{{\rm{g}}_{444}}\sim 27.2$ (28.2). We remind that this is a relative completeness estimate with respect to the deeper JADES catalog, not an absolute completeness curve. However, since the $50 \%$ completeness limit of our catalog is 1.6 magnitude brighter than the limiting magnitude of JADES, the bias caused by the incompleteness of JADES should be small.

\begin{table*}[!tbh]\small
\centering
\begin{minipage}[center]{\textwidth}
\centering
\caption{Key parameters of SExtractor in cold and hot modes detection \label{Tab:Detection}}
\begin{tabular}{lcc}
\hline\hline
  & Cold Mode  & Hot Mode \\
\hline
DETECT\_MINAREA &  10 & 5 \\
DETECT\_THRESH &  18.0 (16.0)$^a$ & 1.5 \\
FILTER\_NAME &  gauss\_5.0\_9x9.conv & gauss\_4.0\_7x7.conv \\
DEBLEND\_NTHRESH &  64 & 64 \\
DEBLEND\_MINCONT &  0.0006 & 0.0001 \\
BACK\_SIZE &  200 & 72 \\
BACK\_FILTERSIZE &  3 & 3 \\
WEIGHT\_TYPE &  MAP\_RMS & MAP\_RMS \\
\hline
\end{tabular}
\begin{flushleft}
{\sc Note.} --- 
($a$) Since the NIRCam observations of Primer-COSMOS are deeper than for Primer-UDS, different cold mode detection thresholds are used for these two fields (18.0 for Primer-COSMOS and 16.0 for Primer-UDS).
\end{flushleft}
\end{minipage}
%\end{center}
\end{table*}
%%%%%%%%%%%%%%%%%%%%%%%%%%%%%%%%%%%%%%%%%%%%%%%%%%%%%%%%%%%%%%%%%%%%%%%%%%%%%%%%
%\section{Multi-wavelength photometry}
%\label{sec:photometry}

\subsection{Aperture photometry for high-resolution data}
\label{sec:aperture_photometry}
For high-resolution images from JWST/NIRCam and HST, we perform aperture photometry with the software APHOT \citep{Merlin2019} using circular apertures within diameters 0.2", 0.3", 0.4", 0.5", 0.7", and 1.0" as well as a Kron elliptical aperture with the Kron parameter $k = 2.5$ \citep{Kron1980}. The Kron apertures are calculated by Source-Extractor on the detection image, and then imported by APHOT. During the aperture photometry, APHOT replaces the bad pixels or the pixels belonging to another source on the segmentation map with the pixels symmetric with respect to the centroid of the source. A local background estimation is also computed and then subtracted by APHOT. If the resolution of the measurement band is better or comparable to that of F444W (as seen in HST/ACS and JWST/NIRCam images), the flux densities are measured on the PSF-matched images (Section~\ref{sec:psf_match}) with the same resolution. Conversely, for HST/WFC3 images, we convolve the stacked detection image (Section~\ref{sec:dual_mode_detection}) to match their PSF. Then, we use the ratio ${f_{\rm{det},{\rm{before\thinspace convolution}}}}/{f_{\rm{det},{\rm{after\thinspace convolution}}}}$ of each source to correct the additional aperture loss due to the lower resolution of HST/WFC3. We determine the error of all aperture fluxes using the root-mean-square (RMS) value of 1000 empty apertures for each band and each aperture size. Then, the uncertainties of flux densities estimated by APHOT are multiplied by the ratio between this RMS and the median uncertainty in the catalog.

%Firstly, all fluxes measured in circular apertures are corrected to Kron fluxes using ${f_{{\rm{aper,measure}}}} \times {f_{{\rm{Kron,det}}}}/{f_{{\rm{aper,det}}}}$. Next, we correct the aperture loss of Kron apertures, 
To obtain the total flux densities, we perform the aperture correction as follows. Firstly, we correct the aperture losses of flux densities measured by Kron apertures, which are necessary since the PSFs of JWST have nonnegligible contributions ($\gtrsim 10\%$) from the outskirts. For faint and compact sources whose Kron apertures are smaller than the Kron aperture calculated for the PSF, the flux losses outside their Kron apertures are determined by measuring the ratio of the PSF outside these apertures ($1-{{\rm{f}}_{{\rm{Kron(source),PSF}}}}/{{\rm{f}}_{{\rm{tot,PSF}}}}$). For the other sources with larger Kron apertures, we directly use the fraction of flux density outside the Kron aperture calculated for the PSF ($1-{{\rm{f}}_{{\rm{Kron(PSF),PSF}}}}/{{\rm{f}}_{{\rm{tot,PSF}}}}$, $\sim 10\%$ for NIRCam) to avoid underestimating the aperture losses for these bright and extended sources. We note that the Kron apertures of some sources can be unreliable due to the enhanced background structure \citep[e.g.][]{Rieke2023} or low SNR. These unreliable Kron apertures are identified using the empirical criteria

\begin{equation}
\label{eq_badKron}
\left\{ {\begin{array}{*{20}{c}}
{{\rm{SNR}_{{\rm{Kron,det}}}} < 3}\\
{{\rm{or}} \thinspace (A > 2 \wedge {f_{{\rm{area}}}} > 18 \wedge {f_{{\rm{flux}}}} > 6)}\\
{{\rm{or}} \thinspace (A > 5 \wedge {f_{{\rm{area}}}} > 6 \wedge {f_{{\rm{flux}}}} > 2)}\\
%{\rm{or} \thinspace (A > 25 \wedge {\rm{hot \thinspace mode \thinspace detection}})}
\end{array}} \right.
\end{equation}
where $A$ is the profile RMS along the major axis of each source ("A\_image" from source-extractor), ${f_{{\rm{area}}}} \equiv {\rm{area}}({\rm{Kron}})/{\rm{area}}({\rm{seg}})$ is the ratio of the area of Kron aperture and segmentation, ${f_{{\rm{flux}}}} \equiv {\rm{flux}}({\rm{Kron}})/{\rm{flux}}({\rm{seg}})$ is the ratio of the flux density within Kron aperture and segmentation. Once the Kron aperture of a source is considered to be unreliable, we set ``flag\_kron = 1" in the catalog. Then, for flux densities measured by circular apertures, we correct them to total fluxes using ${f_{{\rm{aper,measure}}}} \times {f_{{\rm{Kron,det}}}}/{f_{{\rm{aper,det}}}}$, where ${f_{{\rm{aper,measure}}}}$ and ${f_{{\rm{aper,det}}}}$ are the flux density measured by the circular aperture in the measurement and detection band, respectively. If a source has ``flag\_kron = 1", we adopt the point-source assumption to correct the aperture loss of circular apertures instead.

Lastly, among the seven apertures (0.2", 0.3", 0.4", 0.5", 0.7", 1.0", and Kron), we choose the best aperture for each source. If a source has ${\rm{SN}}{{\rm{R}}_{{\rm{Kron,det}}}} > 20$ and is not selected by Equation~\ref{eq_badKron}, we use the Kron aperture as the best aperture. Otherwise, we choose the aperture with the largest ${\rm{SN}}{{\rm{R}}_{{\rm{aper,det}}}} \times {f_{{\rm{aper}},{\rm{det}}}}/{f_{{\rm{Kron}},{\rm{det}}}}$ (when ${f_{{\rm{aper}},{\rm{det}}}}/{f_{{\rm{Kron}},{\rm{det}}}}$ is smaller, the aperture correction can be more uncertain). This best aperture is then used for all HST and JWST/NIRCam bands since the PSFs of all these bands are smaller than our smallest circular aperture with $d=$0.2". We note that we use the same aperture for all bands to obtain more consistent colors and to avoid introducing band-dependent aperture noise. However, this may lead to additional uncertainties for a small fraction of sources whose morphology or centroid varies strongly with wavelength. For these sources, the UV-optical fluxes from TPHOT photometry with iterative shifting of the positions can be more reliable.

\subsection{TPHOT deblending photometry for low-resolution data}
\label{sec:tphot_photometry}

\begin{figure*}[htbp]
\centering
\includegraphics[width=1\textwidth]{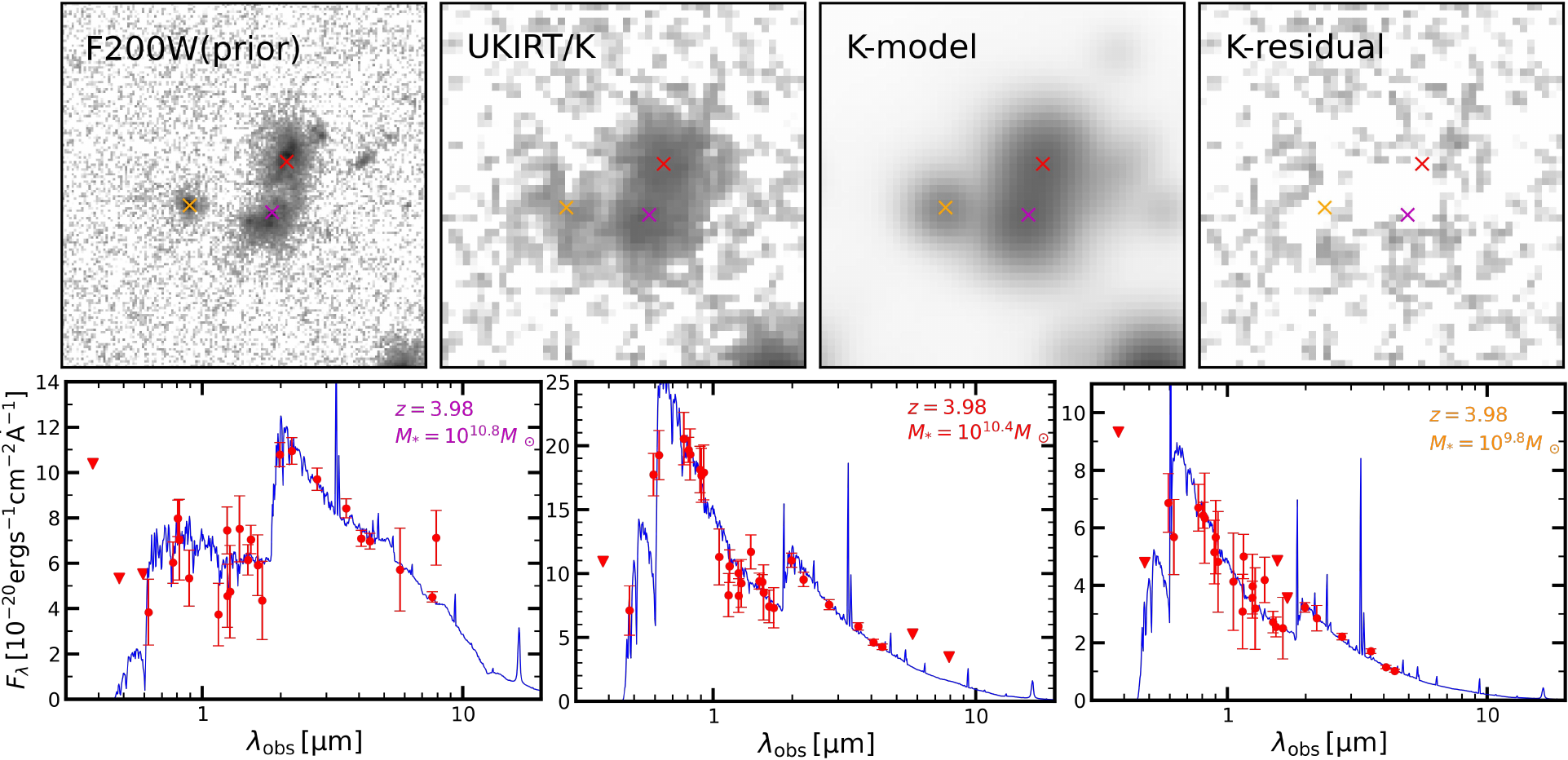}
\caption{\label{Fig:deblending}\textbf{Example of the TPHOT deblending photometry.} The upper panels show the images in the center of the Bigfoot protocluster in PRIMER-UDS \citep{Sun2025}, with locations of the three massive member galaxies in Bigfoot marked as crosses. The first image is the high-resolution image from JWST/NIRCam F200W, which is used to provide high-resolution priors for K-band photometry. The second image is the low-resolution image from UKIRT/K, which is the measurement band in this example. The third and fourth images are the best-fit models and residuals after the deblending photometry. The lower panels show the best-fit SEDs (combining both high-resolution and low-resolution data) of three member galaxies in Bigfoot from Bagpipes.}
\end{figure*}

\begin{figure}[htbp]
\centering
\includegraphics[width=0.45\textwidth]{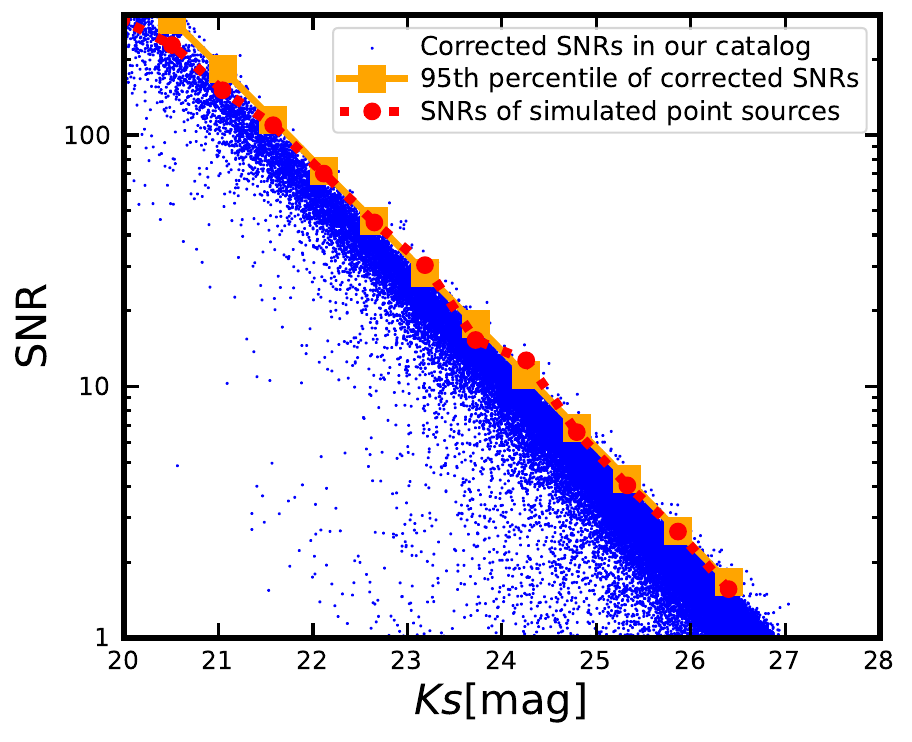}
\caption{\label{Fig:TPHOT_err}\textbf{Example of the uncertainty of TPHOT deblending photometry in the Ks band.} The blue points show the SNRs reported in our catalog, which have been corrected using 1000 empty apertures with $d = 3.5 \times {\rm{FWH}}{{\rm{M}}_{{\rm{PSF}}}}$. The orange line shows the 95th percentile of the corrected SNRs, which represents the SNRs of point sources. The red dotted line shows the SNRs of simulated point sources, which are calculated as ${\rm{flu}}{{\rm{x}}_{{\rm{input}}}}{\rm{/RM}}{{\rm{S}}_{{\rm{output}}}}$ from the simulated TPHOT photometry. The orange line is almost overlapping with the red dotted line, showing that the flux uncertainties corrected by empty apertures are consistent with the results of simulated photometry.}
\end{figure}

For the data from JWST/MIRI, ground-based telescopes (CFHT, Subaru, VISTA, UKIRT, Megellan), and Spitzer/IRAC, we perform deblending photometry with the software TPHOT v2.0 \citep{Merlin2015,Merlin2016}. We use the image of the closest JWST/NIRCam band as the high-resolution prior for TPHOT photometry. Then, TPHOT convolves the high-resolution prior to matching the PSF of the measurement band, and fits the fluxes in the measurement band with these convolved high-resolution priors until the linear ${\chi ^2}$ solution is obtained. During TPHOT photometry, we exclude the faint sources with ${\rm{SNR}} < 3$ on the closest JWST image to avoid unreliable high-resolution priors, and then fit all remaining sources with the cells-on-objects algorithm to split the large mosaic into smaller cells \citep{Merlin2015}. The flux priors of all fitted sources are taken as their flux densities at the closest JWST band, and we input relatively large uncertainties on the flux priors to TPHOT so that the output flux densities will not be biased by the flux priors. During the second pass of TPHOT fitting, we allow shifting of the positions of sources up to 0.3" to account for astrometric differences (although most of the low-resolution mosaics have also been aligned with Gaia, their astrometric accuracy can be poorer due to lower resolution). Figure~\ref{Fig:deblending} shows an example of deblending photometry in the central area of the Bigfoot protocluster at $z=3.98$ in PRIMER-UDS \citep{Sun2025}. We show that even in the core of a protocluster, one of the densest areas in the Universe, we are still able to obtain reliable SEDs with the deblending photometry.

To provide uncertainties for the flux densities measured by TPHOT, we put 1000 empty apertures whose diameter is 3.5 times larger than the FWHM of the PSF of each low-resolution band in the residual map after TPHOT photometry. This factor 3.5 was calibrated using the results of simulated photometry for sources with known fluxes in six bands (CFHT/U; Subaru/B, IB427, and NB2300; UltraVISTA/J, and Ks) during our previous work targeting protocluster J1001 \citep{Sun2024}. Then, we use the RMS values of flux densities measured by these empty apertures divided by the median flux uncertainty output by TPHOT as the correction factor for the flux uncertainties of all sources output by TPHOT. As an example to prove the reliability of the above procedure, we present the results of simulated photometry for point sources with the updated UltraVISTA/Ks band data used by this work (which is 0.7 magnitude deeper than the Ks band data of J1001) in Figure~\ref{Fig:TPHOT_err}. At each input magnitude, we inject 500 point sources into the residual map and run TPHOT photometry for them. The SNR of these simulated point sources can then be calculated as ${\rm{flu}}{{\rm{x}}_{{\rm{input}}}}{\rm{/RM}}{{\rm{S}}_{{\rm{output}}}}$, where ${\rm{/RM}}{{\rm{S}}_{{\rm{output}}}}$ is the $1\sigma$ scatter of the output flux densities of these 500 point sources. In Figure~\ref{Fig:TPHOT_err}, the orange line shows the 95th percentile of the SNRs corrected by the empty apertures, which represents the SNRs of isolated point sources (with the highest SNR at acertain magnitude) in our catalog. This orange line is consistent with the SNRs of simulated point sources (the red dotted line), showing that we can use the method based on empty apertures (which is more efficient) to estimate the uncertainties of TPHOT photometry.

%The errors of flux densities measured by TPHOT are corrected using 1000 empty apertures whose diameter is 3.5 times larger than the FWHM of the PSF of each low-resolution band. As shown in Figure~\ref{Fig:TPHOT_err}, using the RMS values of flux densities measured by these empty apertures, the SNRs reported by our catalog are consistent with our test by injecting simulated point sources with known fluxes into the real image and testing their photometric results from TPHOT. 

After TPHOT photometry, we select the unreliable flux densities measured by TPHOT as follows. Firstly, TPHOT cannot properly fit the sources around bright stars, especially for bright stars saturated on either the measurement image or the JWST image used as the high-resolution prior. These sources affected by bright stars can be selected using 
\begin{equation}
\label{eq_badTPHOT}
d < 50.0 - 2.5 \times R,
\end{equation}
where $d$ is the distance to the bright sources in arcseconds, $R$ is the $R$ band magnitude from Gaia DR3 \citep{Gaia2018}, and the empirical parameters 50 and 2.5 are determined by visual inspection. Sources selected by Equation~\ref{eq_badTPHOT} have "flag\_TPHOT = 1" in our catalog. %Sources in several visually selected regions around bright and extended galaxies are also selected in addition to the sources selected by Equation~\ref{eq_badTPHOT}. 
Secondly, we compare the flux densities from low-resolution images and high-resolution images at similar wavelengths (including four comparisons in total: $i$ compared with F814W, $z$ compared with F090W, J compared with F115W, and H compared with F150W). The outliers from these four comparisons can be selected using
\begin{equation}
\label{eq_badTPHOTB}
\left| {\left. {{\rm{ma}}{{\rm{g}}_{\rm{h}}} - {\rm{ma}}{{\rm{g}}_{\rm{l}}}} \right|} \right. >  - 2.5*\log (1 - \frac{{\Sigma \sqrt {\sigma _{\rm{h}}^2 + \sigma _{\rm{l}}^2} }}{{\max ({f_{\rm{h}}},{f_{\rm{l}}})}}) + 0.3,
\end{equation}
where $\Sigma$ is the confidence level; ${\sigma _{\rm{h}}}$ and ${\sigma _{\rm{l}}}$ are the $1\sigma$ flux uncertainties in the high-resolution band and the low-resolution band; ${f _{\rm{h}}}$ and ${f _{\rm{l}}}$ are the flux densities at the high-resolution band and the low-resolution band, respectively, and the color term 0.3 is a conservative value adopted to consider the intrinsic color between the high-resolution band and low-resolution band. Sources selected with $\Sigma  = 5$ from one comparison or $\Sigma  = 2.5$ from at least two comparisons are considered to have unreliable deblending results, which have "flag\_TPHOT = 2" in our catalog. This can help us exclude all suspicious data, which mainly belong to sources close to a bright and extended galaxy or close mergers.
Thirdly, we select the sources close (distance $<3"$) to the edge of our detection image, and set "flag\_TPHOT = 3".

Combining the above criteria, 25348 sources ($\sim 8.2\%$, 12880 in PRIMER-COSMOS and 12468 in PRIMER-UDS) are considered to have unreliable deblending results. For these sources, we exclude all fluxes measured by TPHOT during SED fitting. For JWST/MIRI data, if the result from TPHOT is unreliable ("flag\_TPHOT $> 0$") or has ${\rm{SNR}{_{{\rm{TPHOT}}}} < 2}$, we use the result from aperture photometry instead. The aperture photometry of JWST/MIRI is performed using the same method as HST/WFC3 (Section~\ref{sec:aperture_photometry}), but we only consider the apertures larger than the PSF of each JWST/MIRI band as the best aperture for MIRI (For example, we only use circular apertures with $d=$0.7", 1.0", and Kron apertures for F1800W).

\subsection{Photometric validations}
\begin{figure*}[]
\centering
\includegraphics[width=0.45\textwidth]{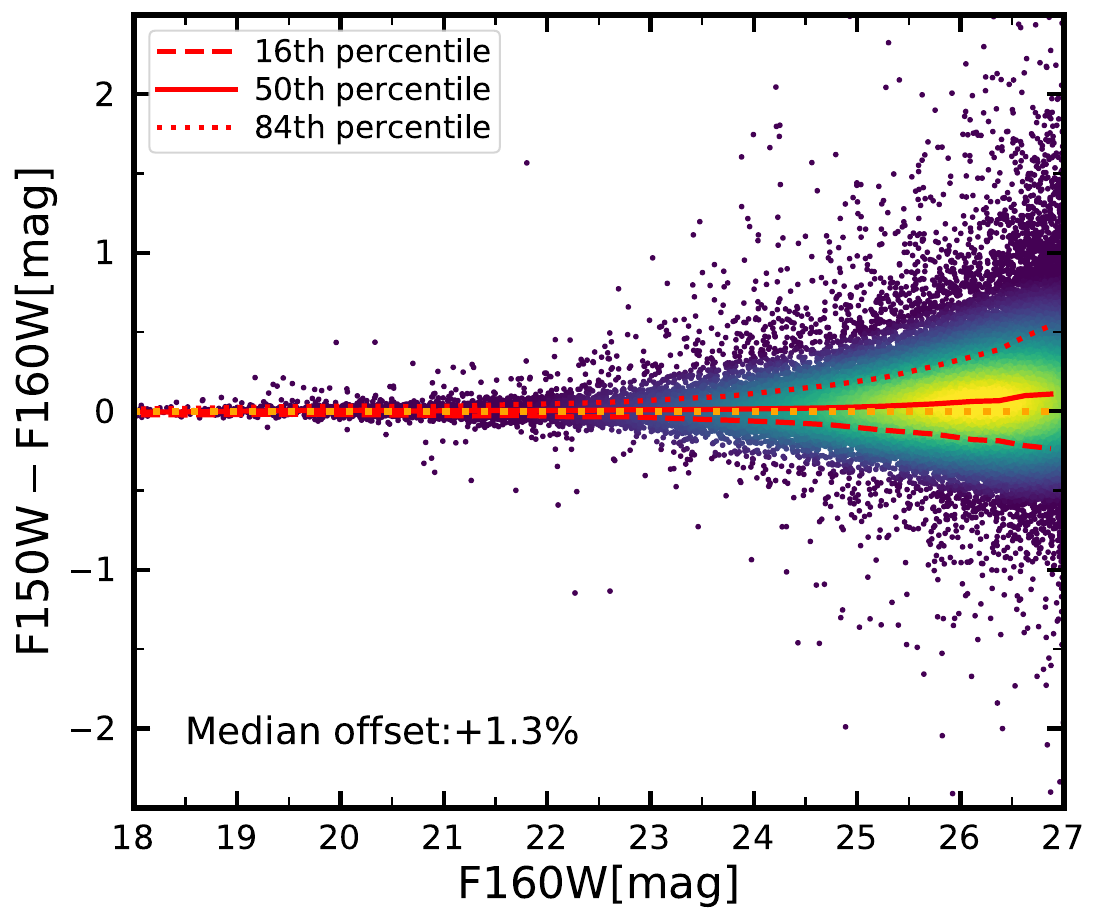}
\includegraphics[width=0.45\textwidth]{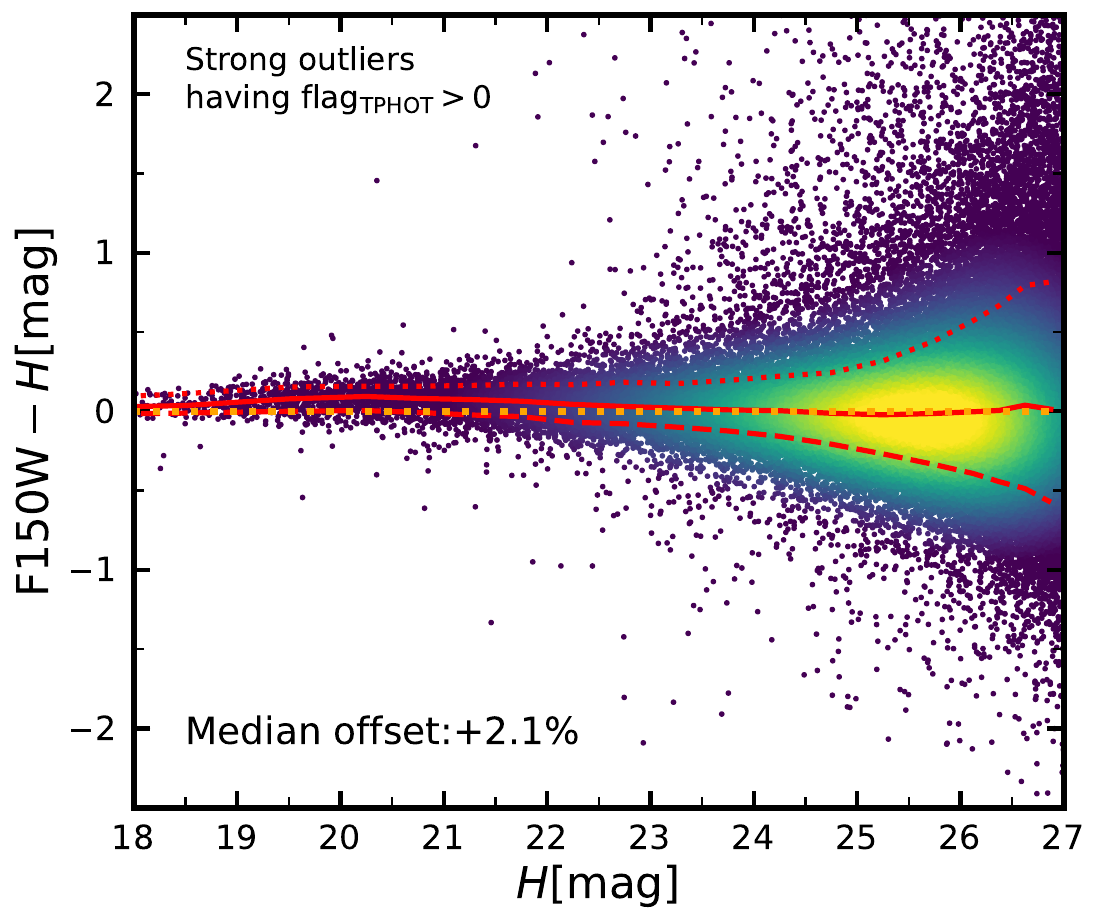}
\caption{\label{Fig:fluxcompare_150} \textbf{Comparison of $\sim 1.5{\rm{\mu m}}$ flux densities measured from different telescopes with different methods.} The left panel compares the photometric results from JWST/NIRCam F150W and HST/WFC3 F160W images, while the right panel compares the results from JWST/NIRCam F150W and VISTA/H. The red dashed, solid, and dotted lines show the 16th, 50th, and 84th percentiles of the colors as functions of magnitudes, respectively. We also report the median offset for all sources with SNR$>7$ in both bands. These results agree well with almost no bias, and we have checked that most of the strong outliers have been selected to have flag\_TPHOT $> 0$ in our catalog (see Section~\ref{sec:tphot_photometry}).}
\end{figure*}

\begin{figure*}[]
\centering
\includegraphics[width=0.98\textwidth]{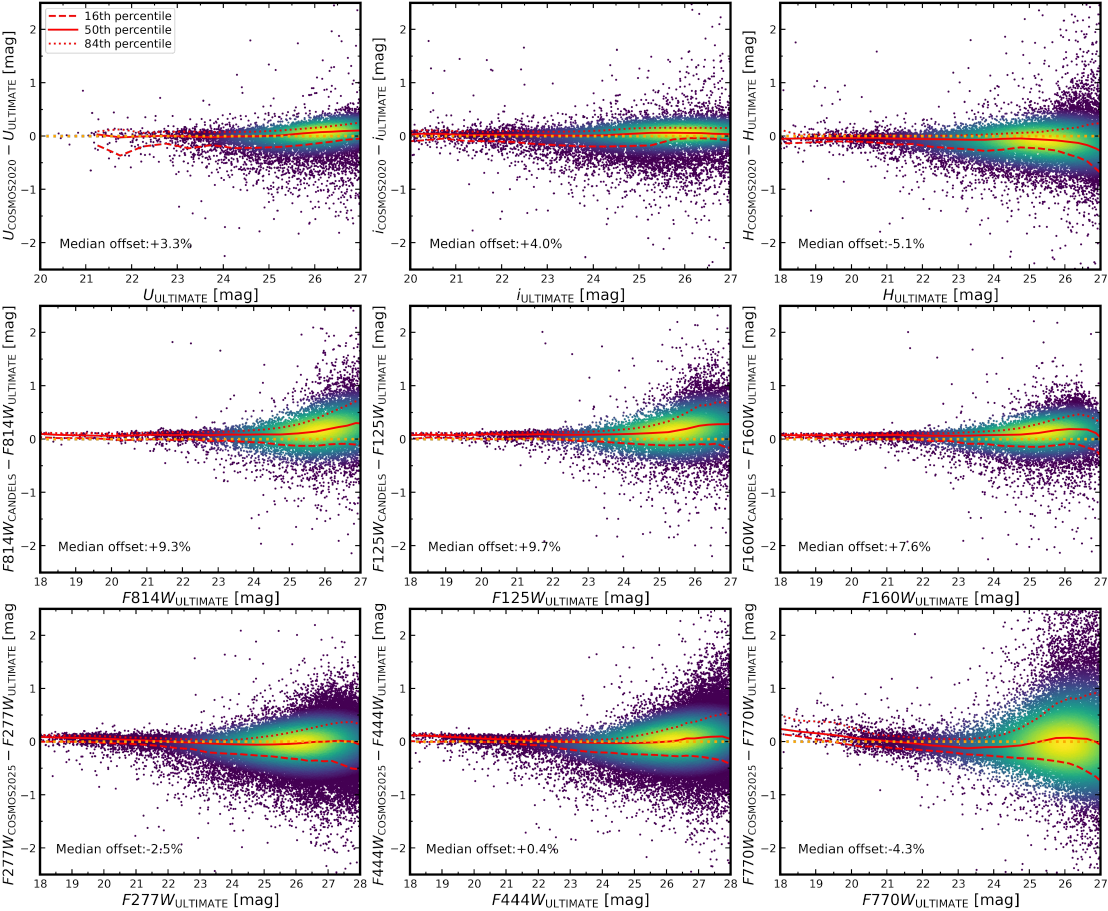}
\caption{\label{Fig:fluxcompare_cosmos} \textbf{Comparison between our catalog and archival results in PRIMER-COSMOS.} The upper panels compare the flux densities observed by ground-based telescopes from our ULTIMATE-deblending catalog and the COSMOS2020 catalog \citep{Weaver2022}. The middle panels compare the HST fluxes from our catalog and the CANDELS catalog \citep{Nayyeri2017}. The CANDELS catalog directly uses the flux densities measured by Kron apertures as total flux densities, which can underestimate the total fluxes of faint sources with small Kron apertures. The lower panels compare the JWST flux densities measured by our catalog and the COSMOS2025 catalog \citep{Shuntov2025}. The median offsets are reported for all sources with SNR$>7$ in each corresponding band.}
\end{figure*}

Our catalog combines flux densities measured by different photometric methods, so we validate their consistency in Figure~\ref{Fig:fluxcompare_150}. The left panel compares the flux densities from JWST/NIRCam F150W and HST/WFC3 F160W images, which are both measured by aperture photometry with APHOT but are processed with different PSF matching programs because of their different resolutions (Section~\ref{sec:aperture_photometry}). Meanwhile, the right panel compares the flux densities from JWST/NIRCam F150W and VISTA/$H$, which are measured by aperture photometry with APHOT and deblending photometry with TPHOT, respectively. For both comparisons, the $\sim 1.5{\rm{\mu m}}$ flux densities measured by the different telescopes using different methods are consistent with almost no systematic offset. This confirms the reliability of the combination of different photometric methods and also validates the accuracy of the photometric calibration of the imaging data. We note that the strong outliers in Figure~\ref{Fig:fluxcompare_150} include saturated stars, sources around bright nearby sources, and sources close to the edge of the images. These unreliable sources have been selected and flagged in our catalog (see Section~\ref{sec:tphot_photometry}).

Figure~\ref{Fig:fluxcompare_cosmos} compares our photometric results with the archival ground-based \citep{Weaver2022}, HST-based \citep{Nayyeri2017}, and JWST-based \citep{Shuntov2025} catalogs in PRIMER-COSMOS to further validate the reliability of our catalog. Our results do not have systematic bias with the COSMOS2020 \citep{Weaver2022} and COSMOS2025 \citep{Shuntov2025} catalogs across all magnitudes. However, the flux densities reported by the CANDELS catalog \citep{Galametz2013,Nayyeri2017} could be systematically underestimated because they do not consider the aperture loss of the Kron apertures (especially for faint sources with smaller Kron apertures). This problem of the CANDELS catalog has also been reported by \citet{Skelton2014}. We note that the scatter between our results and COSMOS2025 can be larger because of the different mosaics used for photometry: \citet{Shuntov2025} only used the JWST data taken by the COSMOS-Web survey \citep{Casey2023}, while our JWST mosaics can be much deeper with data from the PRIMER survey. We have also tested that the flux comparison between our NIRCam fluxes and those reported by the ASTRODEEP-JWST catalog \citep{Merlin2024} has less scatter, but the NIRCam fluxes of faint sources from ASTRODEEP-JWST can also be underestimated (by $\sim 10\%$) due to the lack of aperture loss corrections for Kron apertures.

%%%%%%%%%%%%%%%%%%%%%%%%%%%%%%%%%%%%%%%%%%%%%%%%%%%%%%%%%%%%%%%%%%%%%%%%%%%%%%%%
\section{Spectral Energy Distribution Fitting}
\label{sec:SED}
\subsection{Redshift estimation}
\label{sec:z_estimation}

\begin{figure}[htbp]
\centering
\includegraphics[width=0.45\textwidth]{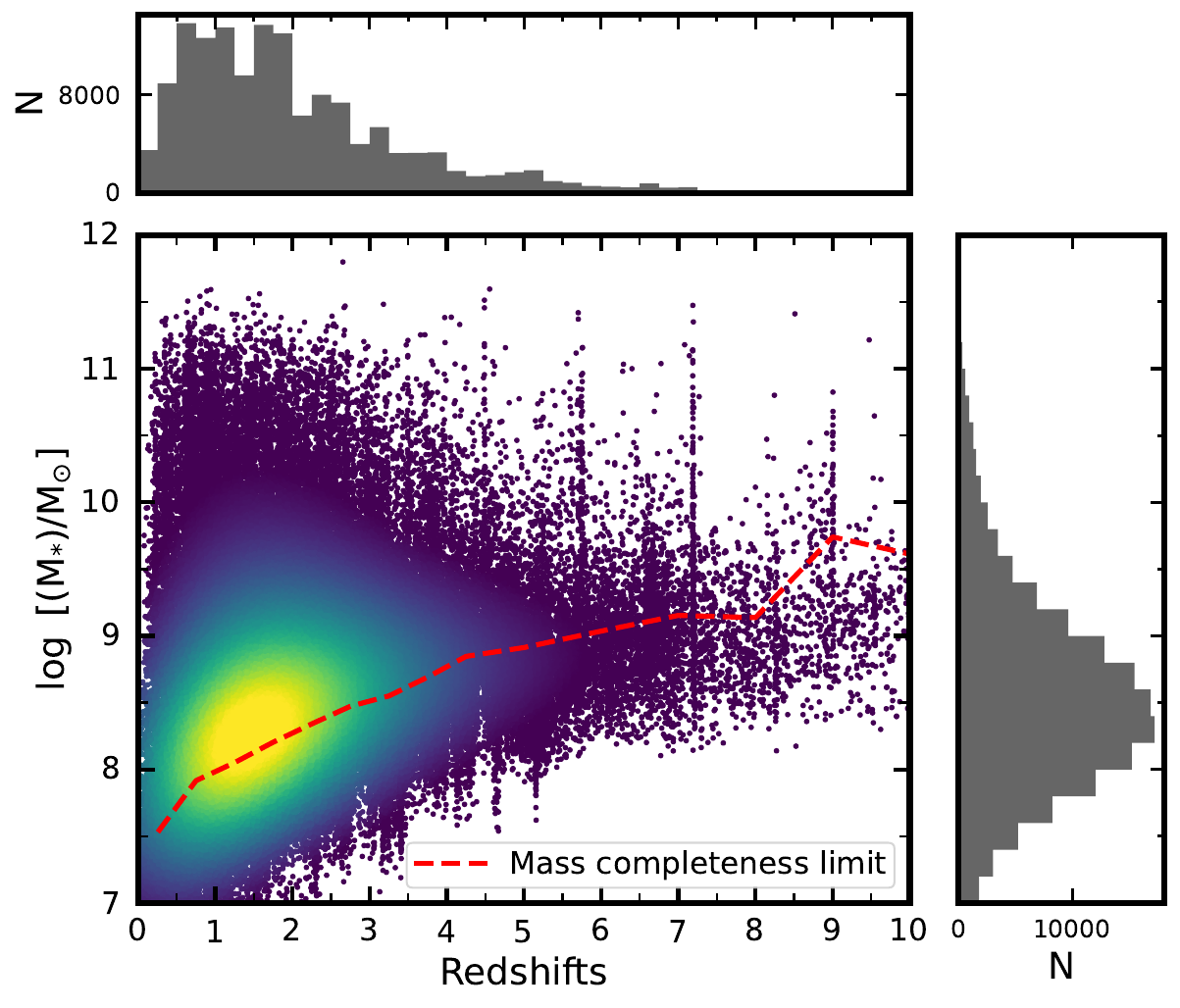}
\caption{\label{Fig:z_m}\textbf{Distribution of the redshifts and galaxy stellar masses in our catalog.} The histograms show the redshift and stellar mass distributions of our sample. The red dashed line shows the stellar mass completeness limit as a function of redshift, which is calculated following~\citet{Pozzetti2010}. Our catalog can provide a mass-complete sample of galaxies at $\log ({M_{\rm{*}}}/{M_ \odot }) > 9$ up to $z\sim8.5$.}
\end{figure}

\begin{figure*}[]
\centering
\includegraphics[width=0.4\textwidth]{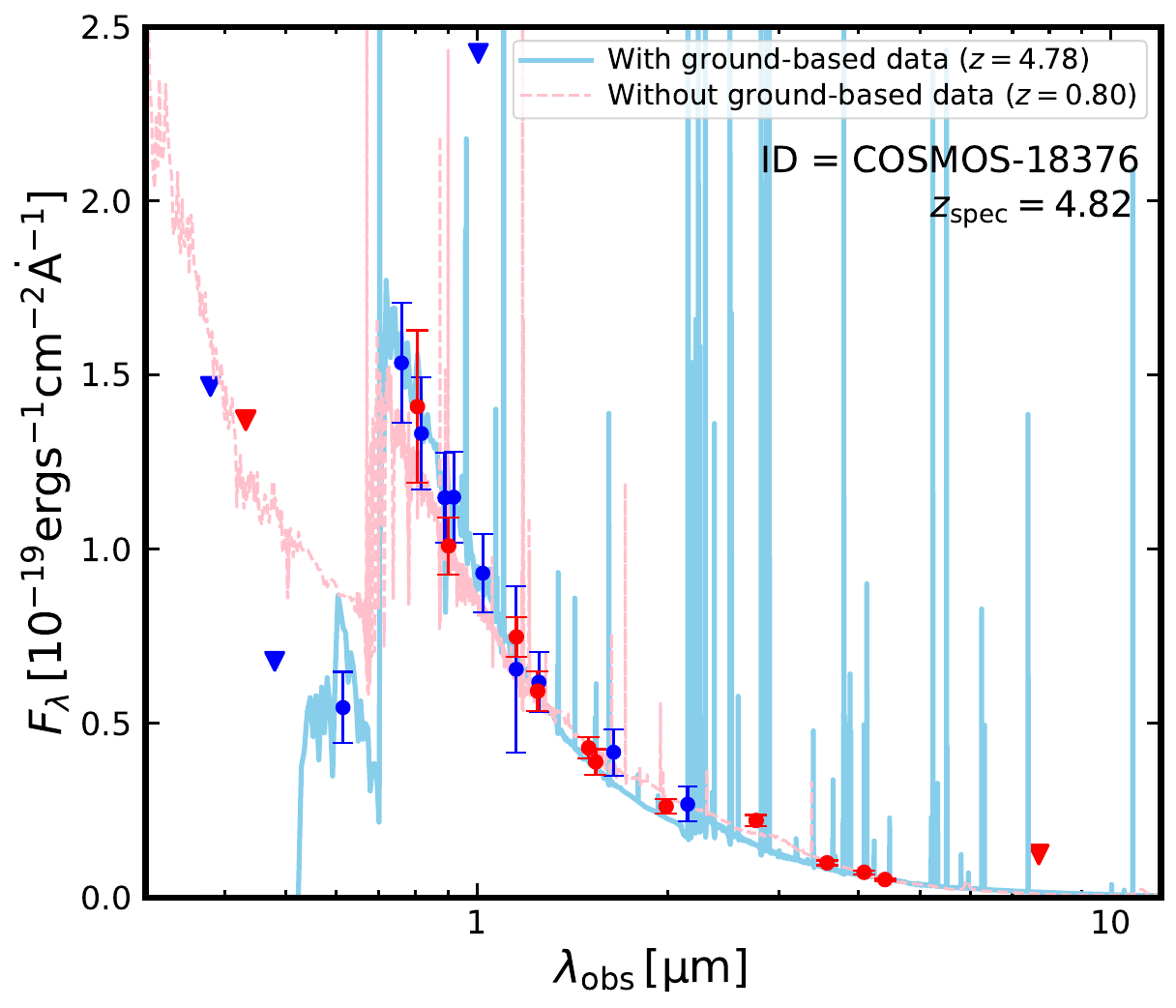}
\includegraphics[width=0.4\textwidth]{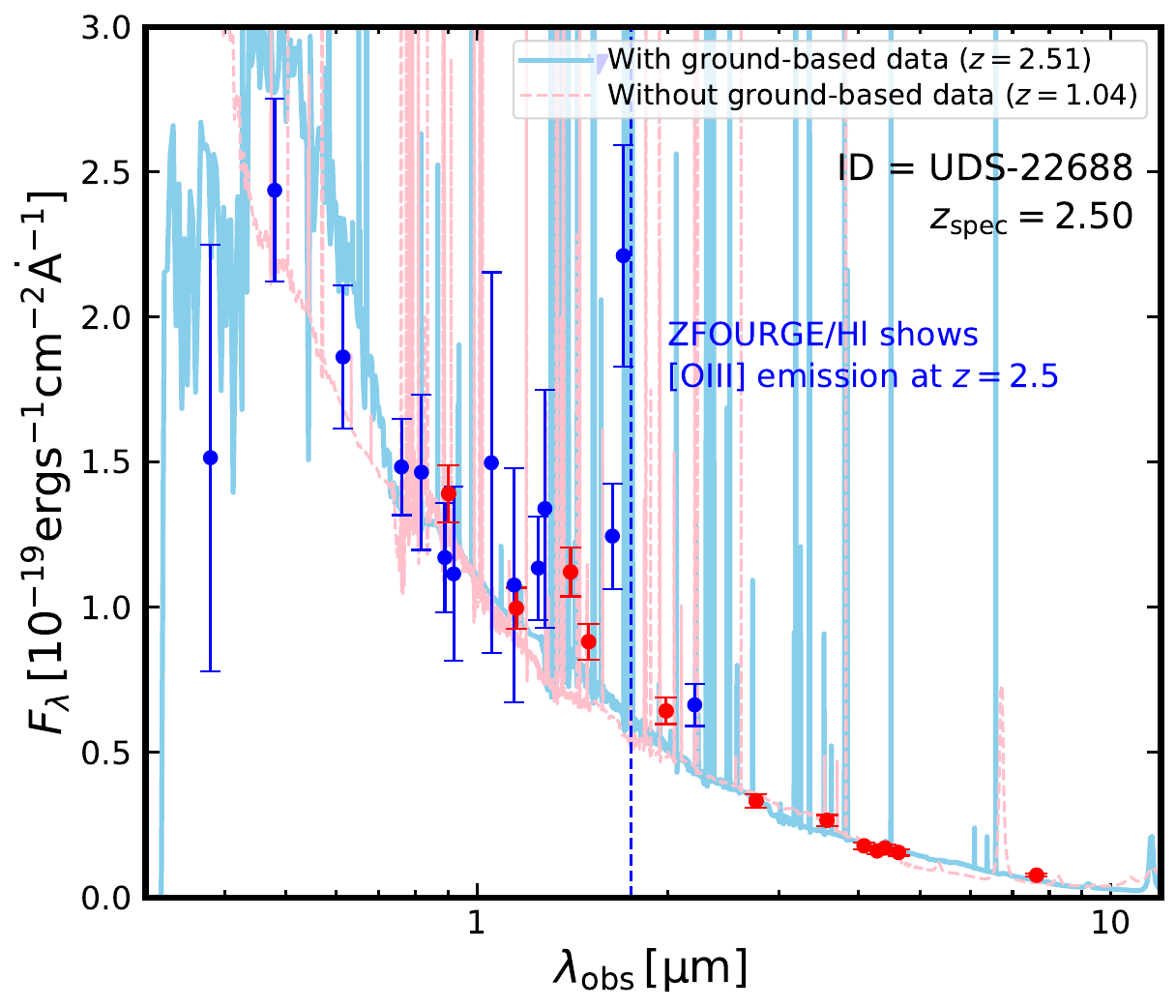}
\caption{\label{Fig:SED_zdiff} \textbf{Examples of galaxies with improved photometric redshifts with low-resolution data} The red and blue points show the photometric results from high-resolution and low-resolution data, respectively. The red lines show the best-fit SED from EAZY when we only use the high-resolution data from HST and JWST. The blue lines show the best-fit SED based on the full multi-wavelength catalog, including the low-resolution ground-based data.}
\end{figure*}

\begin{figure*}[]
\centering
\includegraphics[width=0.95\textwidth]{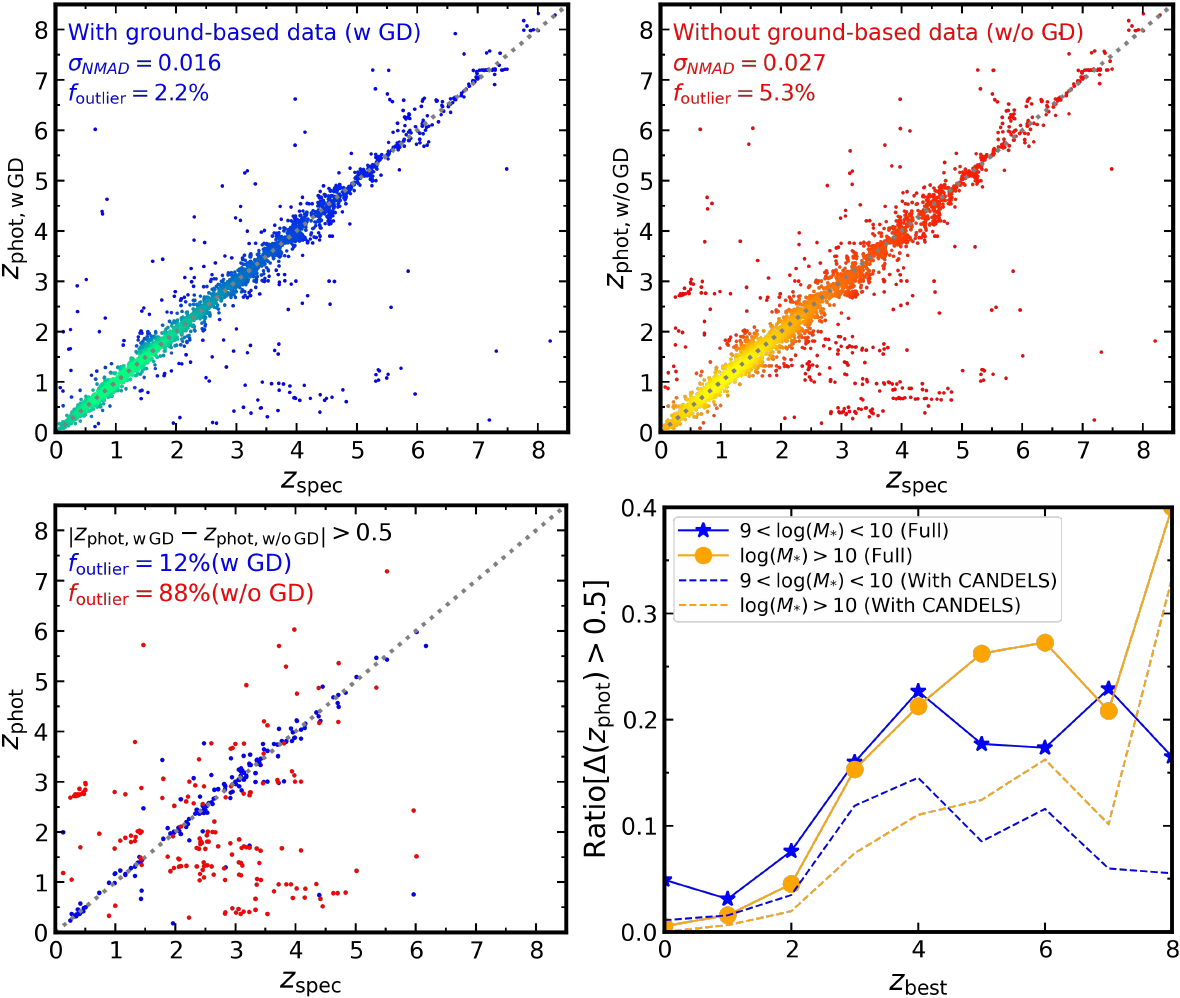}
\caption{\label{Fig:z_compare} \textbf{Validation of photometric redshifts. Upper left:} The comparison between photometric redshifts and spectroscopic redshifts in our final catalog with ground-based data. \textbf{Upper right} The same comparison when we only use the high-resolution data from HST and JWST. \textbf{Lower left:} The photometric and spectroscopic redshift comparison for galaxies that have different photometric redshifts ($\left| {{z_{{\rm{phot,w\thinspace GD}}}} - {z_{{\rm{phot,w/o\thinspace GD}}}}} \right| > 0.5$), the blue and red points represent the results with and without ground-based data, respectively. \textbf{Lower right:} The ratio of galaxies with different photometric redshifts ($\left| {{z_{{\rm{phot,w\thinspace GD}}}} - {z_{{\rm{phot,w/o\thinspace GD}}}}} \right| > 0.5$) as a function of redshift. The blue line and orange line shows the ratio of small galaxies ($9<\log ({M_{\rm{*}}}/{M_ \odot }) <10$) and massive galaxies ($\log ({M_{\rm{*}}}/{M_ \odot }) >10$), respectively. The solid and dashed lines show the ratios in the entire catalog and the region covered by the high-resolution HST data from the CANDELS survey, respectively.}
\end{figure*}
%Merlin+24: NMAD=0.033, f\_outlier = 6.1\%

Based on the UV to MIR photometric catalog, we first estimate the photometric redshifts by running SED fitting with EAZY \citep{Brammer2008}. Before SED fitting, Galactic extinction for all flux measurements is corrected using the dust maps from \citet{Schlafly2011}. For flux densities with SNR$<2$, we consider them to be unreliable and only use their value of $3 \times {f_{{\rm{err}}}}$ as the upper limits. For the other flux densities from high-resolution data and low-resolution data, we apply $5\%$ and $10\%$ systematic error floors, respectively. Then, we adopt the \_blue\_sfhz\_13 template library, which contains models dominated by strong emission lines, to obtain the minimum ${\chi ^2}$ redshifts with EAZY in the extended redshift range $z = 0-16$.

In addition to the photometric redshifts, we collect the available spectroscopic redshifts for our catalog. We first collect all publicly available JWST spectroscopy in the PRIMER-COSMOS and UDS fields from the DAWN JWST Archive (DJA v3.0). These JWST spectra have been reduced with the \textit{MSAEXP} pipeline \citep{msaexp} following the methods described in \citet{degraaff2024} and \citet{Heintz2024}. For the galaxies without JWST spectroscopy, we further search for the spectroscopic redshifts measured by other telescopes. In PRIMER-COSMOS, \citet{Khostovan2025} collected spectroscopic redshifts from 108 different programs observed in the last 20 years. In PRIMER-UDS, we use the spectroscopic redshifts collected by the FENIKS catalog \citep{Zaidi2024} and the VLT/VIMOS spectroscopy from the VANDELS survey \citep{Talia2023}. Combining those surveys, 10551 galaxies are found to have spectroscopic redshifts with high confidence levels, including 5379 in COSMOS and 5172 in UDS.

\subsection{Physical properties from Bagpipes}

Once the redshifts of galaxies are determined, we then perform SED fitting with fixed redshifts using Bagpipes \citep{Carnall2018} to estimate their physical properties, including but not limited to stellar masses, star formation rates, and rest-frame colors. With Bagpipes, we adopt the 2016 revision of the \citet{Bruzual2003} stellar-population synthesis models. Stellar ages are allowed to vary between 0.03-10 Gyr while metallicities span $Z/Z_\odot = 0.01-2.5$. The star-formation history is parameterized as a delayed exponential with a timescale $\rm \tau \in[0.01,10]~Gyr$. Dust attenuation is modeled with the \citet{Calzetti2000} model, which is allowed to be $A_V \in[0,5]$ and has been applied separately to young ($<0.01$Gyr) and old ($>0.01$Gyr) stellar populations.  Nebular emission is added following the \citet{Byler2017} prescription with the ionization parameter constrained to $\log U\in[-5,-2]$. 

Figure~\ref{Fig:z_m} presents the distributions of the redshifts and galaxy stellar masses in our catalog. To estimate the stellar mass completeness limit of our catalog, we adopt the method described by \citet{Pozzetti2010}, which is also used by many other catalogs \citep[e.g.][]{Laigle2016, Weaver2022, Shuntov2025}. In brief, for each source, we calculate its limiting stellar mass at which it can be detected at the magnitude limit
\begin{equation}
\label{eq_M_lim}
\log {M_{*,\lim }} = \log {M_*} - 0.4({\rm{ma}}{{\rm{g}}_{{\rm{F444W,lim}}}} - {\rm{ma}}{{\rm{g}}_{{\rm{F444W}}}}),
\end{equation}
where ${\rm{ma}}{{\rm{g}}_{{\rm{F444W,lim}}}} = 27.2$ is the limiting magnitude of our catalog (see Section~\ref{sec:dual_mode_detection}). Then, we take the 90th percentile of $\log {M_{*,\lim }}$ in each redshift bin as the limiting stellar mass. Based on this estimation, our catalog can provide an unbiased mass-complete census of high-redshift galaxies up to $z \sim 8.5$, while the sample at $z>9$ might be biased to the brightest galaxies. We note that the density peak at $z \sim 7.15$ is because many strong [OIII] emitters at $z = 6.8-7.5$ with flux excess in the F410M medium band are fitted at $z \sim 7.15$ by EAZY, which should be considered during the determination of redshift bins. We also remind that for the area without JWST/MIRI coverage, the stellar masses of massive galaxies at $z>5.5$ might be overestimated as shown in~\citet{WangT2025}.

\section{Discussion: Improved photometric redshifts with low-resolution data}
\begin{figure*}[]
\centering
\includegraphics[width=0.95\textwidth]{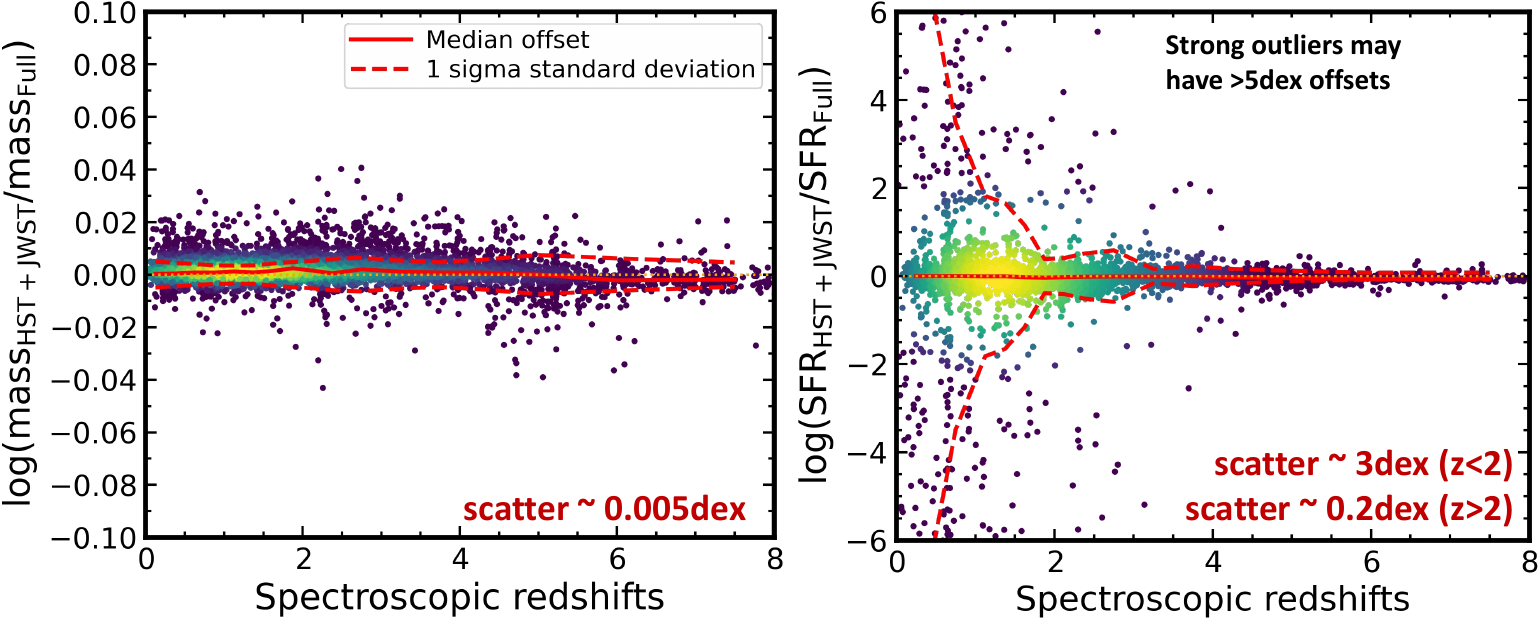}
\caption{\label{Fig:phy_compare} \textbf{Comparison of stellar masses and SFRs with and without ground-based data for the spectroscopic sample in our catalog. Left:} The ratio of stellar masses as a function of redshift. The red solid and dashed line shows the medium offset and the 1 sigma standard deviation, respectively. \textbf{Right:} The ratio of SFRs as a function of redshift. We find that the ground-based data can be important for the measurement of SFRs at $z<2$ because they can provide constraints in the rest-frame UV bands.}
\end{figure*}

\label{sec:discussion}
Compared with the previous JWST-based catalogs, a key improvement of our work is to make use of the abundant archival low-resolution data in the PRIMER-COSMOS and UDS fields. These low-resolution data help to improve the quality of SED fitting in two ways. First, the deep images from CFHT and Subaru provide better constraints on the SED in the UV-to-optical bands, which greatly help to break the degeneracy during SED fitting. Figure~\ref{Fig:SED_zdiff} shows an example of this issue. Using the HST and JWST data, the galaxy COSMOS-18376 is fitted at ${z_{{\rm{phot}}}} = 0.80$ with a Balmer break at ${\lambda _{{\rm{obs}}}}\sim 0.7{\rm{\mu m}}$. Once the low-resolution data are considered, it is clear that this Balmer break is actually a Lyman break. Therefore, its photometric redshift is corrected to be ${z_{{\rm{phot}}}} = 4.78$, which is close to the spectroscopic redshift ${z_{{\rm{spec}}}} = 4.82$. On the other hand, Subaru, VISTA, and Magellan can provide additional medium- and narrow-band data, which helps identify SED features such as emission lines and breaks. For example, the medium-band ${H_{\rm{l}}}$  filter from the ZFOURGE survey \citep{Straatman2016} identified the [O III] emission line for the galaxy UDS-22688 at ${z_{{\rm{spec}}}} = 2.50$, which was previously considered to be a low-redshift galaxy with only the high-resolution data.

In Figure~\ref{Fig:z_compare}, we test the reliability of our photometric redshifts by comparing them to spectroscopic redshifts. Firstly, when we only use the high-resolution data (from HST, JWST/NIRCam and MIRI) without ground-based data (w/o GD), the photometric redshifts are found to have a scatter ${\sigma _{{\rm{NMAD}}}} = 0.027$ and outlier fraction ${f_{{\rm{outlier}}}}(\left| {{z_{{\rm{spec}}}} - {z_{{\rm{phot}}}}} \right|/({z_{{\rm{spec}}}} + 1) > 0.15) = 5.3\% $. Similar to \citet{Merlin2024}, the unreliable sources identified by the flags in our catalog and the active galactic nucleus candidates (e.g., the Little Red Dots selected by \citealt{WangT2025}) are excluded from this comparison. We found that our result without low-resolution data is comparable to the ASTRODEEP-JWST catalog from \citet{Merlin2024}, who reported ${\sigma _{{\rm{NMAD}}}} = 0.033$ (0.033) and ${f_{{\rm{outlier}}}} = 6.1\% (4.5\%)$ in PRIMER-COSMOS (UDS). Once we add the low-resolution data, the quality of photometric redshifts can be improved with ${\sigma _{{\rm{NMAD}}}} = 0.016$ and ${f_{{\rm{outlier}}}} = 2.2\%$. If we only focus on the subsample of galaxies with $\left| {{z_{{\rm{phot,w\thinspace GD}}}} - {z_{{\rm{phot,w/o\thinspace GD}}}}} \right| > 0.5$, the outlier fraction in the JWST+HST catalog reaches ${f_{{\rm{outlier}}}} = 88\%$, while the outlier fraction in the full UV to MIR catalog is only ${f_{{\rm{outlier}}}} = 12\%$. 

In PRIMER-COSMOS, the photometric redshifts reported by our catalog have ${\sigma _{{\rm{NMAD}}}} = 0.013$ and ${f_{{\rm{outlier}}}} = 1.7\%$. After crossmatching this same spectroscopic sample with the COSMOS2025 catalog \citep{Shuntov2025}, we found that the photometric redshifts from COSMOS2025 have ${\sigma _{{\rm{NMAD}}}} = 0.016$ and ${f_{{\rm{outlier}}}} = 5.3\%$. Since COSMOS2025 has also combined the JWST data with the low-resolution ground-based data, our key improvement compared to COSMOS2025 is that the JWST mosaics used by our work are much deeper and have more bands.

In the lower right panel of Figure~\ref{Fig:z_compare}, we remove the requirement for spectroscopic data and check the fraction of galaxies with $\left| {{z_{{\rm{phot,w\thinspace GD}}}} - {z_{{\rm{phot,w/o\thinspace GD}}}}} \right| > 0.5$ in the entire catalog. After excluding the area with only the shallow JWST/NIRCam (F140M, F430M, and F460M) observations from the BEACON survey \citep{Morishita2025} and the faint sources with SNR$_{\rm{F444W}}<7$, we found that $\sim 25-30\%$ of the massive galaxies at $z \gtrsim 4$ could have incorrect photometric redshifts without the low-resolution data. Many of these galaxies are similar to COSMOS-18376 in Figure~\ref{Fig:SED_zdiff}, with their Lyman breaks mistaken as Balmer breaks at lower redshifts. Using the redshifts and stellar masses based on our catalog, the galaxy buildup is smoother across a large redshift range \citep{WangT2025} compared with previous results reporting rapid buildup of massive galaxies at $z = 4-6$ based on catalogs without low-resolution data \citep{Weibel2024}.
We remind that the quantitative results shown in Figure~\ref{Fig:z_compare} are dependent on the quality of the multi-wavelength data collection. As shown by the dashed lines, the ratio of massive galaxies with different photometric redshifts can be reduced from $\sim 25-30\%$ to $\sim 15\%$ when the high-resolution optical HST data are available from the CANDELS survey. This suggests that although the optical data from HST help to improve the photometric redshifts, the low-resolution ground-based data are still required to give the best estimations of photometric redshifts. 

We further test if the ground-based data are necessary for the estimation of physical properties, such as stellar masses and star formation rate (SFRs), for the spectroscopic sample with known redshifts. As shown in Figure~\ref{Fig:phy_compare}, estimation of stellar masses does not require ground-based data because stellar masses are mostly constrained by the JWST data at long wavelength, especially the JWST/MIRI data \citep{WangT2025}. However, the SFRs without ground-based data can be highly uncertain at $z<2$ because JWST can no longer constrain the rest-frame UV emission at this redshift range. For strong outliers at $z<2$, there can be a $>5$dex offset in SFR, showing that we cannot make a reliable distinction between quiescent galaxies and star-forming galaxies at $z<2$ without ground-based data. We note that the ground-based data can be even more important for the sources without spectroscopic redshifts because their physical properties can be directly biased by incorrect luminosity distances based on inaccurate photometric redshifts (Figure~\ref{Fig:z_compare}).

%%%%%%%%%%%%%%%%%%%%%%%%%%%%%%%%%%%%%%%%%%%%%%%%%%%%%%%%%%%%%%%%%%%%%%%%%%%%%%%%

%%%%%%%%%%%%%%%%%%%%%%%%%%%%%%%%%%%%%%%%%%%%%%%%%%%%%%%%%%%%%%%%%%%%%%%%%%%%%%%%
\section{Conclusions}
\label{sec:conclu}
In this paper, we presented the first data release of the ULTIMATE-deblending project: a deblended photometric catalog across UV to MIR based on JWST/PRIMER. We present the reduction of JWST mosaics, the construction of the UV to MIR photometric catalog, and the SED fitting. The JWST mosaics and catalogs listing the results of multiwavelength photometry and SED fitting are publicly available at \dataset[http://www.taoofcosmos.space/ultimate/]{http://www.taoofcosmos.space/ultimate/}.

Based on the deep JWST images of the PRIMER-COSMOS and PRIMER-UDS fields reduced by our modified JWST pipeline, we performed dual-mode source extraction on the combined NIRCam LW image. We conducted aperture photometry with PSF matching for the high-resolution data from JWST/NIRCam and HST, and TPHOT deblending photometry for the other bands with lower resolution. These yield a 50-band UV-to-MIR catalog for 308239 sources in a survey area of 627.1 arcmin$^2$.

Using this comprehensive multi-wavelength catalog, we performed SED fitting to obtain the photometric redshifts and physical properties of the detected sources. We discussed that the additional bands from low-resolution ground-based telescopes improve the accuracy of photometric redshifts by providing extended wavelength coverage in the UV to optical and emphasizing the emission lines and breaks. These accurate photometric redshifts are the foundation of a reliable photometric census of high-redshift galaxies.

In the following works of the ULTIMATE-deblending project, we will further collect new JWST observations in the PRIMER fields,
 %(such as the MINERVA survey \citep{Muzzin2025})
 extend our project to the FIR and radio bands, and then construct UV to Radio catalogs for other JWST deep fields. These will provide a high-quality reference sample of high-redshift galaxies, enabling robust constraints on the stellar mass assembly history of galaxies in the early Universe.

%%%%%%%%%%%%%%%%%%%%%%%%%%%%%%%%%%%%%%%%%%%%%%%%%%%%%%%%%%%%%%%%%%%%%%%%%%%%%%%%

\begin{acknowledgements}
%\subsection*{Acknowledgements}
We thank the anonymous referee and Dr. Mark Dickinson for the helpful comments on this paper, and thank Dr. James Dunlop for his leadership and design of the JWST/PRIMER survey. This work was supported by National Natural Science Foundation of China (Grant No.12525302 and 12141301), Natural Science Foundation of Jiangsu Higher Education Institutions of China(Grant No. BK20250001), National Key R\&D Program of China (Grant no. 2023YFA1605600), Scientific Research Innovation Capability Support Project for Young Faculty (Project No. ZYGXQNJSKYCXNLZCXM-P3), the Fundamental Research Funds for the Central Universities with Grant no.KG202502, and the China Manned Space Program with grant no. CMS-CSST-2025-A04. Y.J.W. acknowledges supports by Jiangsu Natural Science Foundation (Project No. BK20241188) and National Natural Science Foundation of China (Project No. 12403019). V.Sangalli acknowledge the support of France 2030 through the project named Académie Spatiale d’Île-de-France (https://academiespatiale.fr/) managed by the National Research Agency under bearing the reference ANR-23-CMAS-0041. 

This work is based on observations made with the NASA/ESA/CSA James Webb Space Telescope.  The raw JWST imaging data are obtained from the Mikulski Archive for Space Telescopes (MAST) and are available at \href{http://dx.doi.org/10.17909/e1wh-b970}{10.17909/e1wh-b970}. MAST is operated by the Association of Universities for Research in Astronomy, Inc., under NASA contract NAS 5–03127 for JWST. These observations are associated with programmes 1727, 1837, 1840, 2321, 2514, and 3990.

The HST mosaics and the JWST spectroscopic data used in this work are collected from the Dawn JWST Archive (DJA). DJA is an initiative of the Cosmic Dawn Center (DAWN), which is funded by the Danish National Research Foundation under grant DNRF140.

The images of VISTA/VIRCAM are based on observations collected at the European Southern Observatory under ESO programmes 179.A-2005 and 198.A-2003, data obtained from the ESO Science Archive Facility with DOI https://doi.org/10.18727/archive/52, and data products produced by CALET and the Cambridge Astronomy Survey Unit on behalf of the UltraVISTA consortium. 

UKIDSS was undertaken on UKIRT while it was operated by the Joint Astronomy Centre on behalf of the Science and Technology Facilities Council of the UK, which also provided support for the Cambridge Astronomical Survey Unit (CASU) and the Edinburgh Wide Field Astronomy Unit (WFAU) that generated and served the wide-field infrared public surveys from UKIRT.

The images of Subaru/HSC are provided by the Hyper Suprime-Cam (HSC) collaboration, which includes the astronomical communities of Japan and Taiwan, and Princeton University. The HSC instrumentation and software were developed by the National Astronomical Observatory of Japan (NAOJ), the Kavli Institute for the Physics and Mathematics of the Universe (Kavli IPMU), the University of Tokyo, the High Energy Accelerator Research Organization (KEK), the Academia Sinica Institute for Astronomy and Astrophysics in Taiwan (ASIAA), and Princeton University. Funding was contributed by the FIRST program from the Japanese Cabinet Office, the Ministry of Education, Culture, Sports, Science and Technology (MEXT), the Japan Society for the Promotion of Science (JSPS), Japan Science and Technology Agency (JST), the Toray Science Foundation, NAOJ, Kavli IPMU, KEK, ASIAA, and Princeton University. These Subaru/HSC data make use of software developed for the Vera C. Rubin Observatory. We thank the Rubin Observatory for making their code available as free software at http://pipelines.lsst.io/. The Subaru/HSC data were collected at the Subaru Telescope and retrieved from the HSC data archive system, which is operated by the Subaru Telescope and Astronomy Data Center (ADC) at NAOJ. Data analysis was in part carried out with the cooperation of Center for Computational Astrophysics (CfCA), NAOJ. We are honored and grateful for the opportunity of observing the Universe from Maunakea, which has the cultural, historical and natural significance in Hawaii. 

\end{acknowledgements}

%%%%%%%%%%%%%%%%%%%%%%%%%%%%%%%%%%%%%%%%%%%%%%%%%%%%%%%%%%%%%%%%%%%%%%%%%%%%%%%%

\bibliographystyle{aasjournalv7}
\bibliography{reference}

@software{msaexp,
       author = {{Brammer}, Gabriel},
        title = "{msaexp: NIRSpec analyis tools}",
         year = 2023,
        month = sep,
          eid = {10.5281/zenodo.7299500},
          doi = {10.5281/zenodo.7299500},
      version = {0.6.17},
    publisher = {Zenodo},
       adsurl = {https://ui.adsabs.harvard.edu/abs/2022zndo...7299500B}
}

@ARTICLE{Kron1980,
       author = {{Kron}, R.~G.},
        title = "{Photometry of a complete sample of faint galaxies.}",
      journal = {\apjs},
         year = 1980,
        month = jun,
       volume = {43},
        pages = {305-325},
          doi = {10.1086/190669},
       adsurl = {https://ui.adsabs.harvard.edu/abs/1980ApJS...43..305K}
}

@ARTICLE{Oke1983,
       author = {{Oke}, J.~B. and {Gunn}, J.~E.},
        title = "{Secondary standard stars for absolute spectrophotometry.}",
      journal = {\apj},
         year = 1983,
        month = mar,
       volume = {266},
        pages = {713-717},
          doi = {10.1086/160817},
       adsurl = {https://ui.adsabs.harvard.edu/abs/1983ApJ...266..713O}
}

@ARTICLE{Bertin1996,
       author = {{Bertin}, E. and {Arnouts}, S.},
        title = "{SExtractor: Software for source extraction.}",
      journal = {A\&AS},
         year = 1996,
        month = jun,
       volume = {117},
        pages = {393-404},
          doi = {10.1051/aas:1996164},
       adsurl = {https://ui.adsabs.harvard.edu/abs/1996A&AS..117..393B}
}

@ARTICLE{Calzetti2000,
       author = {{Calzetti}, Daniela and {Armus}, Lee and {Bohlin}, Ralph C. and {Kinney}, Anne L. and {Koornneef}, Jan and {Storchi-Bergmann}, Thaisa},
        title = "{The Dust Content and Opacity of Actively Star-forming Galaxies}",
      journal = {\apj},
         year = 2000,
        month = apr,
       volume = {533},
       number = {2},
        pages = {682-695},
          doi = {10.1086/308692},
archivePrefix = {arXiv},
       eprint = {astro-ph/9911459},
 primaryClass = {astro-ph},
       adsurl = {https://ui.adsabs.harvard.edu/abs/2000ApJ...533..682C}
}

@ARTICLE{Kroupa2001,
       author = {{Kroupa}, Pavel},
        title = "{On the variation of the initial mass function}",
      journal = {\mnras},
         year = 2001,
        month = apr,
       volume = {322},
       number = {2},
        pages = {231-246},
          doi = {10.1046/j.1365-8711.2001.04022.x},
archivePrefix = {arXiv},
       eprint = {astro-ph/0009005},
 primaryClass = {astro-ph},
       adsurl = {https://ui.adsabs.harvard.edu/abs/2001MNRAS.322..231K}
}

@ARTICLE{Bruzual2003,
       author = {{Bruzual}, G. and {Charlot}, S.},
        title = "{Stellar population synthesis at the resolution of 2003}",
      journal = {\mnras},
         year = 2003,
        month = oct,
       volume = {344},
       number = {4},
        pages = {1000-1028},
          doi = {10.1046/j.1365-8711.2003.06897.x},
archivePrefix = {arXiv},
       eprint = {astro-ph/0309134},
 primaryClass = {astro-ph},
       adsurl = {https://ui.adsabs.harvard.edu/abs/2003MNRAS.344.1000B}
}

@ARTICLE{Sanders2007,
       author = {{Sanders}, D.~B. and {Salvato}, M. and {Aussel}, H. and {Ilbert}, O. and {Scoville}, N. and {Surace}, J.~A. and {Frayer}, D.~T. and {Sheth}, K. and {Helou}, G. and {Brooke}, T. and {Bhattacharya}, B. and {Yan}, L. and {Kartaltepe}, J.~S. and {Barnes}, J.~E. and {Blain}, A.~W. and {Calzetti}, D. and {Capak}, P. and {Carilli}, C. and {Carollo}, C.~M. and {Comastri}, A. and {Daddi}, E. and {Ellis}, R.~S. and {Elvis}, M. and {Fall}, S.~M. and {Franceschini}, A. and {Giavalisco}, M. and {Hasinger}, G. and {Impey}, C. and {Koekemoer}, A. and {Le F{\`e}vre}, O. and {Lilly}, S. and {Liu}, M.~C. and {McCracken}, H.~J. and {Mobasher}, B. and {Renzini}, A. and {Rich}, M. and {Schinnerer}, E. and {Shopbell}, P.~L. and {Taniguchi}, Y. and {Thompson}, D.~J. and {Urry}, C.~M. and {Williams}, J.~P.},
        title = "{S-COSMOS: The Spitzer Legacy Survey of the Hubble Space Telescope ACS 2 deg$^{2}$ COSMOS Field I: Survey Strategy and First Analysis}",
      journal = {\apjs},
         year = 2007,
        month = sep,
       volume = {172},
       number = {1},
        pages = {86-98},
          doi = {10.1086/517885},
archivePrefix = {arXiv},
       eprint = {astro-ph/0701318},
 primaryClass = {astro-ph},
       adsurl = {https://ui.adsabs.harvard.edu/abs/2007ApJS..172...86S}
}

@ARTICLE{Scoville2007,
       author = {{Scoville}, N. and {Aussel}, H. and {Brusa}, M. and {Capak}, P. and {Carollo}, C.~M. and {Elvis}, M. and {Giavalisco}, M. and {Guzzo}, L. and {Hasinger}, G. and {Impey}, C. and {Kneib}, J. -P. and {LeFevre}, O. and {Lilly}, S.~J. and {Mobasher}, B. and {Renzini}, A. and {Rich}, R.~M. and {Sanders}, D.~B. and {Schinnerer}, E. and {Schminovich}, D. and {Shopbell}, P. and {Taniguchi}, Y. and {Tyson}, N.~D.},
        title = "{The Cosmic Evolution Survey (COSMOS): Overview}",
      journal = {\apjs},
         year = 2007,
        month = sep,
       volume = {172},
       number = {1},
        pages = {1-8},
          doi = {10.1086/516585},
archivePrefix = {arXiv},
       eprint = {astro-ph/0612305},
 primaryClass = {astro-ph},
       absurl = {https://ui.adsabs.harvard.edu/abs/2007ApJS..172....1S}
}

@ARTICLE{Brammer2008,
   author = {{Brammer}, G.~B. and {van Dokkum}, P.~G. and {Coppi}, P.},
    title = "{EAZY: A Fast, Public Photometric Redshift Code}",
  journal = {\apj},
archivePrefix = "arXiv",
   eprint = {0807.1533},
     year = 2008,
    month = oct,
   volume = 686,
    pages = {1503-1513},
      doi = {10.1086/591786},
   adsurl = {http://adsabs.harvard.edu/abs/2008ApJ...686.1503B}
}

@ARTICLE{Pozzetti2010,
       author = {{Pozzetti}, L. and {Bolzonella}, M. and {Zucca}, E. and {Zamorani}, G. and {Lilly}, S. and {Renzini}, A. and {Moresco}, M. and {Mignoli}, M. and {Cassata}, P. and {Tasca}, L. and {Lamareille}, F. and {Maier}, C. and {Meneux}, B. and {Halliday}, C. and {Oesch}, P. and {Vergani}, D. and {Caputi}, K. and {Kova{\v{c}}}, K. and {Cimatti}, A. and {Cucciati}, O. and {Iovino}, A. and {Peng}, Y. and {Carollo}, M. and {Contini}, T. and {Kneib}, J.-P. and {Le F{\'e}vre}, O. and {Mainieri}, V. and {Scodeggio}, M. and {Bardelli}, S. and {Bongiorno}, A. and {Coppa}, G. and {de la Torre}, S. and {de Ravel}, L. and {Franzetti}, P. and {Garilli}, B. and {Kampczyk}, P. and {Knobel}, C. and {Le Borgne}, J.-F. and {Le Brun}, V. and {Pell{\`o}}, R. and {Perez Montero}, E. and {Ricciardelli}, E. and {Silverman}, J.~D. and {Tanaka}, M. and {Tresse}, L. and {Abbas}, U. and {Bottini}, D. and {Cappi}, A. and {Guzzo}, L. and {Koekemoer}, A.~M. and {Leauthaud}, A. and {Maccagni}, D. and {Marinoni}, C. and {McCracken}, H.~J. and {Memeo}, P. and {Porciani}, C. and {Scaramella}, R. and {Scarlata}, C. and {Scoville}, N.},
        title = "{zCOSMOS - 10k-bright spectroscopic sample. The bimodality in the galaxy stellar mass function: exploring its evolution with redshift}",
      journal = {\aap},
         year = 2010,
        month = nov,
       volume = {523},
          eid = {A13},
        pages = {A13},
          doi = {10.1051/0004-6361/200913020},
archivePrefix = {arXiv},
       eprint = {0907.5416},
 primaryClass = {astro-ph.CO},
       adsurl = {https://ui.adsabs.harvard.edu/abs/2010A&A...523A..13P}
}

@ARTICLE{Huang2011,
       author = {{Huang}, J.-S. and {Zheng}, X.~Z. and {Rigopoulou}, D. and {Magdis}, G. and {Fazio}, G.~G. and {Wang}, T.},
        title = "{Four IRAC Sources with an Extremely Red H - [3.6] Color: Passive or Dusty Galaxies at z > 4.5?}",
      journal = {\apjl},
         year = 2011,
        month = nov,
       volume = {742},
       number = {1},
          eid = {L13},
        pages = {L13},
          doi = {10.1088/2041-8205/742/1/L13},
archivePrefix = {arXiv},
       eprint = {1110.4129},
 primaryClass = {astro-ph.CO},
       adsurl = {https://ui.adsabs.harvard.edu/abs/2011ApJ...742L..13H}
}

@ARTICLE{Schlafly2011,
       author = {{Schlafly}, Edward F. and {Finkbeiner}, Douglas P.},
        title = "{Measuring Reddening with Sloan Digital Sky Survey Stellar Spectra and Recalibrating SFD}",
      journal = {\apj},
         year = 2011,
        month = aug,
       volume = {737},
       number = {2},
          eid = {103},
        pages = {103},
          doi = {10.1088/0004-637X/737/2/103},
archivePrefix = {arXiv},
       eprint = {1012.4804},
 primaryClass = {astro-ph.GA},
       adsurl = {https://ui.adsabs.harvard.edu/abs/2011ApJ...737..103S}
}

@ARTICLE{McCracken2012,
       author = {{McCracken}, H.~J. and {Milvang-Jensen}, B. and {Dunlop}, J. and {Franx}, M. and {Fynbo}, J.~P.~U. and {Le F{\`e}vre}, O. and {Holt}, J. and {Caputi}, K.~I. and {Goranova}, Y. and {Buitrago}, F. and {Emerson}, J.~P. and {Freudling}, W. and {Hudelot}, P. and {L{\'o}pez-Sanjuan}, C. and {Magnard}, F. and {Mellier}, Y. and {M{\o}ller}, P. and {Nilsson}, K.~K. and {Sutherland}, W. and {Tasca}, L. and {Zabl}, J.},
        title = "{UltraVISTA: a new ultra-deep near-infrared survey in COSMOS}",
      journal = {\aap},
         year = 2012,
        month = aug,
       volume = {544},
          eid = {A156},
        pages = {A156},
          doi = {10.1051/0004-6361/201219507},
archivePrefix = {arXiv},
       eprint = {1204.6586},
 primaryClass = {astro-ph.CO},
       adsurl = {https://ui.adsabs.harvard.edu/abs/2012A&A...544A.156M}
}

@ARTICLE{Galametz2013,
       author = {{Galametz}, Audrey and {Grazian}, Andrea and {Fontana}, Adriano and {Ferguson}, Henry C. and {Ashby}, M.~L.~N. and {Barro}, Guillermo and {Castellano}, Marco and {Dahlen}, Tomas and {Donley}, Jennifer L. and {Faber}, Sandy M. and {Grogin}, Norman and {Guo}, Yicheng and {Huang}, Kuang-Han and {Kocevski}, Dale D. and {Koekemoer}, Anton M. and {Lee}, Kyoung-Soo and {McGrath}, Elizabeth J. and {Peth}, Michael and {Willner}, S.~P. and {Almaini}, Omar and {Cooper}, Michael and {Cooray}, Asantha and {Conselice}, Christopher J. and {Dickinson}, Mark and {Dunlop}, James S. and {Fazio}, G.~G. and {Foucaud}, Sebastien and {Gardner}, Jonathan P. and {Giavalisco}, Mauro and {Hathi}, N.~P. and {Hartley}, Will G. and {Koo}, David C. and {Lai}, Kamson and {de Mello}, Duilia F. and {McLure}, Ross J. and {Lucas}, Ray A. and {Paris}, Diego and {Pentericci}, Laura and {Santini}, Paola and {Simpson}, Chris and {Sommariva}, Veronica and {Targett}, Thomas and {Weiner}, Benjamin J. and {Wuyts}, Stijn and {CANDELS Team}},
        title = "{CANDELS Multiwavelength Catalogs: Source Identification and Photometry in the CANDELS UKIDSS Ultra-deep Survey Field}",
      journal = {\apjs},
         year = 2013,
        month = jun,
       volume = {206},
       number = {2},
          eid = {10},
        pages = {10},
          doi = {10.1088/0067-0049/206/2/10},
archivePrefix = {arXiv},
       eprint = {1305.1823},
 primaryClass = {astro-ph.CO},
       adsurl = {https://ui.adsabs.harvard.edu/abs/2013ApJS..206...10G}
}

@ARTICLE{Guo2013,
       author = {{Guo}, Yicheng and {Ferguson}, Henry C. and {Giavalisco}, Mauro and {Barro}, Guillermo and {Willner}, S.~P. and {Ashby}, Matthew L.~N. and {Dahlen}, Tomas and {Donley}, Jennifer L. and {Faber}, Sandra M. and {Fontana}, Adriano and {Galametz}, Audrey and {Grazian}, Andrea and {Huang}, Kuang-Han and {Kocevski}, Dale D. and {Koekemoer}, Anton M. and {Koo}, David C. and {McGrath}, Elizabeth J. and {Peth}, Michael and {Salvato}, Mara and {Wuyts}, Stijn and {Castellano}, Marco and {Cooray}, Asantha R. and {Dickinson}, Mark E. and {Dunlop}, James S. and {Fazio}, G.~G. and {Gardner}, Jonathan P. and {Gawiser}, Eric and {Grogin}, Norman A. and {Hathi}, Nimish P. and {Hsu}, Li-Ting and {Lee}, Kyoung-Soo and {Lucas}, Ray A. and {Mobasher}, Bahram and {Nandra}, Kirpal and {Newman}, Jeffery A. and {van der Wel}, Arjen},
        title = "{CANDELS Multi-wavelength Catalogs: Source Detection and Photometry in the GOODS-South Field}",
      journal = {\apjs},
         year = 2013,
        month = aug,
       volume = {207},
       number = {2},
          eid = {24},
        pages = {24},
          doi = {10.1088/0067-0049/207/2/24},
archivePrefix = {arXiv},
       eprint = {1308.4405},
 primaryClass = {astro-ph.CO},
       adsurl = {https://ui.adsabs.harvard.edu/abs/2013ApJS..207...24G}
}

@ARTICLE{Jarvis2013,
       author = {{Jarvis}, Matt J. and {Bonfield}, D.~G. and {Bruce}, V.~A. and {Geach}, J.~E. and {McAlpine}, K. and {McLure}, R.~J. and {Gonz{\'a}lez-Solares}, E. and {Irwin}, M. and {Lewis}, J. and {Yoldas}, A. Kupcu and {Andreon}, S. and {Cross}, N.~J.~G. and {Emerson}, J.~P. and {Dalton}, G. and {Dunlop}, J.~S. and {Hodgkin}, S.~T. and {Le}, F{\`e}vre O. and {Karouzos}, M. and {Meisenheimer}, K. and {Oliver}, S. and {Rawlings}, S. and {Simpson}, C. and {Smail}, I. and {Smith}, D.~J.~B. and {Sullivan}, M. and {Sutherland}, W. and {White}, S.~V. and {Zwart}, J.~T.~L.},
        title = "{The VISTA Deep Extragalactic Observations (VIDEO) survey}",
      journal = {\mnras},
         year = 2013,
        month = jan,
       volume = {428},
       number = {2},
        pages = {1281-1295},
          doi = {10.1093/mnras/sts118},
archivePrefix = {arXiv},
       eprint = {1206.4263},
 primaryClass = {astro-ph.CO},
       adsurl = {https://ui.adsabs.harvard.edu/abs/2013MNRAS.428.1281J}
}

@ARTICLE{Skelton2014,
       author = {{Skelton}, Rosalind E. and {Whitaker}, Katherine E. and {Momcheva}, Ivelina G. and {Brammer}, Gabriel B. and {van Dokkum}, Pieter G. and {Labb{\'e}}, Ivo and {Franx}, Marijn and {van der Wel}, Arjen and {Bezanson}, Rachel and {Da Cunha}, Elisabete and {Fumagalli}, Mattia and {F{\"o}rster Schreiber}, Natascha and {Kriek}, Mariska and {Leja}, Joel and {Lundgren}, Britt F. and {Magee}, Daniel and {Marchesini}, Danilo and {Maseda}, Michael V. and {Nelson}, Erica J. and {Oesch}, Pascal and {Pacifici}, Camilla and {Patel}, Shannon G. and {Price}, Sedona and {Rix}, Hans-Walter and {Tal}, Tomer and {Wake}, David A. and {Wuyts}, Stijn},
        title = "{3D-HST WFC3-selected Photometric Catalogs in the Five CANDELS/3D-HST Fields: Photometry, Photometric Redshifts, and Stellar Masses}",
      journal = {\apjs},
         year = 2014,
        month = oct,
       volume = {214},
       number = {2},
          eid = {24},
        pages = {24},
          doi = {10.1088/0067-0049/214/2/24},
archivePrefix = {arXiv},
       eprint = {1403.3689},
 primaryClass = {astro-ph.GA},
       adsurl = {https://ui.adsabs.harvard.edu/abs/2014ApJS..214...24S}
}

@INPROCEEDINGS{Koekemoer2014AAS,
       author = {{Koekemoer}, Anton M. and {Avila}, R.~J. and {Hammer}, D. and {Mack}, J. and {Ogaz}, S. and {Anderson}, J. and {Barker}, E.~A. and {Hilbert}, B. and {Gonzaga}, S. and {Grogin}, N.~A. and {Fruchter}, A.~S. and {Lotz}, J. and {Lucas}, R.~A. and {Mountain}, M. and {Sokol}, J.},
        title = "{The HST Frontier Fields: Science Data Pipeline, Products, and First Data Release}",
    booktitle = {American Astronomical Society Meeting Abstracts \#223},
         year = 2014,
       series = {American Astronomical Society Meeting Abstracts},
       volume = {223},
        month = jan,
          eid = {254.02},
        pages = {254.02},
       adsurl = {https://ui.adsabs.harvard.edu/abs/2014AAS...22325402K}
}

@INPROCEEDINGS{Lotz2014,
       author = {{Lotz}, Jennifer and {Mountain}, M. and {Grogin}, N.~A. and {Koekemoer}, A.~M. and {Capak}, P.~L. and {Mack}, J. and {Coe}, D.~A. and {Barker}, E.~A. and {Adler}, D.~S. and {Avila}, R.~J. and {Anderson}, J. and {Casertano}, S. and {Christian}, C.~A. and {Gonzaga}, S. and {Ferguson}, H.~C. and {Fruchter}, A.~S. and {Jenkner}, H. and {Jordan}, I.~J. and {Hammer}, D. and {Hilbert}, B. and {Lawton}, B.~L. and {Lee}, J.~C. and {Lucas}, R.~A. and {MacKenty}, J.~W. and {Mutchler}, M.~J. and {Ogaz}, S. and {Reid}, I.~N. and {Royle}, P. and {Robberto}, M. and {Sembach}, K. and {Smith}, L.~J. and {Sokol}, J. and {Surace}, J.~A. and {Taylor}, D. and {Tumlinson}, J. and {Viana}, A. and {Williams}, R.~E. and {Workman}, W.},
        title = "{The HST Frontier Fields}",
    booktitle = {American Astronomical Society Meeting Abstracts \#223},
         year = 2014,
       series = {American Astronomical Society Meeting Abstracts},
       volume = {223},
        month = jan,
          eid = {254.01},
        pages = {254.01},
       adsurl = {https://ui.adsabs.harvard.edu/abs/2014AAS...22325401L}
}

@ARTICLE{Steinhardt2014,
       author = {{Steinhardt}, Charles L. and {Speagle}, Josh S. and {Capak}, Peter and {Silverman}, John D. and {Carollo}, Marcella and {Dunlop}, James and {Hashimoto}, Yasuhiro and {Hsieh}, Bau-Ching and {Ilbert}, Olivier and {Le Fevre}, Olivier and {Le Floc'h}, Emeric and {Lee}, Nicholas and {Lin}, Lihwai and {Lin}, Yen-Ting and {Masters}, Dan and {McCracken}, Henry J. and {Nagao}, Tohru and {Petric}, Andreea and {Salvato}, Mara and {Sanders}, Dave and {Scoville}, Nick and {Sheth}, Kartik and {Strauss}, Michael A. and {Taniguchi}, Yoshiaki},
        title = "{Star Formation at 4 < z < 6 from the Spitzer Large Area Survey with Hyper-Suprime-Cam (SPLASH)}",
      journal = {\apjl},
         year = 2014,
        month = aug,
       volume = {791},
       number = {2},
          eid = {L25},
        pages = {L25},
          doi = {10.1088/2041-8205/791/2/L25},
archivePrefix = {arXiv},
       eprint = {1407.7030},
 primaryClass = {astro-ph.GA},
       adsurl = {https://ui.adsabs.harvard.edu/abs/2014ApJ...791L..25S}
}

@software{Perrin2015,
       author = {{Perrin}, Marshall D. and {Long}, Joseph and {Sivaramakrishnan}, Anand and {Lajoie}, Charles-Phillipe and {Elliot}, Erin and {Pueyo}, Laurent and {Albert}, Loic},
        title = "{WebbPSF: James Webb Space Telescope PSF Simulation Tool}",
 howpublished = {Astrophysics Source Code Library, record ascl:1504.007},
         year = 2015,
        month = apr,
          eid = {ascl:1504.007},
       adsurl = {https://ui.adsabs.harvard.edu/abs/2015ascl.soft04007P}
}

@ARTICLE{Masters2015,
       author = {{Masters}, Daniel and {Capak}, Peter and {Stern}, Daniel and {Ilbert}, Olivier and {Salvato}, Mara and {Schmidt}, Samuel and {Longo}, Giuseppe and {Rhodes}, Jason and {Paltani}, Stephane and {Mobasher}, Bahram and {Hoekstra}, Henk and {Hildebrandt}, Hendrik and {Coupon}, Jean and {Steinhardt}, Charles and {Speagle}, Josh and {Faisst}, Andreas and {Kalinich}, Adam and {Brodwin}, Mark and {Brescia}, Massimo and {Cavuoti}, Stefano},
        title = "{Mapping the Galaxy Color-Redshift Relation: Optimal Photometric Redshift Calibration Strategies for Cosmology Surveys}",
      journal = {\apj},
         year = 2015,
        month = nov,
       volume = {813},
       number = {1},
          eid = {53},
        pages = {53},
          doi = {10.1088/0004-637X/813/1/53},
archivePrefix = {arXiv},
       eprint = {1509.03318},
 primaryClass = {astro-ph.CO},
       adsurl = {https://ui.adsabs.harvard.edu/abs/2015ApJ...813...53M}
}

@ARTICLE{Merlin2015,
       author = {{Merlin}, E. and {Fontana}, A. and {Ferguson}, H.~C. and {Dunlop}, J.~S. and {Elbaz}, D. and {Bourne}, N. and {Bruce}, V.~A. and {Buitrago}, F. and {Castellano}, M. and {Schreiber}, C. and {Grazian}, A. and {McLure}, R.~J. and {Okumura}, K. and {Shu}, X. and {Wang}, T. and {Amor{\'\i}n}, R. and {Boutsia}, K. and {Cappelluti}, N. and {Comastri}, A. and {Derriere}, S. and {Faber}, S.~M. and {Santini}, P.},
        title = "{T-PHOT: A new code for PSF-matched, prior-based, multiwavelength extragalactic deconfusion photometry}",
      journal = {\aap},
         year = 2015,
        month = oct,
       volume = {582},
          eid = {A15},
        pages = {A15},
          doi = {10.1051/0004-6361/201526471},
archivePrefix = {arXiv},
       eprint = {1505.02516},
 primaryClass = {astro-ph.IM},
       adsurl = {https://ui.adsabs.harvard.edu/abs/2015A&A...582A..15M}
}

@ARTICLE{Laigle2016,
       author = {{Laigle}, C. and {McCracken}, H.~J. and {Ilbert}, O. and {Hsieh}, B.~C. and {Davidzon}, I. and {Capak}, P. and {Hasinger}, G. and {Silverman}, J.~D. and {Pichon}, C. and {Coupon}, J. and {Aussel}, H. and {Le Borgne}, D. and {Caputi}, K. and {Cassata}, P. and {Chang}, Y.-Y. and {Civano}, F. and {Dunlop}, J. and {Fynbo}, J. and {Kartaltepe}, J.~S. and {Koekemoer}, A. and {Le F{\`e}vre}, O. and {Le Floc'h}, E. and {Leauthaud}, A. and {Lilly}, S. and {Lin}, L. and {Marchesi}, S. and {Milvang-Jensen}, B. and {Salvato}, M. and {Sanders}, D.~B. and {Scoville}, N. and {Smolcic}, V. and {Stockmann}, M. and {Taniguchi}, Y. and {Tasca}, L. and {Toft}, S. and {Vaccari}, Mattia and {Zabl}, J.},
        title = "{The COSMOS2015 Catalog: Exploring the 1 < z < 6 Universe with Half a Million Galaxies}",
      journal = {\apjs},
         year = 2016,
        month = jun,
       volume = {224},
       number = {2},
          eid = {24},
        pages = {24},
          doi = {10.3847/0067-0049/224/2/24},
archivePrefix = {arXiv},
       eprint = {1604.02350},
 primaryClass = {astro-ph.GA},
       adsurl = {https://ui.adsabs.harvard.edu/abs/2016ApJS..224...24L}
}

@ARTICLE{Merlin2016,
       author = {{Merlin}, E. and {Bourne}, N. and {Castellano}, M. and {Ferguson}, H.~C. and {Wang}, T. and {Derriere}, S. and {Dunlop}, J.~S. and {Elbaz}, D. and {Fontana}, A.},
        title = "{T-PHOT version 2.0: Improved algorithms for background subtraction, local convolution, kernel registration, and new options}",
      journal = {\aap},
         year = 2016,
        month = nov,
       volume = {595},
          eid = {A97},
        pages = {A97},
          doi = {10.1051/0004-6361/201628751},
archivePrefix = {arXiv},
       eprint = {1609.00146},
 primaryClass = {astro-ph.IM},
       adsurl = {https://ui.adsabs.harvard.edu/abs/2016A&A...595A..97M}
}

@ARTICLE{Merlin2016B,
       author = {{Merlin}, E. and {Amor{\'\i}n}, R. and {Castellano}, M. and {Fontana}, A. and {Buitrago}, F. and {Dunlop}, J.~S. and {Elbaz}, D. and {Boucaud}, A. and {Bourne}, N. and {Boutsia}, K. and {Brammer}, G. and {Bruce}, V.~A. and {Capak}, P. and {Cappelluti}, N. and {Ciesla}, L. and {Comastri}, A. and {Cullen}, F. and {Derriere}, S. and {Faber}, S.~M. and {Ferguson}, H.~C. and {Giallongo}, E. and {Grazian}, A. and {Lotz}, J. and {Micha{\l}owski}, M.~J. and {Paris}, D. and {Pentericci}, L. and {Pilo}, S. and {Santini}, P. and {Schreiber}, C. and {Shu}, X. and {Wang}, T.},
        title = "{The ASTRODEEP Frontier Fields catalogues. I. Multiwavelength photometry of Abell-2744 and MACS-J0416}",
      journal = {\aap},
         year = 2016,
        month = may,
       volume = {590},
          eid = {A30},
        pages = {A30},
          doi = {10.1051/0004-6361/201527513},
archivePrefix = {arXiv},
       eprint = {1603.02460},
 primaryClass = {astro-ph.GA},
       adsurl = {https://ui.adsabs.harvard.edu/abs/2016A&A...590A..30M}
}

@ARTICLE{Criscienzo2017,
       author = {{Di Criscienzo}, M. and {Merlin}, E. and {Castellano}, M. and {Santini}, P. and {Fontana}, A. and {Amorin}, R. and {Boutsia}, K. and {Derriere}, S. and {Dunlop}, J.~S. and {Elbaz}, D. and {Grazian}, A. and {McLure}, R.~J. and {M{\'a}rmol-Queralt{\'o}}, E. and {Michalowski}, M.~J. and {Mortlock}, S. and {Parsa}, S. and {Pentericci}, L.},
        title = "{The ASTRODEEP Frontier Fields catalogues. III. Multiwavelength photometry and rest-frame properties of MACS-J0717 and MACS-J1149}",
      journal = {\aap},
         year = 2017,
        month = oct,
       volume = {607},
          eid = {A30},
        pages = {A30},
          doi = {10.1051/0004-6361/201731172},
archivePrefix = {arXiv},
       eprint = {1706.03790},
 primaryClass = {astro-ph.GA},
       adsurl = {https://ui.adsabs.harvard.edu/abs/2017A&A...607A..30D}
}

@ARTICLE{Straatman2016,
       author = {{Straatman}, Caroline M.~S. and {Spitler}, Lee R. and {Quadri}, Ryan F. and {Labb{\'e}}, Ivo and {Glazebrook}, Karl and {Persson}, S. Eric and {Papovich}, Casey and {Tran}, Kim-Vy H. and {Brammer}, Gabriel B. and {Cowley}, Michael and {Tomczak}, Adam and {Nanayakkara}, Themiya and {Alcorn}, Leo and {Allen}, Rebecca and {Broussard}, Adam and {van Dokkum}, Pieter and {Forrest}, Ben and {van Houdt}, Josha and {Kacprzak}, Glenn G. and {Kawinwanichakij}, Lalitwadee and {Kelson}, Daniel D. and {Lee}, Janice and {McCarthy}, Patrick J. and {Mehrtens}, Nicola and {Monson}, Andrew and {Murphy}, David and {Rees}, Glen and {Tilvi}, Vithal and {Whitaker}, Katherine E.},
        title = "{The FourStar Galaxy Evolution Survey (ZFOURGE): Ultraviolet to Far-infrared Catalogs, Medium-bandwidth Photometric Redshifts with Improved Accuracy, Stellar Masses, and Confirmation of Quiescent Galaxies to z {\ensuremath{\sim}} 3.5}",
      journal = {\apj},
         year = 2016,
        month = oct,
       volume = {830},
       number = {1},
          eid = {51},
        pages = {51},
          doi = {10.3847/0004-637X/830/1/51},
archivePrefix = {arXiv},
       eprint = {1608.07579},
 primaryClass = {astro-ph.GA},
       adsurl = {https://ui.adsabs.harvard.edu/abs/2016ApJ...830...51S}
}

@ARTICLE{Byler2017,
       author = {{Byler}, Nell and {Dalcanton}, Julianne J. and {Conroy}, Charlie and {Johnson}, Benjamin D.},
        title = "{Nebular Continuum and Line Emission in Stellar Population Synthesis Models}",
      journal = {\apj},
         year = 2017,
        month = may,
       volume = {840},
       number = {1},
          eid = {44},
        pages = {44},
          doi = {10.3847/1538-4357/aa6c66},
archivePrefix = {arXiv},
       eprint = {1611.08305},
 primaryClass = {astro-ph.GA},
       adsurl = {https://ui.adsabs.harvard.edu/abs/2017ApJ...840...44B}
}

@ARTICLE{Nayyeri2017,
       author = {{Nayyeri}, H. and {Hemmati}, S. and {Mobasher}, B. and {Ferguson}, H.~C. and {Cooray}, A. and {Barro}, G. and {Faber}, S.~M. and {Dickinson}, M. and {Koekemoer}, A.~M. and {Peth}, M. and {Salvato}, M. and {Ashby}, M.~L.~N. and {Darvish}, B. and {Donley}, J. and {Durbin}, M. and {Finkelstein}, S. and {Fontana}, A. and {Grogin}, N.~A. and {Gruetzbauch}, R. and {Huang}, K. and {Khostovan}, A.~A. and {Kocevski}, D. and {Kodra}, D. and {Lee}, B. and {Newman}, J. and {Pacifici}, C. and {Pforr}, J. and {Stefanon}, M. and {Wiklind}, T. and {Willner}, S.~P. and {Wuyts}, S. and {Castellano}, M. and {Conselice}, C. and {Dolch}, T. and {Dunlop}, J.~S. and {Galametz}, A. and {Hathi}, N.~P. and {Lucas}, R.~A. and {Yan}, H.},
        title = "{CANDELS Multi-wavelength Catalogs: Source Identification and Photometry in the CANDELS COSMOS Survey Field}",
      journal = {\apjs},
         year = 2017,
        month = jan,
       volume = {228},
       number = {1},
          eid = {7},
        pages = {7},
          doi = {10.3847/1538-4365/228/1/7},
archivePrefix = {arXiv},
       eprint = {1612.07364},
 primaryClass = {astro-ph.GA},
       adsurl = {https://ui.adsabs.harvard.edu/abs/2017ApJS..228....7N}
}

@ARTICLE{WangWH2017,
       author = {{Wang}, Wei-Hao and {Lin}, Wei-Ching and {Lim}, Chen-Fatt and {Smail}, Ian and {Chapman}, Scott C. and {Zheng}, Xian Zhong and {Shim}, Hyunjin and {Kodama}, Tadayuki and {Almaini}, Omar and {Ao}, Yiping and {Blain}, Andrew W. and {Bourne}, Nathan and {Bunker}, Andrew J. and {Chang}, Yu-Yen and {Chao}, Dani C. -Y. and {Chen}, Chian-Chou and {Clements}, David L. and {Conselice}, Christopher J. and {Cowley}, William I. and {Dannerbauer}, Helmut and {Dunlop}, James S. and {Geach}, James E. and {Goto}, Tomotsugu and {Jiang}, Linhua and {Ivison}, Rob J. and {Jeong}, Woong-Seob and {Kohno}, Kotaro and {Kong}, Xu and {Lee}, Chien-Hsu and {Lee}, Hyung Mok and {Lee}, Minju and {Micha{\l}owski}, Micha{\l} J. and {Oteo}, Iv{\'a}n and {Sawicki}, Marcin and {Scott}, Douglas and {Shu}, Xin Wen and {Simpson}, James M. and {Tee}, Wei-Leong and {Toba}, Yoshiki and {Valiante}, Elisabetta and {Wang}, Jun-Xian and {Wang}, Ran and {Wardlow}, Julie L.},
        title = "{SCUBA-2 Ultra Deep Imaging EAO Survey (STUDIES): Faint-end Counts at 450 {\ensuremath{\mu}}m}",
      journal = {\apj},
         year = 2017,
        month = nov,
       volume = {850},
       number = {1},
          eid = {37},
        pages = {37},
          doi = {10.3847/1538-4357/aa911b},
archivePrefix = {arXiv},
       eprint = {1707.00990},
 primaryClass = {astro-ph.GA},
       adsurl = {https://ui.adsabs.harvard.edu/abs/2017ApJ...850...37W}
}

@ARTICLE{Smolvcic2017,
       author = {{Smol{\v{c}}i{\'c}}, V. and {Novak}, M. and {Bondi}, M. and {Ciliegi}, P. and {Mooley}, K.~P. and {Schinnerer}, E. and {Zamorani}, G. and {Navarrete}, F. and {Bourke}, S. and {Karim}, A. and {Vardoulaki}, E. and {Leslie}, S. and {Delhaize}, J. and {Carilli}, C.~L. and {Myers}, S.~T. and {Baran}, N. and {Delvecchio}, I. and {Miettinen}, O. and {Banfield}, J. and {Balokovi{\'c}}, M. and {Bertoldi}, F. and {Capak}, P. and {Frail}, D.~A. and {Hallinan}, G. and {Hao}, H. and {Herrera Ruiz}, N. and {Horesh}, A. and {Ilbert}, O. and {Intema}, H. and {Jeli{\'c}}, V. and {Kl{\"o}ckner}, H.-R. and {Krpan}, J. and {Kulkarni}, S.~R. and {McCracken}, H. and {Laigle}, C. and {Middleberg}, E. and {Murphy}, E.~J. and {Sargent}, M. and {Scoville}, N.~Z. and {Sheth}, K.},
        title = "{The VLA-COSMOS 3 GHz Large Project: Continuum data and source catalog release}",
      journal = {\aap},
         year = 2017,
        month = jun,
       volume = {602},
          eid = {A1},
        pages = {A1},
          doi = {10.1051/0004-6361/201628704},
archivePrefix = {arXiv},
       eprint = {1703.09713},
 primaryClass = {astro-ph.GA},
       adsurl = {https://ui.adsabs.harvard.edu/abs/2017A&A...602A...1S}
}

@ARTICLE{Carnall2018,
       author = {{Carnall}, A.~C. and {McLure}, R.~J. and {Dunlop}, J.~S. and {Dav{\'e}}, R.},
        title = "{Inferring the star formation histories of massive quiescent galaxies with BAGPIPES: evidence for multiple quenching mechanisms}",
      journal = {\mnras},
         year = 2018,
        month = nov,
       volume = {480},
       number = {4},
        pages = {4379-4401},
          doi = {10.1093/mnras/sty2169},
archivePrefix = {arXiv},
       eprint = {1712.04452},
 primaryClass = {astro-ph.GA},
       adsurl = {https://ui.adsabs.harvard.edu/abs/2018MNRAS.480.4379C}
}

@ARTICLE{Gaia2018,
       author = {{Gaia Collaboration} and {Brown}, A.~G.~A. and {Vallenari}, A. and {Prusti}, T. and {de Bruijne}, J.~H.~J. and {Babusiaux}, C. and {Bailer-Jones}, C.~A.~L. and {Biermann}, M. and {Evans}, D.~W. and {Eyer}, L. and {Jansen}, F. and {Jordi}, C. and {Klioner}, S.~A. and {Lammers}, U. and {Lindegren}, L. and {Luri}, X. and {Mignard}, F. and {Panem}, C. and {Pourbaix}, D. and {Randich}, S. and {Sartoretti}, P. and {Siddiqui}, H.~I. and {Soubiran}, C. and {van Leeuwen}, F. and {Walton}, N.~A. and {Arenou}, F. and {Bastian}, U. and {Cropper}, M. and {Drimmel}, R. and {Katz}, D. and {Lattanzi}, M.~G. and {Bakker}, J. and {Cacciari}, C. and {Casta{\~n}eda}, J. and {Chaoul}, L. and {Cheek}, N. and {De Angeli}, F. and {Fabricius}, C. and {Guerra}, R. and {Holl}, B. and {Masana}, E. and {Messineo}, R. and {Mowlavi}, N. and {Nienartowicz}, K. and {Panuzzo}, P. and {Portell}, J. and {Riello}, M. and {Seabroke}, G.~M. and {Tanga}, P. and {Th{\'e}venin}, F. and {Gracia-Abril}, G. and {Comoretto}, G. and {Garcia-Reinaldos}, M. and {Teyssier}, D. and {Altmann}, M. and {Andrae}, R. and {Audard}, M. and {Bellas-Velidis}, I. and {Benson}, K. and {Berthier}, J. and {Blomme}, R. and {Burgess}, P. and {Busso}, G. and {Carry}, B. and {Cellino}, A. and {Clementini}, G. and {Clotet}, M. and {Creevey}, O. and {Davidson}, M. and {De Ridder}, J. and {Delchambre}, L. and {Dell'Oro}, A. and {Ducourant}, C. and {Fern{\'a}ndez-Hern{\'a}ndez}, J. and {Fouesneau}, M. and {Fr{\'e}mat}, Y. and {Galluccio}, L. and {Garc{\'\i}a-Torres}, M. and {Gonz{\'a}lez-N{\'u}{\~n}ez}, J. and {Gonz{\'a}lez-Vidal}, J.~J. and {Gosset}, E. and {Guy}, L.~P. and {Halbwachs}, J. -L. and {Hambly}, N.~C. and {Harrison}, D.~L. and {Hern{\'a}ndez}, J. and {Hestroffer}, D. and {Hodgkin}, S.~T. and {Hutton}, A. and {Jasniewicz}, G. and {Jean-Antoine-Piccolo}, A. and {Jordan}, S. and {Korn}, A.~J. and {Krone-Martins}, A. and {Lanzafame}, A.~C. and {Lebzelter}, T. and {L{\"o}ffler}, W. and {Manteiga}, M. and {Marrese}, P.~M. and {Mart{\'\i}n-Fleitas}, J.~M. and {Moitinho}, A. and {Mora}, A. and {Muinonen}, K. and {Osinde}, J. and {Pancino}, E. and {Pauwels}, T. and {Petit}, J. -M. and {Recio-Blanco}, A. and {Richards}, P.~J. and {Rimoldini}, L. and {Robin}, A.~C. and {Sarro}, L.~M. and {Siopis}, C. and {Smith}, M. and {Sozzetti}, A. and {S{\"u}veges}, M. and {Torra}, J. and {van Reeven}, W. and {Abbas}, U. and {Abreu Aramburu}, A. and {Accart}, S. and {Aerts}, C. and {Altavilla}, G. and {{\'A}lvarez}, M.~A. and {Alvarez}, R. and {Alves}, J. and {Anderson}, R.~I. and {Andrei}, A.~H. and {Anglada Varela}, E. and {Antiche}, E. and {Antoja}, T. and {Arcay}, B. and {Astraatmadja}, T.~L. and {Bach}, N. and {Baker}, S.~G. and {Balaguer-N{\'u}{\~n}ez}, L. and {Balm}, P. and {Barache}, C. and {Barata}, C. and {Barbato}, D. and {Barblan}, F. and {Barklem}, P.~S. and {Barrado}, D. and {Barros}, M. and {Barstow}, M.~A. and {Bartholom{\'e} Mu{\~n}oz}, S. and {Bassilana}, J. -L. and {Becciani}, U. and {Bellazzini}, M. and {Berihuete}, A. and {Bertone}, S. and {Bianchi}, L. and {Bienaym{\'e}}, O. and {Blanco-Cuaresma}, S. and {Boch}, T. and {Boeche}, C. and {Bombrun}, A. and {Borrachero}, R. and {Bossini}, D. and {Bouquillon}, S. and {Bourda}, G. and {Bragaglia}, A. and {Bramante}, L. and {Breddels}, M.~A. and {Bressan}, A. and {Brouillet}, N. and {Br{\"u}semeister}, T. and {Brugaletta}, E. and {Bucciarelli}, B. and {Burlacu}, A. and {Busonero}, D. and {Butkevich}, A.~G. and {Buzzi}, R. and {Caffau}, E. and {Cancelliere}, R. and {Cannizzaro}, G. and {Cantat-Gaudin}, T. and {Carballo}, R. and {Carlucci}, T. and {Carrasco}, J.~M. and {Casamiquela}, L. and {Castellani}, M. and {Castro-Ginard}, A. and {Charlot}, P. and {Chemin}, L. and {Chiavassa}, A. and {Cocozza}, G. and {Costigan}, G. and {Cowell}, S. and {Crifo}, F. and {Crosta}, M. and {Crowley}, C. and {Cuypers}, J. and {Dafonte}, C. and {Damerdji}, Y. and {Dapergolas}, A. and {David}, P. and {David}, M. and {de Laverny}, P. and {De Luise}, F. and {De March}, R. and {de Martino}, D. and {de Souza}, R. and {de Torres}, A. and {Debosscher}, J. and {del Pozo}, E. and {Delbo}, M. and {Delgado}, A. and {Delgado}, H.~E. and {Di Matteo}, P. and {Diakite}, S. and {Diener}, C. and {Distefano}, E. and {Dolding}, C. and {Drazinos}, P. and {Dur{\'a}n}, J. and {Edvardsson}, B. and {Enke}, H. and {Eriksson}, K. and {Esquej}, P. and {Eynard Bontemps}, G. and {Fabre}, C. and {Fabrizio}, M. and {Faigler}, S. and {Falc{\~a}o}, A.~J. and {Farr{\`a}s Casas}, M. and {Federici}, L. and {Fedorets}, G. and {Fernique}, P. and {Figueras}, F. and {Filippi}, F. and {Findeisen}, K. and {Fonti}, A. and {Fraile}, E. and {Fraser}, M. and {Fr{\'e}zouls}, B. and {Gai}, M. and {Galleti}, S. and {Garabato}, D. and {Garc{\'\i}a-Sedano}, F. and {Garofalo}, A. and {Garralda}, N. and {Gavel}, A. and {Gavras}, P. and {Gerssen}, J. and {Geyer}, R. and {Giacobbe}, P. and {Gilmore}, G. and {Girona}, S. and {Giuffrida}, G. and {Glass}, F. and {Gomes}, M. and {Granvik}, M. and {Gueguen}, A. and {Guerrier}, A. and {Guiraud}, J. and {Guti{\'e}rrez-S{\'a}nchez}, R. and {Haigron}, R. and {Hatzidimitriou}, D. and {Hauser}, M. and {Haywood}, M. and {Heiter}, U. and {Helmi}, A. and {Heu}, J. and {Hilger}, T. and {Hobbs}, D. and {Hofmann}, W. and {Holland}, G. and {Huckle}, H.~E. and {Hypki}, A. and {Icardi}, V. and {Jan{\ss}en}, K. and {Jevardat de Fombelle}, G. and {Jonker}, P.~G. and {Juh{\'a}sz}, {\'A}. L. and {Julbe}, F. and {Karampelas}, A. and {Kewley}, A. and {Klar}, J. and {Kochoska}, A. and {Kohley}, R. and {Kolenberg}, K. and {Kontizas}, M. and {Kontizas}, E. and {Koposov}, S.~E. and {Kordopatis}, G. and {Kostrzewa-Rutkowska}, Z. and {Koubsky}, P. and {Lambert}, S. and {Lanza}, A.~F. and {Lasne}, Y. and {Lavigne}, J. -B. and {Le Fustec}, Y. and {Le Poncin-Lafitte}, C. and {Lebreton}, Y. and {Leccia}, S. and {Leclerc}, N. and {Lecoeur-Taibi}, I. and {Lenhardt}, H. and {Leroux}, F. and {Liao}, S. and {Licata}, E. and {Lindstr{\o}m}, H.~E.~P. and {Lister}, T.~A. and {Livanou}, E. and {Lobel}, A. and {L{\'o}pez}, M. and {Managau}, S. and {Mann}, R.~G. and {Mantelet}, G. and {Marchal}, O. and {Marchant}, J.~M. and {Marconi}, M. and {Marinoni}, S. and {Marschalk{\'o}}, G. and {Marshall}, D.~J. and {Martino}, M. and {Marton}, G. and {Mary}, N. and {Massari}, D. and {Matijevi{\v{c}}}, G. and {Mazeh}, T. and {McMillan}, P.~J. and {Messina}, S. and {Michalik}, D. and {Millar}, N.~R. and {Molina}, D. and {Molinaro}, R. and {Moln{\'a}r}, L. and {Montegriffo}, P. and {Mor}, R. and {Morbidelli}, R. and {Morel}, T. and {Morris}, D. and {Mulone}, A.~F. and {Muraveva}, T. and {Musella}, I. and {Nelemans}, G. and {Nicastro}, L. and {Noval}, L. and {O'Mullane}, W. and {Ord{\'e}novic}, C. and {Ord{\'o}{\~n}ez-Blanco}, D. and {Osborne}, P. and {Pagani}, C. and {Pagano}, I. and {Pailler}, F. and {Palacin}, H. and {Palaversa}, L. and {Panahi}, A. and {Pawlak}, M. and {Piersimoni}, A.~M. and {Pineau}, F. -X. and {Plachy}, E. and {Plum}, G. and {Poggio}, E. and {Poujoulet}, E. and {Pr{\v{s}}a}, A. and {Pulone}, L. and {Racero}, E. and {Ragaini}, S. and {Rambaux}, N. and {Ramos-Lerate}, M. and {Regibo}, S. and {Reyl{\'e}}, C. and {Riclet}, F. and {Ripepi}, V. and {Riva}, A. and {Rivard}, A. and {Rixon}, G. and {Roegiers}, T. and {Roelens}, M. and {Romero-G{\'o}mez}, M. and {Rowell}, N. and {Royer}, F. and {Ruiz-Dern}, L. and {Sadowski}, G. and {Sagrist{\`a} Sell{\'e}s}, T. and {Sahlmann}, J. and {Salgado}, J. and {Salguero}, E. and {Sanna}, N. and {Santana-Ros}, T. and {Sarasso}, M. and {Savietto}, H. and {Schultheis}, M. and {Sciacca}, E. and {Segol}, M. and {Segovia}, J.~C. and {S{\'e}gransan}, D. and {Shih}, I. -C. and {Siltala}, L. and {Silva}, A.~F. and {Smart}, R.~L. and {Smith}, K.~W. and {Solano}, E. and {Solitro}, F. and {Sordo}, R. and {Soria Nieto}, S. and {Souchay}, J. and {Spagna}, A. and {Spoto}, F. and {Stampa}, U. and {Steele}, I.~A. and {Steidelm{\"u}ller}, H. and {Stephenson}, C.~A. and {Stoev}, H. and {Suess}, F.~F. and {Surdej}, J. and {Szabados}, L. and {Szegedi-Elek}, E. and {Tapiador}, D. and {Taris}, F. and {Tauran}, G. and {Taylor}, M.~B. and {Teixeira}, R. and {Terrett}, D. and {Teyssandier}, P. and {Thuillot}, W. and {Titarenko}, A. and {Torra Clotet}, F. and {Turon}, C. and {Ulla}, A. and {Utrilla}, E. and {Uzzi}, S. and {Vaillant}, M. and {Valentini}, G. and {Valette}, V. and {van Elteren}, A. and {Van Hemelryck}, E. and {van Leeuwen}, M. and {Vaschetto}, M. and {Vecchiato}, A. and {Veljanoski}, J. and {Viala}, Y. and {Vicente}, D. and {Vogt}, S. and {von Essen}, C. and {Voss}, H. and {Votruba}, V. and {Voutsinas}, S. and {Walmsley}, G. and {Weiler}, M. and {Wertz}, O. and {Wevers}, T. and {Wyrzykowski}, {\L}. and {Yoldas}, A. and {{\v{Z}}erjal}, M. and {Ziaeepour}, H. and {Zorec}, J. and {Zschocke}, S. and {Zucker}, S. and {Zurbach}, C. and {Zwitter}, T.},
        title = "{Gaia Data Release 2. Summary of the contents and survey properties}",
      journal = {\aap},
         year = 2018,
        month = aug,
       volume = {616},
          eid = {A1},
        pages = {A1},
          doi = {10.1051/0004-6361/201833051},
archivePrefix = {arXiv},
       eprint = {1804.09365},
 primaryClass = {astro-ph.GA},
       adsurl = {https://ui.adsabs.harvard.edu/abs/2018A&A...616A...1G}
}

@ARTICLE{Franco2018,
       author = {{Franco}, M. and {Elbaz}, D. and {B{\'e}thermin}, M. and {Magnelli}, B. and {Schreiber}, C. and {Ciesla}, L. and {Dickinson}, M. and {Nagar}, N. and {Silverman}, J. and {Daddi}, E. and {Alexander}, D.~M. and {Wang}, T. and {Pannella}, M. and {Le Floc'h}, E. and {Pope}, A. and {Giavalisco}, M. and {Maury}, A.~J. and {Bournaud}, F. and {Chary}, R. and {Demarco}, R. and {Ferguson}, H. and {Finkelstein}, S.~L. and {Inami}, H. and {Iono}, D. and {Juneau}, S. and {Lagache}, G. and {Leiton}, R. and {Lin}, L. and {Magdis}, G. and {Messias}, H. and {Motohara}, K. and {Mullaney}, J. and {Okumura}, K. and {Papovich}, C. and {Pforr}, J. and {Rujopakarn}, W. and {Sargent}, M. and {Shu}, X. and {Zhou}, L.},
        title = "{GOODS-ALMA: 1.1 mm galaxy survey. I. Source catalog and optically dark galaxies}",
      journal = {\aap},
         year = 2018,
        month = dec,
       volume = {620},
          eid = {A152},
        pages = {A152},
          doi = {10.1051/0004-6361/201832928},
archivePrefix = {arXiv},
       eprint = {1803.00157},
 primaryClass = {astro-ph.GA},
       adsurl = {https://ui.adsabs.harvard.edu/abs/2018A&A...620A.152F}
}

@ARTICLE{Jin2018,
       author = {{Jin}, Shuowen and {Daddi}, Emanuele and {Liu}, Daizhong and {Smol{\v{c}}i{\'c}}, Vernesa and {Schinnerer}, Eva and {Calabr{\`o}}, Antonello and {Gu}, Qiusheng and {Delhaize}, Jacinta and {Delvecchio}, Ivan and {Gao}, Yu and {Salvato}, Mara and {Puglisi}, Annagrazia and {Dickinson}, Mark and {Bertoldi}, Frank and {Sargent}, Mark and {Novak}, Mladen and {Magdis}, Georgios and {Aretxaga}, Itziar and {Wilson}, Grant W. and {Capak}, Peter},
        title = "{{\textquotedblleft}Super-deblended{\textquotedblright} Dust Emission in Galaxies. II. Far-IR to (Sub)millimeter Photometry and High-redshift Galaxy Candidates in the Full COSMOS Field}",
      journal = {\apj},
         year = 2018,
        month = sep,
       volume = {864},
       number = {1},
          eid = {56},
        pages = {56},
          doi = {10.3847/1538-4357/aad4af},
archivePrefix = {arXiv},
       eprint = {1807.04697},
 primaryClass = {astro-ph.GA},
       adsurl = {https://ui.adsabs.harvard.edu/abs/2018ApJ...864...56J}
}

@ARTICLE{Liu2018,
       author = {{Liu}, Daizhong and {Daddi}, Emanuele and {Dickinson}, Mark and {Owen}, Frazer and {Pannella}, Maurilio and {Sargent}, Mark and {B{\'e}thermin}, Matthieu and {Magdis}, Georgios and {Gao}, Yu and {Shu}, Xinwen and {Wang}, Tao and {Jin}, Shuowen and {Inami}, Hanae},
        title = "{{\textquotedblleft}Super-deblended{\textquotedblright} Dust Emission in Galaxies. I. The GOODS-North Catalog and the Cosmic Star Formation Rate Density out to Redshift 6}",
      journal = {\apj},
         year = 2018,
        month = feb,
       volume = {853},
       number = {2},
          eid = {172},
        pages = {172},
          doi = {10.3847/1538-4357/aaa600},
archivePrefix = {arXiv},
       eprint = {1703.05281},
 primaryClass = {astro-ph.GA},
       adsurl = {https://ui.adsabs.harvard.edu/abs/2018ApJ...853..172L}
}

@ARTICLE{Mehta2018,
       author = {{Mehta}, Vihang and {Scarlata}, Claudia and {Capak}, Peter and {Davidzon}, Iary and {Faisst}, Andreas and {Hsieh}, Bau Ching and {Ilbert}, Olivier and {Jarvis}, Matt and {Laigle}, Clotilde and {Phillips}, John and {Silverman}, John and {Strauss}, Michael A. and {Tanaka}, Masayuki and {Bowler}, Rebecca and {Coupon}, Jean and {Foucaud}, S{\'e}bastien and {Hemmati}, Shoubaneh and {Masters}, Daniel and {McCracken}, Henry Joy and {Mobasher}, Bahram and {Ouchi}, Masami and {Shibuya}, Takatoshi and {Wang}, Wei-Hao},
        title = "{SPLASH-SXDF Multi-wavelength Photometric Catalog}",
      journal = {\apjs},
         year = 2018,
        month = apr,
       volume = {235},
       number = {2},
          eid = {36},
        pages = {36},
          doi = {10.3847/1538-4365/aab60c},
archivePrefix = {arXiv},
       eprint = {1711.05280},
 primaryClass = {astro-ph.GA},
       adsurl = {https://ui.adsabs.harvard.edu/abs/2018ApJS..235...36M}
}

@INPROCEEDINGS{Osborne2018,
       author = {{Osborne}, Shannon and {Perrin}, Marshall D. and {Melendez Hernandez}, Marcio},
        title = "{WebbPSF: Updated PSF Models Based on JWST Ground Testing Results}",
    booktitle = {American Astronomical Society Meeting Abstracts \#232},
         year = 2018,
       series = {American Astronomical Society Meeting Abstracts},
       volume = {232},
        month = jun,
          eid = {119.12},
        pages = {119.12},
       adsurl = {https://ui.adsabs.harvard.edu/abs/2018AAS...23211912O}
}

@ARTICLE{Merlin2019,
       author = {{Merlin}, E. and {Pilo}, S. and {Fontana}, A. and {Castellano}, M. and {Paris}, D. and {Roscani}, V. and {Santini}, P. and {Torelli}, M.},
        title = "{A-PHOT: a new, versatile code for precision aperture photometry}",
      journal = {\aap},
         year = 2019,
        month = feb,
       volume = {622},
          eid = {A169},
        pages = {A169},
          doi = {10.1051/0004-6361/201833991},
archivePrefix = {arXiv},
       eprint = {1812.00727},
 primaryClass = {astro-ph.IM},
       adsurl = {https://ui.adsabs.harvard.edu/abs/2019A&A...622A.169M}
}

@ARTICLE{WangT2019,
       author = {{Wang}, T. and {Schreiber}, C. and {Elbaz}, D. and {Yoshimura}, Y. and {Kohno}, K. and {Shu}, X. and {Yamaguchi}, Y. and {Pannella}, M. and {Franco}, M. and {Huang}, J. and {Lim}, C.-F. and {Wang}, W.-H.},
        title = "{A dominant population of optically invisible massive galaxies in the early Universe}",
      journal = {\nat},
         year = 2019,
        month = aug,
       volume = {572},
       number = {7768},
        pages = {211-214},
          doi = {10.1038/s41586-019-1452-4},
archivePrefix = {arXiv},
       eprint = {1908.02372},
 primaryClass = {astro-ph.GA},
       adsurl = {https://ui.adsabs.harvard.edu/abs/2019Natur.572..211W}
}

@ARTICLE{Yamaguchi2019,
       author = {{Yamaguchi}, Yuki and {Kohno}, Kotaro and {Hatsukade}, Bunyo and {Wang}, Tao and {Yoshimura}, Yuki and {Ao}, Yiping and {Caputi}, Karina I. and {Dunlop}, James S. and {Egami}, Eiichi and {Espada}, Daniel and {Fujimoto}, Seiji and {Hayatsu}, Natsuki H. and {Ivison}, Rob J. and {Kodama}, Tadayuki and {Kusakabe}, Haruka and {Nagao}, Tohru and {Ouchi}, Masami and {Rujopakarn}, Wiphu and {Tadaki}, Ken-ichi and {Tamura}, Yoichi and {Ueda}, Yoshihiro and {Umehata}, Hideki and {Wang}, Wei-Hao and {Yun}, Min S.},
        title = "{ALMA 26 arcmin$^{2}$ Survey of GOODS-S at 1 mm (ASAGAO): Near-infrared-dark Faint ALMA Sources}",
      journal = {\apj},
         year = 2019,
        month = jun,
       volume = {878},
       number = {1},
          eid = {73},
        pages = {73},
          doi = {10.3847/1538-4357/ab0d22},
archivePrefix = {arXiv},
       eprint = {1903.02744},
 primaryClass = {astro-ph.GA},
       adsurl = {https://ui.adsabs.harvard.edu/abs/2019ApJ...878...73Y}
}

@dataset{SpUDS2020,
       author = {{SpUDS Team}},
        title = "{Spitzer Public Legacy Survey of the UKIDSS Ultra Deep Survey}",
 howpublished = {NASA IPAC DataSet, IRSA403},
         year = 2020,
        month = jan,
          doi = {10.26131/IRSA403},
       adsurl = {https://ui.adsabs.harvard.edu/abs/2020ipac.data.I403S}
}

@ARTICLE{Wright2020,
       author = {{Wright}, Angus H. and {Hildebrandt}, Hendrik and {van den Busch}, Jan Luca and {Heymans}, Catherine},
        title = "{Photometric redshift calibration with self-organising maps}",
      journal = {\aap},
         year = 2020,
        month = may,
       volume = {637},
          eid = {A100},
        pages = {A100},
          doi = {10.1051/0004-6361/201936782},
archivePrefix = {arXiv},
       eprint = {1909.09632},
 primaryClass = {astro-ph.CO},
       adsurl = {https://ui.adsabs.harvard.edu/abs/2020A&A...637A.100W}
}

@MISC{Alvarez-Marquez2021,
       author = {{Alvarez-Marquez}, Javier and {Hashimoto}, Takuya and {Arribas}, Santiago and {Bakx}, Tom and {Ceverino}, Daniel and {Colina Robledo}, Luis and {Inoue}, Akio and {Marques-Chaves}, Rui and {Matsuo}, Hiroshi and {Mawatari}, Ken and {Pereira Santaella}, Miguel and {Tamura}, Yoichi and {Yoshida}, Naoki},
        title = "{ALMA [OIII]88um Emitters. Signpost of Early Stellar Buildup and Reionization in the Universe}",
 howpublished = {JWST Proposal. Cycle 1, ID. \#1840},
         year = 2021,
        month = mar,
        pages = {1840},
       adsurl = {https://ui.adsabs.harvard.edu/abs/2021jwst.prop.1840A}
}

@MISC{Dunlop2021,
       author = {{Dunlop}, James S. and {Abraham}, Roberto G. and {Ashby}, Matthew L.~N. and {Bagley}, Micaela and {Best}, Philip N. and {Bongiorno}, Angela and {Bouwens}, Rychard and {Bowler}, Rebecca A.~A. and {Brammer}, Gabriel and {Bremer}, Malcolm and {Calabro'}, Antonello and {Carnall}, Adam and {Castellano}, Marco and {Cirasuolo}, Michele and {Conselice}, Christopher and {Cullen}, Fergus and {Dave}, Romeel and {Dayal}, Pratika and {Dekel}, Avishai and {Dickinson}, Mark and {Duncan}, Kenneth James and {Elbaz}, David and {Ellis}, Richard S. and {Ferguson}, Harry C. and {Ferrara}, Andrea and {Finkelstein}, Steven L. and {Fontana}, Adriano and {Furlanetto}, Steven and {Fynbo}, Johan P.~U. and {Gallerani}, Simona and {Gardner}, Jonathan P. and {Giavalisco}, Mauro and {Grazian}, Andrea and {Grogin}, Norman and {Harikane}, Yuichi and {Hopkins}, Philip F. and {Ilbert}, Olivier and {Illingworth}, Garth D. and {Juneau}, Stephanie and {Jung}, Intae and {Kartaltepe}, Jeyhan and {Kassin}, Susan and {Kauffmann}, Olivier Benjamin and {Khochfar}, Sadegh and {Kirkpatrick}, Allison and {Kocevski}, Dale D. and {Koekemoer}, Anton M. and {Labbe}, Ivo and {Laporte}, Nicolas and {Larson}, Rebecca L. and {Lucas}, Ray A. and {Magee}, Daniel K. and {Mason}, Charlotte and {McCracken}, Henry Joy and {McLeod}, Derek and {McLure}, Ross and {Merlin}, Emiliano and {Mesinger}, Andrei and {Milvang-Jensen}, Bo and {Newman}, Jeffrey Allen and {Oesch}, Pascal and {Ouchi}, Masami and {Pacifici}, Camilla and {Papovich}, Casey and {Peacock}, John and {Peeples}, Molly and {Pentericci}, Laura and {Perez-Gonzalez}, Pablo G. and {Pirzkal}, Norbert and {Pope}, Alexandra and {Pye}, John P. and {Reddy}, Naveen A. and {Robertson}, Brant and {Salvato}, Mara and {Santini}, Paola and {Schaerer}, Daniel and {Shapley}, Alice E. and {Simons}, Raymond and {Smit}, Renske and {Smith}, Britton D. and {Snyder}, Greg and {Somerville}, Rachel S. and {Stanway}, Elizabeth R. and {Stefanon}, Mauro and {Tasca}, Lidia and {Tikkanen}, Tuomo and {Tresse}, Laurence and {Trump}, Jonathan R. and {Whitaker}, Katherine E. and {Wilkins}, Stephen Matthew and {Wright}, Gillian and {Wyithe}, J. Stuart B. and {van Dokkum}, Pieter and {van der Werf}, Paul},
        title = "{PRIMER: Public Release IMaging for Extragalactic Research}",
 howpublished = {JWST Proposal. Cycle 1, ID. \#1837},
         year = 2021,
        month = mar,
        pages = {1837},
       adsurl = {https://ui.adsabs.harvard.edu/abs/2021jwst.prop.1837D}
}

@ARTICLE{Merlin2021,
       author = {{Merlin}, E. and {Castellano}, M. and {Santini}, P. and {Cipolletta}, G. and {Boutsia}, K. and {Schreiber}, C. and {Buitrago}, F. and {Fontana}, A. and {Elbaz}, D. and {Dunlop}, J. and {Grazian}, A. and {McLure}, R. and {McLeod}, D. and {Nonino}, M. and {Milvang-Jensen}, B. and {Derriere}, S. and {Hathi}, N.~P. and {Pentericci}, L. and {Fortuni}, F. and {Calabr{\`o}}, A.},
        title = "{The ASTRODEEP-GS43 catalogue: New photometry and redshifts for the CANDELS GOODS-South field}",
      journal = {\aap},
         year = 2021,
        month = may,
       volume = {649},
          eid = {A22},
        pages = {A22},
          doi = {10.1051/0004-6361/202140310},
archivePrefix = {arXiv},
       eprint = {2103.09246},
 primaryClass = {astro-ph.GA},
       adsurl = {https://ui.adsabs.harvard.edu/abs/2021A&A...649A..22M}
}

@ARTICLE{Aihara2022,
       author = {{Aihara}, Hiroaki and {AlSayyad}, Yusra and {Ando}, Makoto and {Armstrong}, Robert and {Bosch}, James and {Egami}, Eiichi and {Furusawa}, Hisanori and {Furusawa}, Junko and {Harasawa}, Sumiko and {Harikane}, Yuichi and {Hsieh}, Bau-Ching and {Ikeda}, Hiroyuki and {Ito}, Kei and {Iwata}, Ikuru and {Kodama}, Tadayuki and {Koike}, Michitaro and {Kokubo}, Mitsuru and {Komiyama}, Yutaka and {Li}, Xiangchong and {Liang}, Yongming and {Lin}, Yen-Ting and {Lupton}, Robert H. and {Lust}, Nate B. and {MacArthur}, Lauren A. and {Mawatari}, Ken and {Mineo}, Sogo and {Miyatake}, Hironao and {Miyazaki}, Satoshi and {More}, Surhud and {Morishima}, Takahiro and {Murayama}, Hitoshi and {Nakajima}, Kimihiko and {Nakata}, Fumiaki and {Nishizawa}, Atsushi J. and {Oguri}, Masamune and {Okabe}, Nobuhiro and {Okura}, Yuki and {Ono}, Yoshiaki and {Osato}, Ken and {Ouchi}, Masami and {Pan}, Yen-Chen and {Plazas Malag{\'o}n}, Andr{\'e}s A. and {Price}, Paul A. and {Reed}, Sophie L. and {Rykoff}, Eli S. and {Shibuya}, Takatoshi and {Simunovic}, Mirko and {Strauss}, Michael A. and {Sugimori}, Kanako and {Suto}, Yasushi and {Suzuki}, Nao and {Takada}, Masahiro and {Takagi}, Yuhei and {Takata}, Tadafumi and {Takita}, Satoshi and {Tanaka}, Masayuki and {Tang}, Shenli and {Taranu}, Dan S. and {Terai}, Tsuyoshi and {Toba}, Yoshiki and {Turner}, Edwin L. and {Uchiyama}, Hisakazu and {Vijarnwannaluk}, Bovornpratch and {Waters}, Christopher Z. and {Yamada}, Yoshihiko and {Yamamoto}, Naoaki and {Yamashita}, Takuji},
        title = "{Third data release of the Hyper Suprime-Cam Subaru Strategic Program}",
      journal = {\pasj},
         year = 2022,
        month = apr,
       volume = {74},
       number = {2},
        pages = {247-272},
          doi = {10.1093/pasj/psab122},
archivePrefix = {arXiv},
       eprint = {2108.13045},
 primaryClass = {astro-ph.IM},
       adsurl = {https://ui.adsabs.harvard.edu/abs/2022PASJ...74..247A}
}

@ARTICLE{Astropy2022,
       author = {{Astropy Collaboration} and {Price-Whelan}, Adrian M. and {Lim}, Pey Lian and {Earl}, Nicholas and {Starkman}, Nathaniel and {Bradley}, Larry and {Shupe}, David L. and {Patil}, Aarya A. and {Corrales}, Lia and {Brasseur}, C.~E. and {N{\"o}the}, Maximilian and {Donath}, Axel and {Tollerud}, Erik and {Morris}, Brett M. and {Ginsburg}, Adam and {Vaher}, Eero and {Weaver}, Benjamin A. and {Tocknell}, James and {Jamieson}, William and {van Kerkwijk}, Marten H. and {Robitaille}, Thomas P. and {Merry}, Bruce and {Bachetti}, Matteo and {G{\"u}nther}, H. Moritz and {Aldcroft}, Thomas L. and {Alvarado-Montes}, Jaime A. and {Archibald}, Anne M. and {B{\'o}di}, Attila and {Bapat}, Shreyas and {Barentsen}, Geert and {Baz{\'a}n}, Juanjo and {Biswas}, Manish and {Boquien}, M{\'e}d{\'e}ric and {Burke}, D.~J. and {Cara}, Daria and {Cara}, Mihai and {Conroy}, Kyle E. and {Conseil}, Simon and {Craig}, Matthew W. and {Cross}, Robert M. and {Cruz}, Kelle L. and {D'Eugenio}, Francesco and {Dencheva}, Nadia and {Devillepoix}, Hadrien A.~R. and {Dietrich}, J{\"o}rg P. and {Eigenbrot}, Arthur Davis and {Erben}, Thomas and {Ferreira}, Leonardo and {Foreman-Mackey}, Daniel and {Fox}, Ryan and {Freij}, Nabil and {Garg}, Suyog and {Geda}, Robel and {Glattly}, Lauren and {Gondhalekar}, Yash and {Gordon}, Karl D. and {Grant}, David and {Greenfield}, Perry and {Groener}, Austen M. and {Guest}, Steve and {Gurovich}, Sebastian and {Handberg}, Rasmus and {Hart}, Akeem and {Hatfield-Dodds}, Zac and {Homeier}, Derek and {Hosseinzadeh}, Griffin and {Jenness}, Tim and {Jones}, Craig K. and {Joseph}, Prajwel and {Kalmbach}, J. Bryce and {Karamehmetoglu}, Emir and {Ka{\l}uszy{\'n}ski}, Miko{\l}aj and {Kelley}, Michael S.~P. and {Kern}, Nicholas and {Kerzendorf}, Wolfgang E. and {Koch}, Eric W. and {Kulumani}, Shankar and {Lee}, Antony and {Ly}, Chun and {Ma}, Zhiyuan and {MacBride}, Conor and {Maljaars}, Jakob M. and {Muna}, Demitri and {Murphy}, N.~A. and {Norman}, Henrik and {O'Steen}, Richard and {Oman}, Kyle A. and {Pacifici}, Camilla and {Pascual}, Sergio and {Pascual-Granado}, J. and {Patil}, Rohit R. and {Perren}, Gabriel I. and {Pickering}, Timothy E. and {Rastogi}, Tanuj and {Roulston}, Benjamin R. and {Ryan}, Daniel F. and {Rykoff}, Eli S. and {Sabater}, Jose and {Sakurikar}, Parikshit and {Salgado}, Jes{\'u}s and {Sanghi}, Aniket and {Saunders}, Nicholas and {Savchenko}, Volodymyr and {Schwardt}, Ludwig and {Seifert-Eckert}, Michael and {Shih}, Albert Y. and {Jain}, Anany Shrey and {Shukla}, Gyanendra and {Sick}, Jonathan and {Simpson}, Chris and {Singanamalla}, Sudheesh and {Singer}, Leo P. and {Singhal}, Jaladh and {Sinha}, Manodeep and {Sip{\H{o}}cz}, Brigitta M. and {Spitler}, Lee R. and {Stansby}, David and {Streicher}, Ole and {{\v{S}}umak}, Jani and {Swinbank}, John D. and {Taranu}, Dan S. and {Tewary}, Nikita and {Tremblay}, Grant R. and {de Val-Borro}, Miguel and {Van Kooten}, Samuel J. and {Vasovi{\'c}}, Zlatan and {Verma}, Shresth and {de Miranda Cardoso}, Jos{\'e} Vin{\'\i}cius and {Williams}, Peter K.~G. and {Wilson}, Tom J. and {Winkel}, Benjamin and {Wood-Vasey}, W.~M. and {Xue}, Rui and {Yoachim}, Peter and {Zhang}, Chen and {Zonca}, Andrea and {Astropy Project Contributors}},
        title = "{The Astropy Project: Sustaining and Growing a Community-oriented Open-source Project and the Latest Major Release (v5.0) of the Core Package}",
      journal = {\apj},
         year = 2022,
        month = aug,
       volume = {935},
       number = {2},
          eid = {167},
        pages = {167},
          doi = {10.3847/1538-4357/ac7c74},
archivePrefix = {arXiv},
       eprint = {2206.14220},
 primaryClass = {astro-ph.IM},
       adsurl = {https://ui.adsabs.harvard.edu/abs/2022ApJ...935..167A}
}

@MISC{Bradley2022,
       author = {{Bradley}, Larry and {Sip{\H{o}}cz}, Brigitta and {Robitaille}, Thomas and {Tollerud}, Erik and {Vin{\'\i}cius}, Z{\'e} and {Deil}, Christoph and {Barbary}, Kyle and {Wilson}, Tom J and {Busko}, Ivo and {Donath}, Axel and {G{\"u}nther}, Hans Moritz and {Cara}, Mihai and {Lim}, P.~L. and {Me{\ss}linger}, Sebastian and {Conseil}, Simon and {Bostroem}, Azalee and {Droettboom}, Michael and {Bray}, E.~M. and {Andersen Bratholm}, Lars and {Barentsen}, Geert and {Craig}, Matt and {Rathi}, Shivangee and {Pascual}, Sergio and {Perren}, Gabriel and {Georgiev}, Iskren Y. and {De Val-Borro}, Miguel and {Kerzendorf}, Wolfgang and {Bach}, Yoonsoo P. and {Quint}, Bruno and {Souchereau}, Harrison},
        title = "{astropy/photutils: 1.5.0}",
         year = 2022,
        month = jul,
          eid = {10.5281/zenodo.6825092},
          doi = {10.5281/zenodo.6825092},
      version = {1.5.0},
    publisher = {Zenodo},
       adsurl = {https://ui.adsabs.harvard.edu/abs/2022zndo...6825092B}
}

@ARTICLE{WangWH2022,
       author = {{Wang}, Wei-Hao and {Foucaud}, Sebastien and {Hsieh}, Bau-Ching and {Jian}, Hung-Yu and {Lin}, Lihwai and {Lin}, Yen-Ting and {Coupon}, Jean and {Hashimoto}, Yasuhiro and {Ouchi}, Masami and {Shimasaku}, Kazuhiro and {Ohyama}, Youichi and {Umetsu}, Keiichi and {Wang}, Shiang-Yu and {Chang}, Tzu-Ching},
        title = "{MUSUBI (MegaCam Ultra-deep Survey: u*-band Imaging) Data for the COSMOS and SXDS Fields}",
      journal = {\apjs},
         year = 2022,
        month = jun,
       volume = {260},
       number = {2},
          eid = {54},
        pages = {54},
          doi = {10.3847/1538-4365/ac729e},
archivePrefix = {arXiv},
       eprint = {2205.11546},
 primaryClass = {astro-ph.GA},
       adsurl = {https://ui.adsabs.harvard.edu/abs/2022ApJS..260...54W}
}

@ARTICLE{Weaver2022,
       author = {{Weaver}, J.~R. and {Kauffmann}, O.~B. and {Ilbert}, O. and {McCracken}, H.~J. and {Moneti}, A. and {Toft}, S. and {Brammer}, G. and {Shuntov}, M. and {Davidzon}, I. and {Hsieh}, B.~C. and {Laigle}, C. and {Anastasiou}, A. and {Jespersen}, C.~K. and {Vinther}, J. and {Capak}, P. and {Casey}, C.~M. and {McPartland}, C.~J.~R. and {Milvang-Jensen}, B. and {Mobasher}, B. and {Sanders}, D.~B. and {Zalesky}, L. and {Arnouts}, S. and {Aussel}, H. and {Dunlop}, J.~S. and {Faisst}, A. and {Franx}, M. and {Furtak}, L.~J. and {Fynbo}, J.~P.~U. and {Gould}, K.~M.~L. and {Greve}, T.~R. and {Gwyn}, S. and {Kartaltepe}, J.~S. and {Kashino}, D. and {Koekemoer}, A.~M. and {Kokorev}, V. and {Le F{\`e}vre}, O. and {Lilly}, S. and {Masters}, D. and {Magdis}, G. and {Mehta}, V. and {Peng}, Y. and {Riechers}, D.~A. and {Salvato}, M. and {Sawicki}, M. and {Scarlata}, C. and {Scoville}, N. and {Shirley}, R. and {Silverman}, J.~D. and {Sneppen}, A. and {Smolc̆i{\'c}}, V. and {Steinhardt}, C. and {Stern}, D. and {Tanaka}, M. and {Taniguchi}, Y. and {Teplitz}, H.~I. and {Vaccari}, M. and {Wang}, W. -H. and {Zamorani}, G.},
        title = "{COSMOS2020: A Panchromatic View of the Universe to z{\ensuremath{\sim}}10 from Two Complementary Catalogs}",
      journal = {\apjs},
         year = 2022,
        month = jan,
       volume = {258},
       number = {1},
          eid = {11},
        pages = {11},
          doi = {10.3847/1538-4365/ac3078},
archivePrefix = {arXiv},
       eprint = {2110.13923},
 primaryClass = {astro-ph.GA},
       adsurl = {https://ui.adsabs.harvard.edu/abs/2022ApJS..258...11W}
}

@ARTICLE{Bagley2023,
       author = {{Bagley}, Micaela B. and {Finkelstein}, Steven L. and {Koekemoer}, Anton M. and {Ferguson}, Henry C. and {Arrabal Haro}, Pablo and {Dickinson}, Mark and {Kartaltepe}, Jeyhan S. and {Papovich}, Casey and {P{\'e}rez-Gonz{\'a}lez}, Pablo G. and {Pirzkal}, Nor and {Somerville}, Rachel S. and {Willmer}, Christopher N.~A. and {Yang}, Guang and {Yung}, L.~Y. Aaron and {Fontana}, Adriano and {Grazian}, Andrea and {Grogin}, Norman A. and {Hirschmann}, Michaela and {Kewley}, Lisa J. and {Kirkpatrick}, Allison and {Kocevski}, Dale D. and {Lotz}, Jennifer M. and {Medrano}, Aubrey and {Morales}, Alexa M. and {Pentericci}, Laura and {Ravindranath}, Swara and {Trump}, Jonathan R. and {Wilkins}, Stephen M. and {Calabr{\`o}}, Antonello and {Cooper}, M.~C. and {Costantin}, Luca and {de la Vega}, Alexander and {Hilbert}, Bryan and {Hutchison}, Taylor A. and {Larson}, Rebecca L. and {Lucas}, Ray A. and {McGrath}, Elizabeth J. and {Ryan}, Russell and {Wang}, Xin and {Wuyts}, Stijn},
        title = "{CEERS Epoch 1 NIRCam Imaging: Reduction Methods and Simulations Enabling Early JWST Science Results}",
      journal = {\apjl},
         year = 2023,
        month = mar,
       volume = {946},
       number = {1},
          eid = {L12},
        pages = {L12},
          doi = {10.3847/2041-8213/acbb08},
archivePrefix = {arXiv},
       eprint = {2211.02495},
 primaryClass = {astro-ph.IM},
       adsurl = {https://ui.adsabs.harvard.edu/abs/2023ApJ...946L..12B}
}

@ARTICLE{Casey2023,
       author = {{Casey}, Caitlin M. and {Kartaltepe}, Jeyhan S. and {Drakos}, Nicole E. and {Franco}, Maximilien and {Harish}, Santosh and {Paquereau}, Louise and {Ilbert}, Olivier and {Rose}, Caitlin and {Cox}, Isabella G. and {Nightingale}, James W. and {Robertson}, Brant E. and {Silverman}, John D. and {Koekemoer}, Anton M. and {Massey}, Richard and {McCracken}, Henry Joy and {Rhodes}, Jason and {Akins}, Hollis B. and {Allen}, Natalie and {Amvrosiadis}, Aristeidis and {Arango-Toro}, Rafael C. and {Bagley}, Micaela B. and {Bongiorno}, Angela and {Capak}, Peter L. and {Champagne}, Jaclyn B. and {Chartab}, Nima and {Ch{\'a}vez Ortiz}, {\'O}scar A. and {Chworowsky}, Katherine and {Cooke}, Kevin C. and {Cooper}, Olivia R. and {Darvish}, Behnam and {Ding}, Xuheng and {Faisst}, Andreas L. and {Finkelstein}, Steven L. and {Fujimoto}, Seiji and {Gentile}, Fabrizio and {Gillman}, Steven and {Gould}, Katriona M.~L. and {Gozaliasl}, Ghassem and {Hayward}, Christopher C. and {He}, Qiuhan and {Hemmati}, Shoubaneh and {Hirschmann}, Michaela and {Jahnke}, Knud and {Jin}, Shuowen and {Khostovan}, Ali Ahmad and {Kokorev}, Vasily and {Lambrides}, Erini and {Laigle}, Clotilde and {Larson}, Rebecca L. and {Leung}, Gene C.~K. and {Liu}, Daizhong and {Liaudat}, Tobias and {Long}, Arianna S. and {Magdis}, Georgios and {Mahler}, Guillaume and {Mainieri}, Vincenzo and {Manning}, Sinclaire M. and {Maraston}, Claudia and {Martin}, Crystal L. and {McCleary}, Jacqueline E. and {McKinney}, Jed and {McPartland}, Conor J.~R. and {Mobasher}, Bahram and {Pattnaik}, Rohan and {Renzini}, Alvio and {Rich}, R. Michael and {Sanders}, David B. and {Sattari}, Zahra and {Scognamiglio}, Diana and {Scoville}, Nick and {Sheth}, Kartik and {Shuntov}, Marko and {Sparre}, Martin and {Suzuki}, Tomoko L. and {Talia}, Margherita and {Toft}, Sune and {Trakhtenbrot}, Benny and {Urry}, C. Megan and {Valentino}, Francesco and {Vanderhoof}, Brittany N. and {Vardoulaki}, Eleni and {Weaver}, John R. and {Whitaker}, Katherine E. and {Wilkins}, Stephen M. and {Yang}, Lilan and {Zavala}, Jorge A.},
        title = "{COSMOS-Web: An Overview of the JWST Cosmic Origins Survey}",
      journal = {\apj},
         year = 2023,
        month = sep,
       volume = {954},
       number = {1},
          eid = {31},
        pages = {31},
          doi = {10.3847/1538-4357/acc2bc},
archivePrefix = {arXiv},
       eprint = {2211.07865},
 primaryClass = {astro-ph.GA},
       adsurl = {https://ui.adsabs.harvard.edu/abs/2023ApJ...954...31C}
}

@ARTICLE{Eisenstein2023,
       author = {{Eisenstein}, Daniel J. and {Willott}, Chris and {Alberts}, Stacey and {Arribas}, Santiago and {Bonaventura}, Nina and {Bunker}, Andrew J. and {Cameron}, Alex J. and {Carniani}, Stefano and {Charlot}, Stephane and {Curtis-Lake}, Emma and {D'Eugenio}, Francesco and {Ferruit}, Pierre and {Giardino}, Giovanna and {Hainline}, Kevin and {Hausen}, Ryan and {Jakobsen}, Peter and {Johnson}, Benjamin D. and {Maiolino}, Roberto and {Rauscher}, Bernard J. and {Rieke}, Marcia and {Rieke}, George and {Rix}, Hans-Walter and {Robertson}, Brant and {Stark}, Daniel P. and {Tacchella}, Sandro and {Williams}, Christina C. and {Willmer}, Christopher N.~A. and {Baker}, William M. and {Baum}, Stefi and {Bhatawdekar}, Rachana and {Boyett}, Kristan and {Chen}, Zuyi and {Chevallard}, Jacopo and {Circosta}, Chiara and {Curti}, Mirko and {Danhaive}, A. Lola and {DeCoursey}, Christa and {Endsley}, Ryan and {de Graaff}, Anna and {Dressler}, Alan and {Egami}, Eiichi and {Helton}, Jakob M. and {Hviding}, Raphael E. and {Ji}, Zhiyuan and {Jones}, Gareth C. and {Kumari}, Nimisha and {L{\"u}tzgendorf}, Nora and {Laseter}, Isaac and {Looser}, Tobias J. and {Lyu}, Jianwei and {Maseda}, Michael V. and {Nelson}, Erica and {Parlanti}, Eleonora and {Perna}, Michele and {Pusk{\'a}s}, D{\'a}vid and {Rawle}, Tim and {Rodr{\'\i}guez Del Pino}, Bruno and {Rujopakarn}, Wiphu and {Sandles}, Lester and {Saxena}, Aayush and {Scholtz}, Jan and {Sharpe}, Katherine and {Shivaei}, Irene and {Silcock}, Maddie S. and {Simmonds}, Charlotte and {Skarbinski}, Maya and {Smit}, Renske and {Stone}, Meredith and {Suess}, Katherine A. and {Sun}, Fengwu and {Tang}, Mengtao and {Topping}, Michael W. and {{\"U}bler}, Hannah and {Villanueva}, Natalia C. and {Wallace}, Imaan E.~B. and {Whitler}, Lily and {Witstok}, Joris and {Woodrum}, Charity},
        title = "{Overview of the JWST Advanced Deep Extragalactic Survey (JADES)}",
      journal = {\apjs},
         year = 2026,
        month = mar,
       volume = {283},
       number = {1},
          eid = {6},
        pages = {6},
          doi = {10.3847/1538-4365/ae3163},
archivePrefix = {arXiv},
       eprint = {2306.02465},
 primaryClass = {astro-ph.GA},
       adsurl = {https://ui.adsabs.harvard.edu/abs/2026ApJS..283....6E}
}

@ARTICLE{Papovich2023,
       author = {{Papovich}, Casey and {Cole}, Justin W. and {Yang}, Guang and {Finkelstein}, Steven L. and {Barro}, Guillermo and {Buat}, V{\'e}ronique and {Burgarella}, Denis and {P{\'e}rez-Gonz{\'a}lez}, Pablo G. and {Santini}, Paola and {Seill{\'e}}, Lise-Marie and {Shen}, Lu and {Arrabal Haro}, Pablo and {Bagley}, Micaela B. and {Bell}, Eric F. and {Bisigello}, Laura and {Calabr{\`o}}, Antonello and {Casey}, Caitlin M. and {Castellano}, Marco and {Chworowsky}, Katherine and {Cleri}, Nikko J. and {Costantin}, Luca and {Cooper}, M.~C. and {Dickinson}, Mark and {Ferguson}, Henry C. and {Fontana}, Adriano and {Giavalisco}, Mauro and {Grazian}, Andrea and {Grogin}, Norman A. and {Hathi}, Nimish P. and {Holwerda}, Benne W. and {Hutchison}, Taylor A. and {Kartaltepe}, Jeyhan S. and {Kewley}, Lisa J. and {Kirkpatrick}, Allison and {Kocevski}, Dale D. and {Koekemoer}, Anton M. and {Larson}, Rebecca L. and {Long}, Arianna S. and {Lucas}, Ray A. and {Pentericci}, Laura and {Pirzkal}, Nor and {Ravindranath}, Swara and {Somerville}, Rachel S. and {Trump}, Jonathan R. and {Urbano Stawinski}, Stephanie M. and {Weiner}, Benjamin J. and {Wilkins}, Stephen M. and {Yung}, L.~Y. Aaron and {Zavala}, Jorge A.},
        title = "{CEERS Key Paper. V. Galaxies at 4 < z < 9 Are Bluer than They Appear{\textendash}Characterizing Galaxy Stellar Populations from Rest-frame {\ensuremath{\sim}}1 {\ensuremath{\mu}}m Imaging}",
      journal = {\apjl},
         year = 2023,
        month = jun,
       volume = {949},
       number = {2},
          eid = {L18},
        pages = {L18},
          doi = {10.3847/2041-8213/acc948},
archivePrefix = {arXiv},
       eprint = {2301.00027},
 primaryClass = {astro-ph.GA},
       adsurl = {https://ui.adsabs.harvard.edu/abs/2023ApJ...949L..18P}
}

@ARTICLE{Rieke2023,
       author = {{Rieke}, Marcia J. and {Robertson}, Brant and {Tacchella}, Sandro and {Hainline}, Kevin and {Johnson}, Benjamin D. and {Hausen}, Ryan and {Ji}, Zhiyuan and {Willmer}, Christopher N.~A. and {Eisenstein}, Daniel J. and {Pusk{\'a}s}, D{\'a}vid and {Alberts}, Stacey and {Arribas}, Santiago and {Baker}, William M. and {Baum}, Stefi and {Bhatawdekar}, Rachana and {Bonaventura}, Nina and {Boyett}, Kristan and {Bunker}, Andrew J. and {Cameron}, Alex J. and {Carniani}, Stefano and {Charlot}, Stephane and {Chevallard}, Jacopo and {Chen}, Zuyi and {Curti}, Mirko and {Curtis-Lake}, Emma and {Danhaive}, A. Lola and {DeCoursey}, Christa and {Dressler}, Alan and {Egami}, Eiichi and {Endsley}, Ryan and {Helton}, Jakob M. and {Hviding}, Raphael E. and {Kumari}, Nimisha and {Looser}, Tobias J. and {Lyu}, Jianwei and {Maiolino}, Roberto and {Maseda}, Michael V. and {Nelson}, Erica J. and {Rieke}, George and {Rix}, Hans-Walter and {Sandles}, Lester and {Saxena}, Aayush and {Sharpe}, Katherine and {Shivaei}, Irene and {Skarbinski}, Maya and {Smit}, Renske and {Stark}, Daniel P. and {Stone}, Meredith and {Suess}, Katherine A. and {Sun}, Fengwu and {Topping}, Michael and {{\"U}bler}, Hannah and {Villanueva}, Natalia C. and {Wallace}, Imaan E.~B. and {Williams}, Christina C. and {Willott}, Chris and {Whitler}, Lily and {Witstok}, Joris and {Woodrum}, Charity},
        title = "{JADES Initial Data Release for the Hubble Ultra Deep Field: Revealing the Faint Infrared Sky with Deep JWST NIRCam Imaging}",
      journal = {\apjs},
         year = 2023,
        month = nov,
       volume = {269},
       number = {1},
          eid = {16},
        pages = {16},
          doi = {10.3847/1538-4365/acf44d},
archivePrefix = {arXiv},
       eprint = {2306.02466},
 primaryClass = {astro-ph.GA},
       adsurl = {https://ui.adsabs.harvard.edu/abs/2023ApJS..269...16R}
}

@ARTICLE{Finkelstein2023,
       author = {{Finkelstein}, Steven L. and {Bagley}, Micaela B. and {Ferguson}, Henry C. and {Wilkins}, Stephen M. and {Kartaltepe}, Jeyhan S. and {Papovich}, Casey and {Yung}, L.~Y. Aaron and {Arrabal Haro}, Pablo and {Behroozi}, Peter and {Dickinson}, Mark and {Kocevski}, Dale D. and {Koekemoer}, Anton M. and {Larson}, Rebecca L. and {Le Bail}, Aur{\'e}lien and {Morales}, Alexa M. and {P{\'e}rez-Gonz{\'a}lez}, Pablo G. and {Burgarella}, Denis and {Dav{\'e}}, Romeel and {Hirschmann}, Michaela and {Somerville}, Rachel S. and {Wuyts}, Stijn and {Bromm}, Volker and {Casey}, Caitlin M. and {Fontana}, Adriano and {Fujimoto}, Seiji and {Gardner}, Jonathan P. and {Giavalisco}, Mauro and {Grazian}, Andrea and {Grogin}, Norman A. and {Hathi}, Nimish P. and {Hutchison}, Taylor A. and {Jha}, Saurabh W. and {Jogee}, Shardha and {Kewley}, Lisa J. and {Kirkpatrick}, Allison and {Long}, Arianna S. and {Lotz}, Jennifer M. and {Pentericci}, Laura and {Pierel}, Justin D.~R. and {Pirzkal}, Nor and {Ravindranath}, Swara and {Ryan}, Russell E. and {Trump}, Jonathan R. and {Yang}, Guang and {Bhatawdekar}, Rachana and {Bisigello}, Laura and {Buat}, V{\'e}ronique and {Calabr{\`o}}, Antonello and {Castellano}, Marco and {Cleri}, Nikko J. and {Cooper}, M.~C. and {Croton}, Darren and {Daddi}, Emanuele and {Dekel}, Avishai and {Elbaz}, David and {Franco}, Maximilien and {Gawiser}, Eric and {Holwerda}, Benne W. and {Huertas-Company}, Marc and {Jaskot}, Anne E. and {Leung}, Gene C.~K. and {Lucas}, Ray A. and {Mobasher}, Bahram and {Pandya}, Viraj and {Tacchella}, Sandro and {Weiner}, Benjamin J. and {Zavala}, Jorge A.},
        title = "{CEERS Key Paper. I. An Early Look into the First 500 Myr of Galaxy Formation with JWST}",
      journal = {\apjl},
         year = 2023,
        month = mar,
       volume = {946},
       number = {1},
          eid = {L13},
        pages = {L13},
          doi = {10.3847/2041-8213/acade4},
archivePrefix = {arXiv},
       eprint = {2211.05792},
 primaryClass = {astro-ph.GA},
       adsurl = {https://ui.adsabs.harvard.edu/abs/2023ApJ...946L..13F}
}

@ARTICLE{Kartaltepe2023,
       author = {{Kartaltepe}, Jeyhan S. and {Rose}, Caitlin and {Vanderhoof}, Brittany N. and {McGrath}, Elizabeth J. and {Costantin}, Luca and {Cox}, Isabella G. and {Yung}, L.~Y. Aaron and {Kocevski}, Dale D. and {Wuyts}, Stijn and {Ferguson}, Henry C. and {Bagley}, Micaela B. and {Finkelstein}, Steven L. and {Amor{\'\i}n}, Ricardo O. and {Andrews}, Brett H. and {Arrabal Haro}, Pablo and {Backhaus}, Bren E. and {Behroozi}, Peter and {Bisigello}, Laura and {Calabr{\`o}}, Antonello and {Casey}, Caitlin M. and {Coogan}, Rosemary T. and {Cooper}, M.~C. and {Croton}, Darren and {de la Vega}, Alexander and {Dickinson}, Mark and {Fontana}, Adriano and {Franco}, Maximilien and {Grazian}, Andrea and {Grogin}, Norman A. and {Hathi}, Nimish P. and {Holwerda}, Benne W. and {Huertas-Company}, Marc and {Iyer}, Kartheik G. and {Jogee}, Shardha and {Jung}, Intae and {Kewley}, Lisa J. and {Kirkpatrick}, Allison and {Koekemoer}, Anton M. and {Liu}, James and {Lotz}, Jennifer M. and {Lucas}, Ray A. and {Newman}, Jeffrey A. and {Pacifici}, Camilla and {Pandya}, Viraj and {Papovich}, Casey and {Pentericci}, Laura and {P{\'e}rez-Gonz{\'a}lez}, Pablo G. and {Petersen}, Jayse and {Pirzkal}, Nor and {Rafelski}, Marc and {Ravindranath}, Swara and {Simons}, Raymond C. and {Snyder}, Gregory F. and {Somerville}, Rachel S. and {Stanway}, Elizabeth R. and {Straughn}, Amber N. and {Tacchella}, Sandro and {Trump}, Jonathan R. and {Vega-Ferrero}, Jes{\'u}s and {Wilkins}, Stephen M. and {Yang}, Guang and {Zavala}, Jorge A.},
        title = "{CEERS Key Paper. III. The Diversity of Galaxy Structure and Morphology at z = 3-9 with JWST}",
      journal = {\apjl},
         year = 2023,
        month = mar,
       volume = {946},
       number = {1},
          eid = {L15},
        pages = {L15},
          doi = {10.3847/2041-8213/acad01},
archivePrefix = {arXiv},
       eprint = {2210.14713},
 primaryClass = {astro-ph.GA},
       adsurl = {https://ui.adsabs.harvard.edu/abs/2023ApJ...946L..15K}
}

@ARTICLE{Song2023,
       author = {{Song}, Jie and {Fang}, GuanWen and {Lin}, Zesen and {Gu}, Yizhou and {Kong}, Xu},
        title = "{Solution to the Conflict between the Estimations of Resolved and Unresolved Galaxy Stellar Mass from the Perspective of JWST}",
      journal = {\apj},
         year = 2023,
        month = nov,
       volume = {958},
       number = {1},
          eid = {82},
        pages = {82},
          doi = {10.3847/1538-4357/ad0365},
archivePrefix = {arXiv},
       eprint = {2310.12228},
 primaryClass = {astro-ph.GA},
       adsurl = {https://ui.adsabs.harvard.edu/abs/2023ApJ...958...82S}
}

@ARTICLE{Talia2023,
       author = {{Talia}, M. and {Schreiber}, C. and {Garilli}, B. and {Pentericci}, L. and {Pozzetti}, L. and {Zamorani}, G. and {Cullen}, F. and {Moresco}, M. and {Calabr{\`o}}, A. and {Castellano}, M. and {Fynbo}, J.~P.~U. and {Guaita}, L. and {Marchi}, F. and {Mascia}, S. and {McLure}, R. and {Mignoli}, M. and {Pompei}, E. and {Vanzella}, E. and {Bongiorno}, A. and {Vietri}, G. and {Amor{\'\i}n}, R.~O. and {Bolzonella}, M. and {Carnall}, A.~C. and {Cimatti}, A. and {Cresci}, G. and {Cristiani}, S. and {Cucciati}, O. and {Dunlop}, J.~S. and {Fontanot}, F. and {Franzetti}, P. and {Gargiulo}, A. and {Hamadouche}, M.~L. and {Hathi}, N.~P. and {Hibon}, P. and {Iovino}, A. and {Koekemoer}, A.~M. and {Mannucci}, F. and {McLeod}, D.~J. and {Saldana-Lopez}, A.},
        title = "{The VANDELS ESO public spectroscopic survey: The spectroscopic measurements catalogue}",
      journal = {\aap},
         year = 2023,
        month = oct,
       volume = {678},
          eid = {A25},
        pages = {A25},
          doi = {10.1051/0004-6361/202346293},
archivePrefix = {arXiv},
       eprint = {2309.14436},
 primaryClass = {astro-ph.GA},
       adsurl = {https://ui.adsabs.harvard.edu/abs/2023A&A...678A..25T}
}

@ARTICLE{YangG2023,
       author = {{Yang}, G. and {Caputi}, K.~I. and {Papovich}, C. and {Arrabal Haro}, P. and {Bagley}, M.~B. and {Behroozi}, P. and {Bell}, E.~F. and {Bisigello}, L. and {Buat}, V. and {Burgarella}, D. and {Cheng}, Y. and {Cleri}, N.~J. and {Dav{\'e}}, R. and {Dickinson}, M. and {Elbaz}, D. and {Ferguson}, H.~C. and {Finkelstein}, S.~L. and {Grogin}, N.~A. and {Hathi}, N.~P. and {Hirschmann}, M. and {Holwerda}, B.~W. and {Huertas-Company}, M. and {Hutchison}, T.~A. and {Iani}, E. and {Kartaltepe}, J.~S. and {Kirkpatrick}, A. and {Kocevski}, D.~D. and {Koekemoer}, A.~M. and {Kokorev}, V. and {Larson}, R.~L. and {Lucas}, R.~A. and {P{\'e}rez-Gonz{\'a}lez}, P.~G. and {Rinaldi}, P. and {Shen}, L. and {Trump}, J.~R. and {de la Vega}, A. and {Yung}, L.~Y.~A. and {Zavala}, J.~A.},
        title = "{CEERS Key Paper. VI. JWST/MIRI Uncovers a Large Population of Obscured AGN at High Redshifts}",
      journal = {\apjl},
         year = 2023,
        month = jun,
       volume = {950},
       number = {1},
          eid = {L5},
        pages = {L5},
          doi = {10.3847/2041-8213/acd639},
archivePrefix = {arXiv},
       eprint = {2303.11736},
 primaryClass = {astro-ph.GA},
       adsurl = {https://ui.adsabs.harvard.edu/abs/2023ApJ...950L...5Y}
}

@software{Bushouse2024,
       author = {{Bushouse}, Howard and {Eisenhamer}, Jonathan and {Dencheva}, Nadia and {Davies}, James and {Greenfield}, Perry and {Morrison}, Jane and {Hodge}, Phil and {Simon}, Bernie and {Grumm}, David and {Droettboom}, Michael and {Slavich}, Edward and {Sosey}, Megan and {Pauly}, Tyler and {Miller}, Todd and {Jedrzejewski}, Robert and {Hack}, Warren and {Davis}, David and {Crawford}, Steven and {Law}, David and {Gordon}, Karl and {Regan}, Michael and {Cara}, Mihai and {MacDonald}, Ken and {Bradley}, Larry and {Shanahan}, Clare and {Jamieson}, William and {Teodoro}, Mairan and {Williams}, Thomas and {Pena-Guerrero}, Maria},
        title = "{JWST Calibration Pipeline}",
         year = 2024,
        month = jan,
          eid = {10.5281/zenodo.10569856},
          doi = {10.5281/zenodo.10569856},
      version = {1.13.4},
    publisher = {Zenodo},
       adsurl = {https://ui.adsabs.harvard.edu/abs/2024zndo..10569856B}
}

@ARTICLE{Heintz2024,
       author = {{Heintz}, Kasper E. and {Watson}, Darach and {Brammer}, Gabriel and {Vejlgaard}, Simone and {Hutter}, Anne and {Strait}, Victoria B. and {Matthee}, Jorryt and {Oesch}, Pascal A. and {Jakobsson}, P{\'a}ll and {Tanvir}, Nial R. and {Laursen}, Peter and {Naidu}, Rohan P. and {Mason}, Charlotte A. and {Killi}, Meghana and {Jung}, Intae and {Hsiao}, Tiger Yu-Yang and {Abdurro'uf} and {Coe}, Dan and {Arrabal Haro}, Pablo and {Finkelstein}, Steven L. and {Toft}, Sune},
        title = "{Strong damped Lyman-{\ensuremath{\alpha}} absorption in young star-forming galaxies at redshifts 9 to 11}",
      journal = {Science},
         year = 2024,
        month = may,
       volume = {384},
       number = {6698},
        pages = {890-894},
          doi = {10.1126/science.adj0343},
archivePrefix = {arXiv},
       eprint = {2306.00647},
 primaryClass = {astro-ph.GA},
       adsurl = {https://ui.adsabs.harvard.edu/abs/2024Sci...384..890H}
}

@ARTICLE{degraaff2024,
       author = {{de Graaff}, Anna and {Brammer}, Gabriel and {Weibel}, Andrea and {Lewis}, Zach and {Maseda}, Michael V. and {Oesch}, Pascal A. and {Bezanson}, Rachel and {Boogaard}, Leindert A. and {Cleri}, Nikko J. and {Cooper}, Olivia R. and {Gottumukkala}, Rashmi and {Greene}, Jenny E. and {Hirschmann}, Michaela and {Hviding}, Raphael E. and {Katz}, Harley and {Labb{\'e}}, Ivo and {Leja}, Joel and {Matthee}, Jorryt and {McConachie}, Ian and {Miller}, Tim B. and {Naidu}, Rohan P. and {Price}, Sedona H. and {Rix}, Hans-Walter and {Setton}, David J. and {Suess}, Katherine A. and {Wang}, Bingjie and {Whitaker}, Katherine E. and {Williams}, Christina C.},
        title = "{RUBIES: A complete census of the bright and red distant Universe with JWST/NIRSpec}",
      journal = {\aap},
         year = 2025,
        month = may,
       volume = {697},
          eid = {A189},
        pages = {A189},
          doi = {10.1051/0004-6361/202452186},
archivePrefix = {arXiv},
       eprint = {2409.05948},
 primaryClass = {astro-ph.GA},
       adsurl = {https://ui.adsabs.harvard.edu/abs/2025A&A...697A.189D}
}

@ARTICLE{GaoZK2024,
       author = {{Gao}, Zhen-Kai and {Lim}, Chen-Fatt and {Wang}, Wei-Hao and {Chen}, Chian-Chou and {Smail}, Ian and {Chapman}, Scott C. and {Zheng}, Xian Zhong and {Shim}, Hyunjin and {Kodama}, Tadayuki and {Ao}, Yiping and {Chang}, Siou-Yu and {Clements}, David L. and {Dunlop}, James S. and {Ho}, Luis C. and {Hsu}, Yun-Hsin and {Hwang}, Chorng-Yuan and {Hwang}, Ho Seong and {Koprowski}, M.~P. and {Scott}, Douglas and {Serjeant}, Stephen and {Toba}, Yoshiki and {Urquhart}, Sheona A.},
        title = "{SCUBA-2 Ultra Deep Imaging EAO Survey (STUDIES). V. Confusion-limited Submillimeter Galaxy Number Counts at 450 {\ensuremath{\mu}}m and Data Release for the COSMOS Field}",
      journal = {\apj},
         year = 2024,
        month = aug,
       volume = {971},
       number = {1},
          eid = {117},
        pages = {117},
          doi = {10.3847/1538-4357/ad53c1},
archivePrefix = {arXiv},
       eprint = {2405.20616},
 primaryClass = {astro-ph.GA},
       adsurl = {https://ui.adsabs.harvard.edu/abs/2024ApJ...971..117G}
}

@ARTICLE{Ji2024,
       author = {{Ji}, Zhiyuan and {Williams}, Christina C. and {Tacchella}, Sandro and {Suess}, Katherine A. and {Baker}, William M. and {Alberts}, Stacey and {Bunker}, Andrew J. and {Johnson}, Benjamin D. and {Robertson}, Brant and {Sun}, Fengwu and {Eisenstein}, Daniel J. and {Rieke}, Marcia and {Maseda}, Michael V. and {Hainline}, Kevin and {Hausen}, Ryan and {Rieke}, George and {Willmer}, Christopher N.~A. and {Egami}, Eiichi and {Shivaei}, Irene and {Carniani}, Stefano and {Charlot}, Stephane and {Chevallard}, Jacopo and {Curtis-Lake}, Emma and {Looser}, Tobias J. and {Maiolino}, Roberto and {Willott}, Chris and {Chen}, Zuyi and {Helton}, Jakob M. and {Lyu}, Jianwei and {Nelson}, Erica and {Bhatawdekar}, Rachana and {Boyett}, Kristan and {Sandles}, Lester},
        title = "{JADES + JEMS: A Detailed Look at the Buildup of Central Stellar Cores and Suppression of Star Formation in Galaxies at Redshifts 3 < z < 4.5}",
      journal = {\apj},
         year = 2024,
        month = oct,
       volume = {974},
       number = {1},
          eid = {135},
        pages = {135},
          doi = {10.3847/1538-4357/ad6e7f},
archivePrefix = {arXiv},
       eprint = {2305.18518},
 primaryClass = {astro-ph.GA},
       adsurl = {https://ui.adsabs.harvard.edu/abs/2024ApJ...974..135J}
}

@ARTICLE{Mehta2024,
       author = {{Mehta}, Vihang and {Rafelski}, Marc and {Sunnquist}, Ben and {Teplitz}, Harry I. and {Scarlata}, Claudia and {Wang}, Xin and {Fontana}, Adriano and {Hathi}, Nimish P. and {Iyer}, Kartheik G. and {Alavi}, Anahita and {Colbert}, James and {Grogin}, Norman and {Koekemoer}, Anton and {Nedkova}, Kalina V. and {Hayes}, Matthew and {Prichard}, Laura and {Siana}, Brian and {Smith}, Brent M. and {Windhorst}, Rogier and {Ashcraft}, Teresa and {Bagley}, Micaela and {Baronchelli}, Ivano and {Barro}, Guillermo and {Blanche}, Alex and {Broussard}, Adam and {Carleton}, Timothy and {Chartab}, Nima and {Codoreanu}, Alex and {Cohen}, Seth and {Conselice}, Christopher and {Dai}, Y. Sophia and {Darvish}, Behnam and {Dav{\'e}}, Romeel and {Degroot}, Laura and {de Mello}, Duilia and {Dickinson}, Mark and {Emami}, Najmeh and {Ferguson}, Henry and {Ferreira}, Leonardo and {Finkelstein}, Keely and {Finkelstein}, Steven and {Gardner}, Jonathan P. and {Gawiser}, Eric and {Gburek}, Timothy and {Giavalisco}, Mauro and {Grazian}, Andrea and {Gronwall}, Caryl and {Guo}, Yicheng and {Arrabal Haro}, Pablo and {Hemmati}, Shoubaneh and {Howell}, Justin and {Jansen}, Rolf A. and {Ji}, Zhiyuan and {Kaviraj}, Sugata and {Kim}, Keunho J. and {Kurczynski}, Peter and {Lazar}, Ilin and {Lucas}, Ray A. and {MacKenty}, John and {Mantha}, Kameswara Bharadwaj and {Martin}, Alec and {Martin}, Garreth and {McCabe}, Tyler and {Mobasher}, Bahram and {Morales}, Alexa M. and {O'Connell}, Robert and {Olsen}, Charlotte and {Otteson}, Lillian and {Ravindranath}, Swara and {Redshaw}, Caleb and {Rutkowski}, Michael and {Robertson}, Brant and {Sattari}, Zahra and {Soto}, Emmaris and {Sun}, Lei and {Taamoli}, Sina and {Vanzella}, Eros and {Yung}, L.~Y. Aaron and {Zabelle}, Bonnabelle and {UVCANDELS Team}},
        title = "{UVCANDELS: Catalogs of Photometric Redshifts and Galaxy Physical Properties}",
      journal = {\apjs},
         year = 2024,
        month = nov,
       volume = {275},
       number = {1},
          eid = {17},
        pages = {17},
          doi = {10.3847/1538-4365/ad7d8f},
archivePrefix = {arXiv},
       eprint = {2410.16404},
 primaryClass = {astro-ph.GA},
       adsurl = {https://ui.adsabs.harvard.edu/abs/2024ApJS..275...17M}
}

@ARTICLE{Merlin2024,
       author = {{Merlin}, E. and {Santini}, P. and {Paris}, D. and {Castellano}, M. and {Fontana}, A. and {Treu}, T. and {Finkelstein}, S.~L. and {Dunlop}, J.~S. and {Arrabal Haro}, P. and {Bagley}, M. and {Boyett}, K. and {Calabr{\`o}}, A. and {Correnti}, M. and {Davis}, K. and {Dickinson}, M. and {Donnan}, C.~T. and {Ferguson}, H.~C. and {Fortuni}, F. and {Giavalisco}, M. and {Glazebrook}, K. and {Grazian}, A. and {Grogin}, N.~A. and {Hathi}, N. and {Hirschmann}, M. and {Kartaltepe}, J.~S. and {Kewley}, L.~J. and {Kirkpatrick}, A. and {Kocevski}, D.~D. and {Koekemoer}, A.~M. and {Leung}, G. and {Lotz}, J.~M. and {Lucas}, R.~A. and {Magee}, D.~K. and {Marchesini}, D. and {Mascia}, S. and {McLeod}, D.~J. and {McLure}, R.~J. and {Nanayakkara}, T. and {Napolitano}, L. and {Nonino}, M. and {Papovich}, C. and {Pentericci}, L. and {P{\'e}rez-Gonz{\'a}lez}, P.~G. and {Pirzkal}, N. and {Ravindranath}, S. and {Roberts-Borsani}, G. and {Somerville}, R.~S. and {Trenti}, M. and {Trump}, J.~R. and {Vulcani}, B. and {Wang}, X. and {Watson}, P.~J. and {Wilkins}, S.~M. and {Yang}, G. and {Yung}, L.~Y.~A.},
        title = "{ASTRODEEP-JWST: NIRCam-HST multi-band photometry and redshifts for half a million sources in six extragalactic deep fields}",
      journal = {\aap},
         year = 2024,
        month = nov,
       volume = {691},
          eid = {A240},
        pages = {A240},
          doi = {10.1051/0004-6361/202451409},
archivePrefix = {arXiv},
       eprint = {2409.00169},
 primaryClass = {astro-ph.GA},
       adsurl = {https://ui.adsabs.harvard.edu/abs/2024A&A...691A.240M}
}

@ARTICLE{Rachel2024,
       author = {{Bezanson}, Rachel and {Labbe}, Ivo and {Whitaker}, Katherine E. and {Leja}, Joel and {Price}, Sedona H. and {Franx}, Marijn and {Brammer}, Gabriel and {Marchesini}, Danilo and {Zitrin}, Adi and {Wang}, Bingjie and {Weaver}, John R. and {Furtak}, Lukas J. and {Atek}, Hakim and {Coe}, Dan and {Cutler}, Sam E. and {Dayal}, Pratika and {van Dokkum}, Pieter and {Feldmann}, Robert and {F{\"o}rster Schreiber}, Natascha M. and {Fujimoto}, Seiji and {Geha}, Marla and {Glazebrook}, Karl and {de Graaff}, Anna and {Greene}, Jenny E. and {Juneau}, St{\'e}phanie and {Kassin}, Susan and {Kriek}, Mariska and {Khullar}, Gourav and {Maseda}, Michael and {Mowla}, Lamiya A. and {Muzzin}, Adam and {Nanayakkara}, Themiya and {Nelson}, Erica J. and {Oesch}, Pascal A. and {Pacifici}, Camilla and {Pan}, Richard and {Papovich}, Casey and {Setton}, David J. and {Shapley}, Alice E. and {Smit}, Renske and {Stefanon}, Mauro and {Taylor}, Edward N. and {Williams}, Christina C.},
        title = "{The JWST UNCOVER Treasury Survey: Ultradeep NIRSpec and NIRCam Observations before the Epoch of Reionization}",
      journal = {\apj},
         year = 2024,
        month = oct,
       volume = {974},
       number = {1},
          eid = {92},
        pages = {92},
          doi = {10.3847/1538-4357/ad66cf},
archivePrefix = {arXiv},
       eprint = {2212.04026},
 primaryClass = {astro-ph.GA},
       adsurl = {https://ui.adsabs.harvard.edu/abs/2024ApJ...974...92B}
}

@ARTICLE{Sun2024,
       author = {{Sun}, Hanwen and {Wang}, Tao and {Xu}, Ke and {Daddi}, Emanuele and {Gu}, Qing and {Kodama}, Tadayuki and {Zanella}, Anita and {Elbaz}, David and {Tanaka}, Ichi and {Gobat}, Raphael and {Guo}, Qi and {Han}, Jiaxin and {Lu}, Shiying and {Zhou}, Luwenjia},
        title = "{JWST's First Glimpse of a $z > 2$ Forming Cluster Reveals a Top-heavy Stellar Mass Function}",
      journal = {\apjl},
         year = 2024,
        month = jun,
       volume = {967},
       number = {2},
          eid = {L34},
        pages = {L34},
          doi = {10.3847/2041-8213/ad4986},
archivePrefix = {arXiv},
       eprint = {2403.05248},
 primaryClass = {astro-ph.GA},
       adsurl = {https://ui.adsabs.harvard.edu/abs/2024ApJ...967L..34S}
}

@ARTICLE{Liu2026,
       author = {{Liu}, Weiyang and {Jiang}, Linhua},
        title = "{A high rate of foreground contaminants toward high-redshift galaxies resolved by JWST}",
      journal = {arXiv e-prints},
         year = 2026,
        month = feb,
          eid = {arXiv:2602.04179},
        pages = {arXiv:2602.04179},
          doi = {10.48550/arXiv.2602.04179},
archivePrefix = {arXiv},
       eprint = {2602.04179},
 primaryClass = {astro-ph.GA},
       adsurl = {https://ui.adsabs.harvard.edu/abs/2026arXiv260204179L}
}

%%%%%%%%%%%%%%%%%%%%%%%%%%%%%%%%%%%%%%%%%%%%%%%%%%%%%%%%%%%%%%%%%%%%%%%%%%%%%%%%
\appendix

\section{Content of Data Release}
\label{sec:release}
The first data release of ULTIMATE deblending includes two catalogs for each field. The first catalog lists the basic information of source detection and the results of multi-wavelength photometry. The parameters listed in this catalog are shown in Table~\ref{Tab:cat_phot}. The second catalog (Table~\ref{Tab:cat_phy}) lists the redshifts and physical properties obtained from SED fitting. In addition to the catalogs, we also provide the JWST mosaics, error maps, PSFs, and the convolution kernels used for PSF match on \dataset[http://www.taoofcosmos.space/ultimate/]{http://www.taoofcosmos.space/ultimate/}.
%These catalogs will be publicly released on \dataset[106.14.163.220/ultimate/]{http://106.14.163.220/ultimate/} after the acceptance of this paper.
\begin{table*}[!tbh]\small
\centering
\begin{minipage}[center]{\textwidth}
\centering
\caption{Parameters listed in the photometric catalog\label{Tab:cat_phot}}
\begin{tabular}{lcc}
\hline\hline
Name  & Units  & Description \\
\hline
ID &  & Identifier of the JWST-detected sources \\
RA & deg & Right ascension \\
Dec & deg & Declination \\
X\_image & pixel & Positions in JWST/NIRCam mosaics \\
Y\_image & pixel & Positions in JWST/NIRCam mosaics \\
A\_image & pixel$^a$ & Profile RMS along major axis \\
B\_image & pixel & Profile RMS along minor axis \\
THETA\_IMAGE & deg & Position angle (counterclockwise/x) \\
KRON\_RADIUS &  & Kron apertures in units of A and B \\
CLASS\_STAR &  & S/G classifier output from source-extractor \\
flag\_sextractor & & Flags given by source-extractor \\
flag\_kron & & 1: Kron apertures considered as unreliable; 0: Normal (see Section~\ref{sec:aperture_photometry})\\
flag\_TPHOT & & $>0$: TPHOT photometry are unreliable; 0: Normal (see Section~\ref{sec:tphot_photometry})\\
flux\_F444W\_0.4/0.2 &  & Ratio of uncorrected F444W flux densities in apertures with $d = 0.4"$ and $d = 0.2"$ $^b$ \\
flux\_Band\_tphot & $\mu$Jy & Total flux densities in the low-resolution band ``Band"$^c$ fitted by TPHOT\\
fluxerr\_Band\_tphot & $\mu$Jy & Corrected flux density uncertainties in the low-resolution band ``Band"\\
flux\_Band\_aper0.2 & $\mu$Jy & Flux densities in circular aperture ($d = 0.2"$) with aperture corrections in the band ``Band"\\
fluxerr\_Band\_aper0.2 & $\mu$Jy & Corrected flux density uncertainties in circular aperture ($d = 0.2"$) in the band ``Band"\\
flux\_Band\_aper0.3 & $\mu$Jy & Flux densities in circular aperture ($d = 0.3"$) with aperture corrections in the band ``Band"\\
fluxerr\_Band\_aper0.3 & $\mu$Jy & Corrected flux density uncertainties in circular aperture ($d = 0.3"$) in the band ``Band"\\
flux\_Band\_aper0.4 & $\mu$Jy & Flux densities in circular aperture ($d = 0.4"$) with aperture corrections in the band ``Band"\\
fluxerr\_Band\_aper0.4 & $\mu$Jy & Corrected flux density uncertainties in circular aperture ($d = 0.4"$) in the band ``Band"\\
flux\_Band\_aper0.5 & $\mu$Jy & Flux densities in circular aperture ($d = 0.5"$) with aperture corrections in the band ``Band"\\
fluxerr\_Band\_aper0.5 & $\mu$Jy & Corrected flux density uncertainties in circular aperture ($d = 0.5"$) in the band ``Band"\\
flux\_Band\_aper0.7 & $\mu$Jy & Flux densities in circular aperture ($d = 0.7"$) with aperture corrections in the band ``Band"\\
fluxerr\_Band\_aper0.7 & $\mu$Jy & Corrected flux density uncertainties in circular aperture ($d = 0.7"$) in the band ``Band"\\
flux\_Band\_aper1.0 & $\mu$Jy & Flux densities in circular aperture ($d = 1.0"$) with aperture corrections in the band ``Band"\\
fluxerr\_Band\_aper1.0 & $\mu$Jy & Corrected flux density uncertainties in circular aperture ($d = 1.0"$) in the band ``Band"\\
flux\_Band\_kron & $\mu$Jy & Flux densities in Kron aperture ($k = 2.5$) with aperture corrections in the band ``Band"\\
fluxerr\_Band\_kron & $\mu$Jy & Corrected flux density uncertainties in Kron aperture ($k = 2.5$) in the band ``Band"\\
flux\_Band\_best & $\mu$Jy & The best estimated total flux densities in the band ``Band"\\
fluxerr\_Band\_best & $\mu$Jy & Uncertainties of the best estimated total flux densities in the band ``Band"\\
\hline
\end{tabular}
\begin{flushleft}
{\sc Note.} --- 
($a$) Pixel scale is 0.04 arcseconds. ($b$) Following \citet{Labbe2025}, sources with flux\_F444W\_0.4/0.2 $<1.7$ can be considered as compact sources. ($c$) ``Band" is the name of each band listed in Table~\ref{tab:band_infos}.
\end{flushleft}
\end{minipage}
\end{table*}

\begin{table*}[!tbh]\small
\centering
\begin{minipage}[center]{\textwidth}
\centering
\caption{Parameters listed in the physical catalog\label{Tab:cat_phy}}
\begin{tabular}{lcc}
\hline\hline
Name  & Units  & Description \\
\hline
ID &  & Identifier of the JWST-detected sources \\
RA & deg & Right ascension \\
Dec & deg & Declination \\
zphot &  & Photometric redshift fitted by Eazy \\
zspec &  & Spectroscopic redshift (see Section~\ref{sec:z_estimation}) \\
zspec\_ref &  & Description of the spectroscopic redshift \\
zbest$^a$ &  & The redshift used to obtain the following parameters \\
delayed:age\_16 &Gyr&the 16th percentile (lower limit) of time since star formation began\\
delayed:age\_50 &Gyr&the 50th percentile (best estimation) of time since star formation began\\
delayed:age\_84 &Gyr&the 84th percentile (upper limit) of time since star formation began\\
delayed:tau\_16 &Gyr&the 16th percentile of the timescale of the delayed star-formation history\\
delayed:tau\_50 &Gyr&the 50th percentile of the timescale of the delayed star-formation history\\
delayed:tau\_84 &Gyr&the 84th percentile of the timescale of the delayed star-formation history\\
dust:Av\_16 &magnitude&the 16th percentile of the global absolute attenuation in the V band\\
dust:Av\_50 &magnitude&the 50th percentile of the global absolute attenuation in the V band\\
dust:Av\_84 &magnitude&the 84th percentile of the global absolute attenuation in the V band\\
dust:Av,b\_16 &magnitude&the 16th percentile of the Av for stars in birth clouds\\
dust:Av,b\_50 &magnitude&the 50th percentile of the Av for stars in birth clouds\\
dust:Av,b\_84 &magnitude&the 84th percentile of the Av for stars in birth clouds\\
stellar\_mass\_16& &the 16th percentile of the log of the stellar mass (in $\rm M_\odot$ unit)\\
stellar\_mass\_50& &the 50th percentile of the log of the stellar mass (in $\rm M_\odot$ unit)\\
stellar\_mass\_84& &the 84th percentile of the log of the stellar mass (in $\rm M_\odot$ unit)\\
sfr\_16&$\rm M_\odot~yr^{-1}$&the 16th percentile of the star-formation rate\\
sfr\_50&$\rm M_\odot~yr^{-1}$&the 50th percentile of the star-formation rate\\
sfr\_84&$\rm M_\odot~yr^{-1}$&the 84th percentile of the star-formation rate\\
UV\_colour\_16& &the 16th percentile of the rest-frame U-V color\\
UV\_colour\_50& &the 50th percentile of the rest-frame U-V color\\
UV\_colour\_84& &the 84th percentile of the rest-frame U-V color\\
VJ\_colour\_16& &the 16th percentile of the rest-frame V-J color\\
VJ\_colour\_50& &the 50th percentile of the rest-frame V-J color\\
VJ\_colour\_84& &the 84th percentile of the rest-frame V-J color\\
%KE
\hline
\end{tabular}
\begin{flushleft}
{\sc Note.} --- 
($a$) When the spectroscopic redshift is available, zbest $=$ zspec, otherwise zbest $=$ zphot.
\end{flushleft}
\end{minipage}
\end{table*}

\end{document}